\documentclass[twoside, a4paper, notoc]{JHEP3}
\usepackage{amsfonts}
\usepackage{amssymb}
\usepackage{comment}
\usepackage{epsfig}
\usepackage{graphicx}
\usepackage{amsmath}

\newcommand{\be}{\begin{equation}}
\newcommand{\ee}{\end{equation}}
\newcommand{\bea}{\begin{eqnarray}}
\newcommand{\eea}{\end{eqnarray}}

\newcommand{\mc}{\mathcal}

\newcommand{\beqa}{\begin{eqnarray}}
\newcommand{\eeqa}{\end{eqnarray}}

\newtheorem{theorem}{Theorem}

\newtheorem{claim}[theorem]{Claim}

\newenvironment{proof}[1][Proof]{\noindent\textbf{#1.} }{\ \rule{0.5em}{0.5em}}

\title{General Analysis of LARGE Volume Scenarios with String Loop Moduli
Stabilisation}
\author{Michele Cicoli, Joseph P. Conlon and Fernando Quevedo \\
DAMTP, Centre for Mathematical Sciences, \\
Wilberforce Road, Cambridge, CB3 0WA, UK. \\
Email: \email{M.Cicoli@damtp.cam.ac.uk,
J.P.Conlon@damtp.cam.ac.uk, F.Quevedo@damtp.cam.ac.uk}}

\abstract{We study the topological conditions for general
Calabi-Yaus to get a non-supersymmetric AdS exponentially large
volume minimum of the scalar potential in flux compactifications
of IIB string theory. We show that negative Euler number and the
existence of at least one blow-up mode resolving point-like
singularities are necessary and sufficient conditions for moduli
stabilisation with exponentially large volumes. We also analyse
the general effects of string loop corrections on this scenario.
While the combination of $\alpha'$ and nonperturbative corrections
are sufficient to stabilise blow-up modes and the overall volume,
quantum corrections are needed to stabilise other directions
transverse to the overall volume. This allows exponentially large
volume minima to be realised for fibration Calabi-Yaus, with the
various moduli of the fibration all being stabilised at
exponentially large values. String loop corrections may also play
a role in stabilising 4-cycles which support chiral matter and
cannot enter directly into the non-perturbative superpotential. We
illustrate these ideas by studying the scalar potential for
various Calabi-Yau three-folds including $K3$ fibrations and
briefly discuss the potential phenomenological and cosmological
implications of our results.}

\preprint{DAMTP-2008-16}


\begin{document}

\tableofcontents

\bigskip

\section{Introduction}

In constructing string models that can be compared to experiment,
it is essential to understand the vacuum state. It is the vacuum
state whose properties will give rise to, for example, low-energy
supersymmetry and the pattern of Yukawa couplings. For a long time
string compactifications were plagued by the problem of moduli.
These were massless scalars that parameterised the vacuum and
whose VEVs entered into the value of physical observables. The
existence of moduli leads to fifth forces and also makes the
vacuum inherently unpredictive. Perturbative and nonperturbative
quantum effects, generically present in $N=1$ supersymmetric
theories, were expected to lead to moduli stabilisation but for a
long time no controlled procedure existed.

This situation improved remarkably after the systematic study of
flux compactifications and their impact on moduli stabilisation
\cite{gkp} (for reviews see \cite{dk, hepth0610327}) Fluxes are
particularly powerful in type IIB string compactifications, as
there the backreaction is well-controlled and gives rise only to
warped Calabi-Yau metrics. Furthermore, the combination of
nonperturbative effects \cite{witten, kklt} with perturbative
$\alpha'$ corrections \cite{hepth0204254, hepth0408054,
hepth0502058} has allowed a general study of the large volume
scalar potential for arbitrary Calabi-Yau compactifications and
has led to the discovery of exponentially large volume
compactifications \cite{hepth0502058, hepth0505076} with very
interesting phenomenological and cosmological implications.

If $\alpha'$ corrections can play a significant role in moduli
stabilisation in the phenomenologically relevant regime of large
volume and weak coupling, it is natural to wonder whether $g_s$
corrections may also have a significant effect. At first sight
this seems unavoidable, as at large volume the corrections to the
K\"{a}hler potential induced by string loops are parametrically
larger than those induced by $\alpha'$ corrections \cite{bhk}.
However, the scalar potential exhibits an extended no-scale
structure, and the loop corrections contribute to the scalar
potential at a level subleading to their contribution to the
K\"{a}hler potential and subleading to the $\alpha'$ corrections
\cite{bhk, hepth0507131, cicq}. \cite{07040737} studied the effect
of loop corrections on the $\mathbb{C}P^4_{[1,1,1,6,9]}$ large
volume model and found it only gave minor corrections to the
moduli stabilisation and sub-sub-leading corrections to the soft
term computation. It is then natural to ask whether loop
corrections to the scalar potential can give a qualitative, rather
than only quantitative, change to moduli stabilisation.

We study this question in this paper and find that the answer is
affirmative. The LARGE Volume Scenario (LVS)\footnote{The
capitalisation of LARGE is a reminder that the volume is
exponentially large, and not just large enough to trust the
supergravity limit.} stabilises the overall volume at an
exponentially large value using $\alpha'$ and $g_s$ corrections.
Most previous work has focused on `Swiss-Cheese' Calabi-Yaus,
where one cycle controls the overall volume (`size of the cheese')
and the remaining moduli control the volume of blow-up cycles
(`holes in the cheese'). However for Calabi-Yaus with a fibration
structure - the torus is the simplest example - multiple moduli
enter into the overall volume. For the overall volume to be made
large in a homogeneous fashion, several moduli must become large.
In these cases, while the existence of at least one blow-up mode
is still necessary, loop corrections turn out to be necessary in
order to realise the LVS and obtain a stable minimum at
exponentially large volume. The loop corrections lift directions
transverse to the overall breathing mode and stabilise these.

The fact that the volume is stabilised large is desirable, first
to trust the perturbative expansion in the low energy effective
field theory and secondly as a powerful tool to generate
hierarchies. Moreover we stress that this minimum is found for
generic values of $W_0$,
 without the need to fine tune
$W_{0}$.

 We will illustrate our results by applying them to some
examples of Calabi-Yau three-folds that are hypersurfaces in
complex weighted projected spaces. From this analysis, it turns
out that a necessary and sufficient condition for LARGE volume is
the presence of blow-up modes resolving point-like singularities.
It seems also that the Calabi-Yaus which have a fibration
structure cannot present the interesting phenomenological
properties of the LVS. However, as we will explain in Sections
\ref{5} and \ref{6}, via the inclusion of the string loop
corrections to the K\"{a}hler potential, those K3 fibrations can
also present an exponentially large volume minimum, provided a
blow-up mode exists.

We will also show that the string loop corrections may play a role
in addressing the problem stressed in \cite{07113389}. The authors
there argued that the 4-cycle on which the Standard Model lives
cannot get non-perturbative corrections since their prefactor is
proportional to the VEV of Standard Model fields which, at this
stage, is required to vanish. However, due to the constraints of
the Standard Model gauge couplings this cycle must still be
stabilised at a relatively small size.

This problem may be cured through having at least two blow-up
modes and then adding $g_{s}$ corrections. The loop corrections
have the ability to stabilise the Standard Model cycle, while the
`transverse' cycle is stabilised non-perturbatively as usual. This
possible solution is discussed for the example
$\mathbb{C}P^{4}_{[1,3,3,3,5]}$, studied in detail in Section
\ref{SM}. We will see that the inclusion of the $g_{s}$
corrections can freeze $\tau_{SM}$ small producing a minimum of
the full scalar potential at exponentially large volume.

This paper is organised as follows. In the next section we briefly
review the LARGE Volume Scenario and state the general conditions
that have to be satisfied in order to have exponentially large
volume. The detailed proof of this general result is left to the
appendix. Section {3} illustrates our general results for several
Calabi-Yaus, including both Swiss cheese models where all
K\"{a}hler moduli other than the overall volume are blow-ups, and
also fibration Calabi-Yaus, such as K3 fibrations. Section {4}
reviews the general structure of string loop corrections and
section {5} illustrates how these corrections affect the results
on the models of section {3}. We finish with a short section on
conclusions and potential applications of our results.

We shall not discuss obtaining be Sitter vacua in this paper. For
a recent analysis of the conditions for de Sitter vacua from
supergravity, see \cite{08041073}.

\section{The LARGE Volume Scenario}
\label{Section3}

Large volume compactifications of string theories are desirable
for several reasons. First, large volume allows massive string
states to be consistently integrated out and makes the effective
field theory description of the compactification more robust.
Second, large volume is desirable for phenomenological reasons.
Given the standard relation $M_{planck}^2 \sim
\frac{M_{string}^2\, {\cal V}}{g_s} $, it is clear that LARGE
volume (or WEAK coupling \cite{08041248}) is necessary to have the
fundamental string scale hierarchically smaller than the Planck
scale. In string units a volume ${\cal V} \sim 10^4$ is needed for
$M_{string}= M_{GUT}$ and much larger volumes are needed for an
intermediate scale $M_{string}\sim 10^{11} \textrm{GeV}$  (${\cal
  V}\sim 10^{15}$), as attractive for
gravity mediated supersymmetry breaking, or $M_{string} \sim 1$
TeV (${\cal V}\sim 10^{30}$) in the extreme case of TeV strings.

Explicitly obtaining exponentially large volume in string theory,
with all the geometric moduli stabilised, goes much farther than
the original large extra dimensions proposals \cite{add} where the
volume was simply assumed to be large. In this section we will
briefly review how moduli are typically stabilised in IIB string
theory and then we will describe the general conditions under
which an exponentially large volume Calabi-Yau compactification is
obtained.

\subsection{Low Energy Limit for Type IIB Flux Compactifications}
\label{2}

To establish notation and conventions we start with a rapid review
of type IIB flux compactifications on Calabi-Yau three-folds $X$
\cite{gkp}, dimensionally reduced to 4D $\mathcal{N}=1$
supergravity. The K\"{a}hler potential takes the form:
\begin{equation}
K=-2\ln \left( \mathcal{V}+\frac{\xi} {2g_{s}^{3/2}}\right) -\ln
\left(S+\bar{S}\right) -\ln \left( -i\int\limits_{X}\Omega \wedge
\bar{\Omega}\right). \label{eq}
\end{equation}
In (\ref{eq}) we have included the leading $\alpha'$ correction
but have not included any string loop corrections. The constant
$\xi$ is given by: $ \xi =-\frac{\chi (X)\zeta (3)}{2(2\pi )^{3}},
$ where $\chi$ is the Euler number of the Calabi-Yau $X$,
$\zeta(3) \approx 1.2$, $S$ is the axio-dilaton, $S=e^{-\varphi
}+iC_{0}$, and $\Omega $ is the holomorphic (3,0)-form which
implicitly depends on the complex structure moduli $U_{\alpha}$,
$\alpha=1,...,h_{2,1}(X)$. $\mathcal{V}$ is the internal volume,
measured with an Einstein frame metric $g_{\mu \nu,E
}=e^{-\varphi/2}g_{\mu \nu ,s}$ and in units of
$l_{s}=2\pi\sqrt{\alpha'}$. It can be expressed in terms of the
K\"{a}hler form $J$ once this is expanded in a base
$\{\hat{D}_{i}\}$ of $H^{1,1}(X,\mathbb{Z})$ as
$J=\sum_{i=1}^{h_{1,1}}t^{i}\hat{D}_{i}$ (we focus on orientifold
projections such that $h_{1,1}^{-}=0\Rightarrow
h_{1,1}^{+}=h_{1,1}$):
\begin{equation}
\mathcal{V}=\frac{1}{6}\int\limits_{X}J\wedge J\wedge
J=\frac{1}{6}k_{ijk}t^{i}t^{j}t^{k}. \label{IlVolume}
\end{equation}
Here $k_{ijk}$ are the triple intersection numbers of $X$ and the
$t^{i}$ are 2-cycle volumes.

The fields entering the $\mathcal{N}=1$ chiral multiplets are the
complexified K\"{a}hler moduli $T_{i}=\tau _{i}+ib_{i}$ where
$\tau _{i}$ is the Einstein frame volume (in units of $l_s$) of
the divisor $D_{i}\in H_{4}(X,\mathbb{Z})$, which is the
Poincar\'{e} dual to $\hat{D}_{i}$. Its axionic partner $b_{i}$ is
the component of the RR 4-form $C_{4}$ along this cycle:
$\int\limits_{D_{i}}C_{4}=b_{i}$. The 4-cycle volumes $\tau _{i}$
are related to the 2-cycle volumes $t^{i}$:
\begin{equation}
\tau _{i}=\frac{\partial \mathcal{V}}{\partial
t^{i}}=\frac{1}{2}\int\limits_{X}\hat{D}_{i}\wedge J\wedge
J=\frac{1}{2} k_{ijk}t^{j}t^{k}. \label{defOfTau}
\end{equation}

The classical superpotential is generated by background fluxes
$G_{3}=F_{3}+iSH_{3}$, where $F_{3}$ and $H_{3}$ are the RR and
NSNS 3-form fluxes respectively. This superpotential also receives
nonperturbative corrections from either brane instantons
($a_{i}=2\pi $) or gaugino condensation ($a_{i}=2\pi /N$).
 The full
superpotential is then:
\begin{equation}
W=W_{tree}+W_{np}=\int\limits_{X}G_{3}\wedge \Omega+
\sum\limits_{i}A_{i}e^{-a_{i}T_{i}}. \label{yuj}
\end{equation}
The sum is over cycles generating nonperturbative contributions to
$W$. The $A_{i}$ correspond to threshold effects and depend on the
complex structure moduli and positions of D3-branes. There may
additionally be higher instanton effects in (\ref{yuj}), but these
can be neglected so long as each $\tau_{i}$ is stabilised such
that $a_{i}\tau_{i}\gg 1$.

The $\mathcal{N}=1$ F-term supergravity scalar potential is given
by:
\begin{equation}
V=e^{K}\left\{
\sum\limits_{i=T,S,U}K_{i\bar{j}}^{-1}D_{i}WD_{\bar{j}}\bar{W}-3\left\vert
W\right\vert ^{2}\right\} ,  \label{b}
\end{equation}
where
\begin{equation}
\left\{
\begin{array}{l}
D_{i}W=\partial_{i}W+W \partial_{i} K,
\\
D_{\bar{j}}\bar{W}=\partial_{\bar{j}} \bar{W}+\bar{W}
\partial_{\bar{j}} K.
\end{array}
\right.
\end{equation}
Classically, the potential for the K\"{a}hler moduli is flat due
to the no-scale structure:
\begin{equation}
\left(\frac{\partial^{2} K_{tree}}{\partial T_{i} \partial \bar{
T_{j}}}\right)^{-1}\frac{\partial K_{tree}}{\partial
T_{i}}\frac{\partial K_{tree}}{\partial \bar{T_{j}}}=3.
\end{equation}
This implies that the dilaton and complex structure moduli are
stabilised supersymmetrically at tree level, $D_S W = D_U W = 0$.

The classical flatness of the potential for the K\"{a}hler moduli
implies that to study K\"{a}hler moduli stabilisation we should
keep all possible quantum corrections, while for the $U$ and $S$
moduli it is sufficient to stabilise them classically:
\begin{equation}
\left\{
\begin{array}{l}
K=K_{cs}-2\ln\left(\frac{2}{g_{s}}\right)
-2\ln \left( \mathcal{V}+\frac{\xi }{2g_{s}^{3/2}}\right), \\
W=W_{0}+\sum\limits_{i}A_{i}e^{-a_{i}T_{i}},
\end{array}
\right.  \label{j}
\end{equation}
where $K_{cs}=\langle  -\ln \left( -i\int\limits_{X}\Omega \wedge
\bar{\Omega}\right) \rangle$ and $W_{0}=\langle W_{tree} \rangle$.
Substituting (\ref{j}) in (\ref{b}), we obtain the following
potential (defining $\hat{\xi}\equiv \xi/g_{s}^{3/2}$):
\begin{eqnarray}
V &=&e^{K}\left[ K_{0}^{jk}\left(
a_{j}A_{j}a_{k}\bar{A}_{k}e^{-\left(
a_{j}T_{j}+a_{k}\bar{T}_{k}\right) }-\left(
a_{j}A_{j}e^{-a_{j}T_{j}}\bar{W}
\partial _{\bar{T}_{k}}K_{0}+a_{k}\bar{A}_{k}e^{-a_{k}\bar{T}_{k}}W\partial
_{T_{j}}K_{0}\right) \right) \right.  \notag \\
&&\left. +3\hat{\xi} \frac{\left( \hat{\xi}^{2}+7\hat{\xi}
\mathcal{V}+\mathcal{V}^{2}\right) }{\left( \mathcal{V}-\hat{\xi}
\right) \left( 2\mathcal{V}+\hat{\xi} \right) ^{2}} \left\vert
W\right\vert ^{2}\right].  \label{scalar}
\end{eqnarray}

The aim of this paper will be a detailed study of the potential
(\ref{scalar}) for various different Calabi-Yaus, incorporating
string loop corrections into the form of the K\"{a}hler potential.
We aim to work out conditions for when the exponentially large
volume minimum of \cite{hepth0502058} is present.

\subsection{General Analysis for the Large Volume Limit}

We now investigate the topological conditions on an arbitrary
Calabi-Yau three-fold under which the scalar potential
(\ref{scalar}) admits an AdS non-supersymmetric minimum at
exponentially large volume deepening the analysis performed in
\cite{hepth0502058}. We will refer to those constructions as LARGE
Volume Scenarios (LVS).

\begin{claim}
(LARGE Volume) Let $X$ be a Calabi-Yau three-fold and let the
large volume limit be taken in the following way:
\begin{equation}
\left\{
\begin{array}{c}
\tau _{j}\text{ remains small, }\forall j=1,...,N_{small}, \\
\mathcal{V}\rightarrow \infty \text{ \ \ for \ }\tau
_{j}\rightarrow \infty ,\text{\ }\forall
j=N_{small}+1,..,h_{1,1}(X),
\end{array}
\right. \label{limit}
\end{equation}
within type IIB $\mathcal{N}=1$ $4D$ supergravity where the
K\"{a}hler potential and the superpotential in Einstein frame take
the form:
\begin{equation}
\left\{
\begin{array}{l}
K=K_{cs}-2\ln \left( \mathcal{V}+\hat{\xi}\right), \\
W= W_{0}+\sum\limits_{j=1}^{N_{small}}A_{j}e^{-a_{j}T_{j}}.
\end{array}
\right.  \label{explicit}
\end{equation}
Then the scalar potential admits a set $H$ of AdS
non-supersymmetric minima at exponentially large volume located at
$\mathcal{V}\sim e^{a_{j}\tau _{j}}$ $\forall j=1,...,N_{small}$
if and only if $h_{2,1}(X)>h_{1,1}(X)>1$, i.e. $\xi>0$ and
$\tau_{j}$ is a local blow-up mode resolving a given point-like
singularity $\forall j=1,...,N_{small}$. In this case
\begin{equation}
\left\{
\begin{array}{c}
\text{if }h_{1,1}(X)=N_{small}+1\text{, }H=\left\{ \text{a point}\right\} , \\
\text{if }h_{1,1}(X)>N_{small}+1\text{, }H=\left\{
(h_{1,1}(X)-N_{small}-1)\text{ flat directions} \right\}.
\end{array}
\right.
\end{equation}
\end{claim}

The proof of the previous Claim is presented in appendix
\ref{Appendix A} where we show also that $\tau_{j}$ is the only
blow-up mode resolving a point-like singularity if and only if
$K^{-1}_{jj}\sim \mathcal{V}\sqrt{\tau_{j}}$. On the contrary when
the same singularity is resolved by several independent blow-ups,
say $\tau_{1}$ and $\tau_{2}$, then $K^{-1}_{11}\sim
\mathcal{V}h^{(1)}_{1/2}(\tau_{1},\tau_{2})$ and $K^{-1}_{22}\sim
\mathcal{V}h^{(2)}_{1/2}(\tau_{1},\tau_{2})$ with $h^{(j)}_{1/2}$
homogeneous function of degree 1/2 such that
$\frac{\partial^{2}h^{(j)}}{\partial\tau_{1}\partial\tau_{2}}\neq
0$ $\forall j=1,2$.

Let us now explain schematically the global picture of LVS for
arbitrary Calabi-Yau manifolds according to the LARGE Volume
Claim:

\begin{enumerate}
\item{}
The Euler number of the Calabi Yau manifold must be negative. More
precisely: $h_{12}> h_{11}>1$. This means that the coefficient
$\hat\xi$ must be positive in order to guarantee that in a
particular direction the potential goes to zero at infinity from
below \cite{hepth0502058}. This is a both sufficient and necessary
condition.

\item{}
The Calabi-Yau manifold must have at least one blow-up mode
corresponding to a 4-cycle modulus that resolves a point-like
singularity. The associated modulus must have an induced
non-perturbative superpotential.
 This is usually guaranteed since these cycles
are rigid cycles of arithmetic genus one, which is precisely the
condition needed for the existence of
 non-perturbative
superpotentials  in the flux-less case \cite{witten}.

\item{}
This 4-cycle, together with other blow-up modes possibly present,
are fixed small by the interplay of non-perturbative and $\alpha'$
corrections, which stabilise also the overall volume mode. Here
small means larger than the string scale but not exponentially
large unlike the volume.

\item{}
All the other 4-cycles, such as those corresponding to fibrations,
cannot be stabilised small even though they may have induced
non-perturbative effects. They are sent large making their
non-perturbative corrections negligible.

\item{}
At this stage, non blow-up K\"{a}hler moduli, except the overall
volume mode, remain unfixed giving rise to essentially flat
directions.

\item{}
It turns out then that in order to freeze these moduli, it is
crucial to study string loop corrections as the leading term in a
$g_{s}$ expansion will be dominant over any potential
 non-perturbative correction.

\end{enumerate}
Notice that these are conditions to find exponentially large
volume minima and our results hold for generic $\mc{O}(1)$ values
of $W_0$. There may exist other minima which do not have
exponentially large volume for which our results do not have
anything to say. For example, $\vert W_0 \vert \ll 1$ may give
rise to KKLT-like minima.

Summarising, if there are $N_{small}$ blow-up modes and
$L=(h_{11}-N_{small}-1)$ modes which do not blow-up point-like
singularities nor correspond to the overall modulus, then our
results state that all the  $N_{small}$ can be fixed at values
large with respect to the string scale but not exponentially
large, the overall volume is exponentially large and the other $L$
K\"{a}hler moduli are not fixed by these effects.

In reality, the directions corresponding to the non blow-up modes,
if they have non-perturbative effects, will be lifted by these
tiny exponential terms, which however we neglect at this level of
approximation. The reason is that, as we will see in the next
sections, those directions will be lifted by the inclusion of
string loop corrections which are always dominant with respect to
the non-perturbative ones.

We would also like to stress that the previous general picture
shows how we need non-perturbative effects only in the blow-up
modes to get an exponentially large volume minimum. As blow-up
modes correspond to rigid exceptional divisors, the corresponding
non-perturbative corrections will be generally present even in the
fluxless case \cite{witten}. They can arise from either gaugino
condensation of the gauge theory living on the stack of branes
wrapping that 4-cycle or from Euclidean D3 brane instantons. On
the contrary, it is not clear if all the other cycles can indeed
get non-perturbative corrections to $W$, but this is not necessary
to obtain LARGE Volume.

\section{Particular Examples}
\label{examples}

Let us illustrate these results in a few explicit examples. At
this stage we ignore string loop corrections but as we will show
in section \ref{5} and \ref{6}, these can in some cases actually
be important and change the configuration of the system studied.

\subsection{The single-hole Swiss cheese:
$\mathbb{C}P_{[1,1,1,6,9]}^{4}$} \label{Swisscheese}

The weighted projective space $\mathbb{C}P^4_{[1,1,1,6,9]}$ is the
Calabi-Yau on which the LVS was originally realised. The overall
volume in terms of 2-cycle volumes is given by
\begin{equation}
\mathcal{V}=\frac{1}{6}\left(
3t_{1}^{2}t_{5}+18t_{1}t_{5}^{2}+36t_{5}^{3}\right).
\end{equation}
The divisor volumes take the form $\tau _{4}=\frac{t_{1}^{2}}{2}$,
$\tau _{5}=\frac{\left( t_{1}+6t_{5}\right) ^{2}}{2}$, from which
it is immediate to see that
\begin{equation}
\mathcal{V}=\frac{1}{9\sqrt{2}}\left( \tau _{5}^{3/2}-\tau
_{4}^{3/2}\right) .  \label{volume form}
\end{equation}
$\xi$ is positive since $h_{1,1}<h_{2,1}$ and the limit
(\ref{limit}) can be correctly performed with $\tau_{5}\rightarrow
\infty $ and $\tau_{4}$ remaining small. Thus $N_{small}=1$ and we
have to check if this case satisfies the condition of the LARGE
Volume Claim which is $K^{-1}_{44}\simeq
\mathcal{V}\sqrt{\tau_{4}}$. This is indeed satisfied as it can be
seen either by direct calculation or by noticing that $\tau_{4}$
is a local blow-up. Omitting numerical factors, the scalar
potential takes the form
\begin{equation}
V\simeq\frac{\sqrt{\tau_{4}}e^{-2a_{4}\tau_{4}}}{\mathcal{V}}
-\frac{W_{0}\tau_{4} e^{-2a_{4}\tau_{4}}}{\mathcal{V}^{2}}
+\frac{\hat{\xi}W_{0}^{2}}{\mathcal{V}^{3}}. \label{mio22}
\end{equation}
As the $\mathbb{C}P_{[1,1,1,6,9]}^{4}$ example is a particular
case of the LARGE Volume Claim, we conclude that the scalar
potential (\ref{mio22}) will admit an AdS minimum at exponentially
large volume with $(h_{1,1}-N_{small}-1)=0$ flat directions. This
is consistent with the original calculation in
\cite{hepth0502058}, which shows that the minimum is located at
\begin{equation}
\langle\tau_{4}\rangle\simeq (4\hat{\xi})^{2/3},\textit{ \ \ \
}\langle\mathcal{V}\rangle\simeq\frac{\hat{\xi}^{1/3}W_{0}}
{a_{4}A_{4}}e^{a_{4}\langle\tau_{4}\rangle}.
\end{equation}

\subsection{The multiple-hole Swiss cheese:
$\mathcal{F}_{11}$ and $\mathbb{C}P_{[1,3,3,3,5]}^{4}$}
\label{Swisscheese2}

It is straightforward to realise that the LARGE Volume Claim can
be used to generalise the previous case by adding several blow-up
modes resolving point-like singularities that will be stabilised
small. In this case the overall volume looks like
\begin{equation}
\mathcal{V}=\alpha \left( \tau
_{big}^{3/2}-\sum\limits_{i=1}^{N_{small}}\lambda _{i}\tau
_{i}^{3/2}\right), \label{cheese}
\end{equation}
where $\alpha $ and $\lambda_{i}$ are positive model-dependent
parameters and the Calabi-Yau manifold presents a typical ``Swiss
cheese'' shape. An explicit example is the Fano three-fold
$\mathcal{F}_{11}$ described in \cite{hepth0404257}, which is
topologically a $\mathbb{Z}_{2}$ quotient of a $CY_{3}$ with Hodge
numbers $h_{1,1}=3$, $h_{2,1}=111$. The total volume of the
$\mathcal{F}_{11}$ reads
\begin{equation}
\mathcal{V}=\frac{t_{1}^{2}t_{2}}{2}+\frac{t_{1}t_{2}^{2}}{2}+\frac{t_{2}^{3}
}{6}+\frac{t_{1}^{2}t_{3}}{2}
+2t_{1}t_{2}t_{3}+t_{2}^{2}t_{3}+t_{1}t_{3}^{2}+2t_{2}t_{3}^{2}+\frac{
2t_{3}^{3}}{3},
\end{equation}
and the 4-cycle moduli are given by
\begin{equation}
\tau _{1}=\frac{t_{2}}{2}\left( 2t_{1}+t_{2}+4t_{3}\right) ,\text{
\ \ \ } \tau _{2}=\frac{t_{1}^{2}}{2},\text{ \ \ \ \ }\tau
_{3}=t_{3}\left( t_{1}+t_{3}\right).
\end{equation}
It is then possible to express $\mathcal{V}$ in terms of the
$\tau$-moduli as
\begin{equation}
\mathcal{V}=\frac{1}{3\sqrt{2}}\left( 2\left( \tau _{1}+\tau
_{2}+2\tau _{3}\right) ^{3/2}-\left( \tau _{2}+2\tau _{3}\right)
^{3/2}-\tau _{2}^{3/2}\right).
\end{equation}
The resemblance with the general ``Swiss cheese'' picture
(\ref{cheese}) is now manifest. Two further Calabi-Yau
realisations of this Swiss-Cheese structure have been presented in
\cite{07113389}. They are the $h_{1,1}=3$ degree 15 hypersurface
embedded in $\mathbb{C}P_{[1,3,3,3,5]}^{4}$ and the $h_{1,1}=5$
degree 30 hypersurface in $\mathbb{C}P_{[1,1,3,10,15]}^{4}$. All
these many moduli ``Swiss cheese'' Calabi-Yaus will admit a LARGE
Volume minimum located at
\begin{equation}
\langle\mathcal{V}\rangle\sim
e^{a_{i}\langle\tau_{i}\rangle},\textit{ \ }\forall
i=1,...,N_{small},
\end{equation}
with no orthogonal flat directions. Let us briefly review the
geometric data of the resolution of the
$\mathbb{C}P_{[1,3,3,3,5]}^{4}$ manifold \cite{07113389}, since it
turns out to be an interesting case in which loop corrections may
potentially stabilise the Standard Model cycle that does not admit
non-perturbative superpotential contributions.
In the diagonal basis the total volume becomes
\begin{equation}
\mathcal{V}=\sqrt{\frac{2}{45}}\left(\tau_{a}^{3/2}-\frac{1}{3}\tau_{b}^{3/2}
-\frac{\sqrt{5}}{3}\tau_{c}^{3/2}\right). \label{NuovoVolume}
\end{equation}
A Euclidean D3-brane instanton wraps the rigid 4-cycle
$D_{E3}=\frac{1}{3}\left(D_{b}+D_{c}\right)$, giving a
non-perturbative superpotential term
$W_{np}=e^{-\frac{2\pi}{3}\left(\tau_{b}+\tau_{c}\right)}$. There
are also two stacks of D7-branes wrapping the rigid four cycles
$D_{D7A}=\frac{1}{3}\left(D_{b}-2D_{c}\right)$ and $D_{D7B}=D_{c}$
with line bundles
$\mathcal{L}_{A}=\frac{1}{3}\left(2D_{b}+5D_{c}\right)$ and
$\mathcal{L}_{B}=\mathcal{O}$. This choice guarantees that there
are no chiral zero modes on the D7-E3 intersections. The
``Standard Model'' is part of the $U(N_{A})$ gauge group on the
stack $A$ of D7-branes, with SM matter obtained from the
intersections $AA'$ and $AB$ where the prime denotes the
orientifold image.

Neglecting the D-term part of the scalar potential we obtain
\begin{equation}
V=\frac{\lambda_{1}\left(\sqrt{5\tau_{b}}+\sqrt{\tau_{c}}\right)
e^{-\frac{4\pi}{3}\left(\tau_{b}+\tau_{c}\right)}}{\mathcal{V}}
-\frac{\lambda_{2}\left(\tau_{b}+\tau_{c}\right)
e^{-\frac{2\pi}{3}\left(\tau_{b}+\tau_{c}\right)}}{\mathcal{V}^{2}}
+\frac{\lambda_{3}}{\mathcal{V}^{3}}, \label{mio}
\end{equation}
where $\lambda_{i}>0$, $\forall i=1,2,3$ are unimportant numerical
factors. Now to make the study of the scalar potential (\ref{mio})
simpler, we perform the change of coordinates $\tau _{b}=2\tau
_{E3}+\tau _{SM}, \tau _{c}=\tau _{E3}-\tau _{SM}$, bringing
(\ref{mio}) to the form
\begin{equation}
V=\frac{\lambda_{1}\left(\sqrt{5\left(2\tau_{E3}+\tau_{SM}\right)}
+\sqrt{\tau_{E3}-\tau_{SM}}\right)
e^{-4\pi\tau_{E3}}}{\mathcal{V}} -\frac{3\lambda_{2}\tau_{E3}
e^{-2\pi\tau_{E3}}}{\mathcal{V}^{2}}
+\frac{\lambda_{3}}{\mathcal{V}^{3}}. \label{mio2}
\end{equation}
The scalar potential (\ref{mio2}) then has a critical point at
$\tau_{E3}=2\tau_{SM}$. However, this is not a minimum of the full
scalar potential but is actually a saddle point along $\tau_{SM}$
at fixed $\tau_{E3}$ and $\mathcal{V}$. In subsection \ref{SM} we
will show how string loop corrections may give rise to a stable
LVS even though no non-perturbative corrections in $\tau_{SM}$ are
included (see \cite{07113389} for a discussion of freezing
$\tau_{SM}$ by including D-terms with (\ref{mio2})).

\subsection{2-Parameter K3 Fibration: $\mathbb{C}P_{[1,1,2,2,6]}^{4}$}
\label{2modK3noLoop}

Our next example is a fibration Calabi-Yau,
$\mathbb{C}P_{[1,1,2,2,6]}^{4}$. The overall volume in terms of
2-cycle volumes is given by
\begin{equation}
\mathcal{V}=t_{1}t_{2}^{2}+\frac{2}{3}t_{2}^{3}. \label{linearity}
\end{equation}
The 4-cycle volumes take the form $\tau _{1}=t_{2}^{2}$, $\tau
_{2}=2t_{2}\left( t_{1}+t_{2}\right)$, yielding
\begin{equation}
\mathcal{V}=\frac{1}{2}\sqrt{\tau _{1}}\left( \tau
_{2}-\frac{2}{3}\tau _{1}\right).  \label{vol11226}
\end{equation}
It is possible to invert the relations
$\tau_i=\partial{\mathcal{V}}/\partial t_i$  to produce
\begin{equation}
t_{2}=\sqrt{\tau _{1}}\text{, \ \ \ \ \ }t_{1}=\frac{\tau
_{2}-2\tau _{1}}{2 \sqrt{\tau _{1}}}. \label{ecco}
\end{equation}
The Euler characteristic of the Calabi-Yau is negative and the
limit (\ref{limit}) can be performed only with
$\tau_{2}\rightarrow \infty $ and keeping $\tau_{1}$ small. This
corresponds to $t_{1}\rightarrow \infty $ and $t_{2}$ small. In
this limit the volume becomes
\begin{equation}
\mathcal{V}=\frac{1}{2}\sqrt{\tau _{1}}\tau _{2}\simeq
t_{1}t_{2}^{2}\simeq t_{1}\tau_{1}. \label{lavel}
\end{equation}
Thus $N_{small}=1$ again and we need to check the condition of the
LARGE Volume Claim: $K^{-1}_{11}\simeq
\mathcal{V}\sqrt{\tau_{1}}$. However this is clearly not
satisfied, as $\tau_{1}$ is a fibration over the base $t_{1}$.

This is therefore a situation where no exponentially large volume
minimum is present, as can be confirmed by the explicit
calculation below.

\subsubsection{Explicit Calculation} \label{11226Sec}

Here we verify that the $\mathbb{C}P^4_{[1,1,2,2,6]}$ model does
not give a realisation of the LVS. We take the large volume limit
in the following way
\begin{equation}
\left\{
\begin{array}{c}
\tau _{1}\text{ small}, \\
\text{\ }\tau _{2}\gg 1,
\end{array}
\right.  \label{ooooo}
\end{equation}which, after the axion minimisation ($W_{0}>0$), gives a
scalar potential of the form
\begin{eqnarray}
V &=&V_{np}+V_{\alpha'}=\frac{4}{\mathcal{V}^{2}}\left[
a_{1}A_{1}^{2}\tau _{1}\left( a_{1}\tau _{1}+1\right)
e^{-2a_{1}\tau
_{1}}-a_{1}A_{1}\tau _{1}W_{0}e^{-a_{1}\tau _{1}}\right]  \notag \\
&&+\frac{3}{4}\frac{\xi }{\mathcal{V}^{3}}\left(
W_{0}^{2}+A_{1}^{2}e^{-2a_{1}\tau _{1}}-2A_{1}W_{0}e^{-a_{1}\tau
_{1}}\right).  \label{111}
\end{eqnarray}
We set $A_{1}=1$ and recall that to neglect higher order instanton
corrections we need $a_{1}\tau _{1}\gg 1$. (\ref{111}) becomes
\begin{equation}
V =\frac{4}{\mathcal{V}^{2}}\left[ \left( a_{1}\tau
_{1}e^{-a_{1}\tau _{1}}-W_{0}\right) a_{1}\tau _{1}e^{-a_{1}\tau
_{1}}\right]+\frac{3}{4}\frac{\xi }{\mathcal{V}^{3}}\left[
W_{0}^{2}+\left( e^{-a_{1}\tau _{1}}-2W_{0}\right) e^{-a_{1}\tau
_{1}}\right]. \label{1111}
\end{equation}
The previous expression (\ref{1111}) can be rewritten as
\begin{eqnarray}
V &=&\frac{e^{-2a_{1}\tau _{1}}}{\mathcal{V}^{2}}\left(
4a_{1}^{2}\tau _{1}^{2}+\frac{3}{4}\frac{\xi }{\mathcal{V}}\right)
-\frac{ 2W_{0}e^{-a_{1}\tau _{1}}}{\mathcal{V}^{2}}\left(
2a_{1}\tau _{1}+\frac{3}{4} \frac{\xi }{\mathcal{V}}\right)
+\frac{3}{4}\frac{\xi }{\mathcal{V}^{3}}
W_{0}^{2}  \notag \\
&&\underset{\mathcal{V}\gg 1}{\sim
}\frac{4}{\mathcal{V}^{2}}\left[ \left( a_{1}\tau
_{1}e^{-a_{1}\tau _{1}}-W_{0}\right) a_{1}\tau _{1}e^{-a_{1}\tau
_{1}}\right] +\frac{3}{4}\frac{\xi }{\mathcal{V}^{3}}W_{0}^{2}.
\label{1122}
\end{eqnarray}
Assuming a natural value $W_{0}\sim \mathcal{O}(1)$, then
(\ref{1122}) simplifies to
\begin{equation}
V=-\frac{4}{\mathcal{V}^{2}}W_{0}a_{1}\tau _{1}e^{-a_{1}\tau
_{1}}+\frac{3}{4 }\frac{\xi }{\mathcal{V}^{3}}W_{0}^{2}.
\label{oooo}
\end{equation}
Extremising this scalar potential, we get
\begin{equation}
\frac{\partial V}{\partial \tau _{1}}=\frac{4}{\mathcal{V}^{2}}
W_{0}a_{1}e^{-a_{1}\tau _{1}}\left( a_{1}\tau _{1}-1\right) =0,
\end{equation}
whose only possible solution for $W_{0}\neq 0$ is $a_{1}\tau
_{1}=1$, which is not in the controlled regime of parameter space.
However, when $W_{0}=0$, (\ref{1122}) gives
\begin{equation}
V=\left( 4a_{1}^{2}\tau _{1}^{2}+\frac{3}{4}\frac{\xi
}{\mathcal{V}}\right) \frac{e^{-2a_{1}\tau
_{1}}}{\mathcal{V}^{2}}\underset{\mathcal{V}\gg 1}{\sim
}4a_{1}^{2}\tau _{1}^{2}\frac{e^{-2a_{1}\tau
_{1}}}{\mathcal{V}^{2}},
\end{equation}
and the first derivative with respect to $\tau _{1}$ is
\begin{equation}
\frac{\partial V}{\partial \tau _{1}}=8a_{1}^{2}\tau _{1}\frac{
e^{-2a_{1}\tau _{1}}}{\mathcal{V}^{2}}\left( a_{1}\tau
_{1}-1\right) ,
\end{equation}
which also has no minimum. Thus we have shown that for $W_{0}\sim
\mathcal{O}(1)$, the $\mathbb{C}P^{4}_{[1,1,2,2,6]}$ model has no
exponentially large volume minimum. The last hope is to find a
minimum fine tuning $W_{0}\ll 1$. In this case taking the
derivatives of (\ref{1122}), one obtains
\begin{equation}
\frac{\partial V}{\partial \tau _{1}}=\frac{4}{\mathcal{V}^{2}}
a_{1}e^{-2a_{1}\tau _{1}}\left( a_{1}\tau _{1}-1\right) \left(
W_{0}e^{a_{1}\tau _{1}}-2a_{1}\tau _{1}\right) =0,
\end{equation}
whose only possible solution is
\begin{equation}
2a_{1}\left\langle \tau _{1}\right\rangle
=W_{0}e^{a_{1}\left\langle \tau _{1}\right\rangle},  \label{ede}
\end{equation}
but then fixing $\tau_{1}$, the scalar potential (\ref{1122})
along the volume direction looks like
\begin{equation}
V\sim
\frac{W_{0}^{2}}{\mathcal{V}^{2}}\left(-1+\frac{3}{4}\frac{\xi
}{\mathcal{V}}\right)\sim -\frac{W_{0}^{2}}{\mathcal{V}^{2}}.
\label{efe}
\end{equation}
The potential (\ref{efe}) has no LARGE Volume minimum and so we
conclude that the $\mathbb{C}P^{4}_{[1,1,2,2,6]}$ model does not
admit an exponentially large volume minimum for any value of
$W_{0}$.

It is still of course possible to fix the moduli using other
stabilisation schemes - for example KKLT. However, in this case
there will not be a large hierarchy between the two K\"{a}hler
moduli, with instead $\tau_{1}\lesssim \tau_{2}$, and the volume
can never be exponentially large.

\subsection{3-Parameter K3 Fibration}
\label{3modK3noLoop}

In the previous Sections \ref{Swisscheese}, \ref{Swisscheese2} and
\ref{2modK3noLoop}, we have presented three examples which
illustrate two of the three possible situations which the general
analysis determines. We now illustrate the case when an
exponentially large volume minimum can be found, but with flat
directions still present. We will then explain how these can be
lifted using string loop corrections.

This example concerns Calabi-Yau three-folds which are single K3
Fibrations with three K\"{a}hler moduli. We start off with the
following expression for the overall volume in terms of the three
moduli
\begin{equation}
\mathcal{V}=\alpha \left[ \sqrt{\tau _{1}}(\tau _{2}-\beta \tau
_{1})-\gamma \tau _{3}^{3/2}\right] ,  \label{hhh}
\end{equation}
where $\alpha $, $\beta $, $\gamma $ are positive model-dependent
constants. While we do not have any explicit realisation of such
kind of Calabi-Yau manifold, eq. (\ref{hhh}) is simply the
expression for the $\mathbb{C}P^{4}_{[1,1,2,2,6]}$ case
(\ref{vol11226}), augmented by the inclusion of a blow-up mode
$\tau_{3}$. We also assume that $h_{2,1}(X)>h_{1,1}(X)=3$, thus
satisfying the other condition of the LARGE Volume Claim. There
are then two ways to perform the limit (\ref{limit}) without
obtaining an internal volume that is formally negative:

\begin{enumerate}
\item
\begin{equation}
\left\{
\begin{array}{l}
\tau _{i}\rightarrow \infty ,\text{ \ }\forall i=1,2\text{ with
the
constraint }\tau _{1}<\tau _{2}, \\
\tau _{3}\text{\ remains small.}
\end{array}
\right.
\end{equation}
This case keeps both cycles associated with the fibration large,
while the blow-up cycle remains small. Given that
$\tau_{1}\rightarrow\infty$, this situation resembles the "Swiss
cheese" picture
\begin{equation}
\mathcal{V}=\alpha \underset{\tau
_{big}^{3/2}}{[\underbrace{\sqrt{\tau _{1}} (\tau _{2}-\beta \tau
_{1})}}-\gamma \tau _{3}^{3/2}],
\end{equation}
and due to this analogy with the $\mathbb{C}P_{[1,1,1,6,9]}^{4}$
model, the condition $K^{-1}_{33}\simeq\mathcal{V}\sqrt{\tau_{3}}$
is verified. Thus we can apply the LARGE Volume Claim which states
that the scalar potential will have an AdS exponentially large
volume set of minima together with $(h_{1,1}-N_{small}-1)=1$ flat
directions. In the following section we shall confirm this with an
explicit calculation.

\item
\begin{equation}
\left\{
\begin{array}{l}
\tau _{2}\rightarrow \infty, \\
\tau _{1}\text{ and }\tau _{3}\text{\ remain small.}
\end{array}
\right.
\end{equation}
In this case $N_{small}=2$ and according to the LARGE Volume Claim
there will be an exponentially large volume minimum of the scalar
potential if and only if both $\tau_{1}$ and $\tau_{3}$ is a
blow-up mode. As we show in the next Section
\ref{3modK3noLoopCalc},
$K^{-1}_{33}\sim\mathcal{V}\sqrt{\tau_{3}}$, as is suggested by
the volume form (\ref{hhh}). However $K^{-1}_{11}\sim
\tau_{1}^{2}$, as could be guessed from the fact that the overall
volume (\ref{linearity}) in terms of the 2-cycles moduli is linear
in $t_{1}$. Hence $\tau_{1}$ is not a blow-up but a fibration
modulus that does not give rise to LVS.

\end{enumerate}

We now confirm these statements with explicit calculations.

\subsubsection{Explicit Calculation} \label{3modK3noLoopCalc}

We focus on the case in which
\begin{equation}
\mathcal{V}=\alpha \left[ \sqrt{\tau _{1}}(\tau _{2}-\beta \tau
_{1})-\gamma \tau _{3}^{3/2}\right] ,
\end{equation}
where $\alpha $, $\beta $, $\gamma $ are positive model-dependent
constants and the K\"{a}hler potential and the superpotential take
the form (defining $\hat{\xi}\equiv \xi g_{s}^{-3/2}$):
\begin{equation}
K=K_{0}+\delta K_{(\alpha')}=-2\ln \left(
\mathcal{V}+\frac{\hat{\xi}}{2}\right) , \label{3}
\end{equation}
\begin{equation}
W=W_{0}+A_{1}e^{-a_{1}T_{1}}+A_{2}e^{-a_{2}T_{2}}+A_{3}e^{-a_{3}T_{3}}.
\label{sp}
\end{equation}
In the large volume limit the K\"{a}hler matrix and its inverse
look like:
\begin{equation}
K_{ij}^{0}=\frac{1}{4\mathcal{V}^{2}}\left(
\begin{array}{ccc}
\frac{\mathcal{V}^{2}}{\tau _{1}^{2}}+2\alpha ^{2}\beta ^{2}\tau
_{1} & \frac{\alpha ^{2}}{\sqrt{\tau _{1}}}\left( \gamma \tau
_{3}^{3/2}-2\beta \tau _{1}^{3/2}\right) & \frac{3\alpha \gamma
}{2}\frac{\sqrt{\tau _{3}}}{
\tau _{1}}\left( 2\alpha \beta \tau _{1}^{3/2}-\mathcal{V}\right) \\
\frac{\alpha ^{2}}{\sqrt{\tau _{1}}}\left( \gamma \tau
_{3}^{3/2}-2\beta \tau _{1}^{3/2}\right) & 2\alpha ^{2}\tau _{1} &
-3\alpha ^{2}\gamma \sqrt{
\tau _{1}}\sqrt{\tau _{3}} \\
\frac{3\alpha \gamma }{2}\frac{\sqrt{\tau _{3}}}{\tau _{1}}\left(
2\alpha \beta \tau _{1}^{3/2}-\mathcal{V}\right) & -3\alpha
^{2}\gamma \sqrt{\tau _{1}}\sqrt{\tau _{3}} & \frac{3\alpha \gamma
}{2}\frac{\mathcal{V}}{\sqrt{ \tau _{3}}}
\end{array}
\right), \label{LaDiretta}
\end{equation}
and
\begin{equation}
K_{0}^{ij}=4\left(
\begin{array}{ccc}
\tau _{1}^{2} & \beta \tau _{1}^{2}+\gamma \sqrt{\tau _{1}}\tau
_{3}^{3/2} &
\tau _{1}\tau _{3} \\
\beta \tau _{1}^{2}+\gamma \sqrt{\tau _{1}}\tau _{3}^{3/2} &
\frac{\mathcal{V }^{2}}{2\alpha ^{2}\tau _{1}}+\beta ^{2}\tau
_{1}^{2} & \tau _{3}\left(
\frac{\mathcal{V}}{\alpha \sqrt{\tau _{1}}}+\beta \tau _{1}\right) \\
\tau _{1}\tau _{3} & \tau _{3}\left( \frac{\mathcal{V}}{\alpha
\sqrt{\tau _{1}}}+\beta \tau _{1}\right) & \frac{2}{3\alpha \gamma
}\mathcal{V}\sqrt{ \tau _{3}}
\end{array}
\right).  \label{Kinverse}
\end{equation}
Both of the ways outlined above to take the large volume limit
have $\tau _{2}\gg 1$ and so the superpotential (\ref{sp}) can be
simplified as follows:
\begin{equation}
W\simeq W_{0}+A_{1}e^{-a_{1}T_{1}}+A_{3}e^{-a_{3}T_{3}}.
\label{spot}
\end{equation}
The scalar potential takes its general form (\ref{scalar}). In the
large volume limit, the $\alpha'$ leading contribution to the
scalar potential becomes
\begin{equation} V_{\left( \alpha
\prime \right) }=3e^{K}\hat{\xi}\frac{\left( \hat{\xi}
^{2}+7\hat{\xi} \mathcal{V}+\mathcal{V}^{2}\right) }{\left(
\mathcal{V}-\hat{\xi} \right) \left( 2 \mathcal{V}+\hat{\xi}
\right) ^{2}}\left\vert W\right\vert ^{2}\underset{\mathcal{V
}\rightarrow \infty }{\longrightarrow }\frac{3}{4}\frac{\hat{\xi}
}{\mathcal{V}^{3} }\left\vert W\right\vert ^{2}.
\end{equation}
Adding this to the non-perturbative part, we are left with
\begin{eqnarray}
V &=&\frac{1}{\mathcal{V}^{2}}\left[
K_{0}^{ij}a_{i}A_{i}a_{j}\bar{A} _{j}e^{-\left(
a_{i}T_{i}+a_{j}\bar{T}_{j}\right) }+2a_{i}A_{i}\tau
_{i}e^{-a_{i}T_{i}}\bar{W}\right.  \notag \\
&&\left. +2a_{j}\bar{A}_{j}\tau _{j}e^{-a_{j}\bar{T}_{j}}W\right]
+\frac{ 3}{4}\frac{\hat{\xi} }{\mathcal{V}^{3}}\left\vert
W\right\vert ^{2}. \label{v1v}
\end{eqnarray}

We shall focus on the case $W_{0}\sim \mathcal{O}(1)$, (since a
tuned small value of $W_0$ cannot give rise to large volume)
 and
for $a_{1}\tau_{1}\gg 1$, $a_{3}\tau_{3}\gg 1$, after extremising
the axion directions, the scalar potential simplifies to (with
$\lambda\equiv 8/(3\alpha\gamma)$, $\nu\equiv 3\hat{\xi}/4$ and
$A_{1}=A_{3}=1$)
\begin{equation}
V =\frac{\lambda a_{3}^{2}}{\mathcal{V}}\sqrt{\tau
_{3}}e^{-2a_{3}\tau _{3}}-\frac{4}{\mathcal{V}^{2}}W_{0}a_{1}\tau
_{1}e^{-a_{1}\tau _{1}}-\frac{4}{\mathcal{V}^{2}}W_{0}a_{3}\tau
_{3}e^{-a_{3}\tau _{3}}+\frac{\nu }{\mathcal{V}^{3}}W_{0}^{2}.
\label{ygfdo}
\end{equation}
The scalar potential $V$ depends on $\mathcal{V}$, $\tau_{1}$ and
$\tau_{3}$: $V=V(\mathcal{V},\tau _{1},\tau _{3})$, with the
dependence on $\tau_{2}$ implicit in the internal volume
$\mathcal{V}$. The large volume limit can be taken in the two ways
(1 and 2) outlined in the previous section \ref{3modK3noLoop}. The
difference between these two cases is that in limit 1
$\tau_{1}\rightarrow\infty$ whereas in limit 2 $\tau_{1}$ remains
small. Let us now study these two different cases in detail.

\bigskip

\textbf{1) }$\tau_{1}\rightarrow\infty$\textbf{ \
}$\Leftrightarrow \tau_{3}\ll \tau_{1}<\tau_{2}$

In this case, the superpotential (\ref{spot}) obtains
non-perturbative corrections only in $\tau_{3}$:
\begin{equation}
W\simeq W_{0}+A_{3}e^{-a_{3}T_{3}}. \label{ew}
\end{equation}
Since the $A_{1}$ term is not present in (\ref{ew}), we will be
unable to stabilise the corresponding K\"{a}hler modulus
$\tau_{1}$, thereby giving rise to an exactly flat direction. In
this case the scalar potential (\ref{ygfdo}) further reduces to:
\begin{equation}
V =\frac{\lambda a_{3}^{2}}{\mathcal{V}}\sqrt{\tau
_{3}}e^{-2a_{3}\tau _{3}}-\frac{4}{\mathcal{V}^{2}}W_{0}a_{3}\tau
_{3}e^{-a_{3}\tau _{3}}+\frac{\nu }{\mathcal{V}^{3}}W_{0}^{2}.
\label{41}
\end{equation}
and $V$ depends only on $\mathcal{V}$ and $\tau_{3}$: $V=V(V,$\
$\tau_{3})$. The potential (\ref{41}) has the same form as the
scalar potential found in section 3.2 of \cite{hepth0502058} where
the $\mathbb{C}P_{[ 1,1,1,6,9]}^{4}$ case was first discussed.
Following the
 same reasoning, we
look for possible minima of the scalar potential (\ref{41}) by
working out the two minimisation conditions:
\begin{equation}
\frac{\partial V}{\partial \mathcal{V}}=0\text{ \ \
}\Leftrightarrow \text{ \ \ }\left(\lambda a_{3}^{2}\sqrt{\tau
_{3}}e^{-2a_{3}\tau _{3}}\right) \text{\
}\mathcal{V}^{2}-\left(8W_{0}a_{3}\tau _{3}e^{-a_{3}\tau
_{3}}\right) \mathcal{V}+3 \nu W_{0}^{2}=0, \label{III}
\end{equation}
\begin{equation}
\frac{\partial V}{\partial \tau _{3}}=0\text{ \ \ }\Leftrightarrow
\text{ \ \ }\frac{\lambda a_{3}}{2\sqrt{\tau
_{3}}}\mathcal{V}\text{\
}e^{-a_{3}\tau_{3}}\left(1-4a_{3}\tau_{3}\right)+ 4W_{0}\left(
a_{3}\tau_{3}-1\right)=0. \label{IV}
\end{equation}
Equation (\ref{III}) admits a solution of the form
\begin{equation}
\frac{\lambda a_{3}}{4
W_{0}\sqrt{\tau_{3}}}\mathcal{V}e^{-a_{3}\tau_{3}}=1\pm
\sqrt{1-\frac{3\lambda\nu}{16\tau_{3}^{3/2}}}, \label{uno}
\end{equation}
whereas in the approximation $a_{3}\tau_{3}\gg 1$, (\ref{IV})
becomes:
\begin{equation}
\frac{\lambda a_{3}}{2\sqrt{\tau
_{3}}}\mathcal{V}e^{-a_{3}\tau_{3}}= W_{0}. \label{due}
\end{equation}
Combining (\ref{uno}) and (\ref{due}), we find $\frac{1}{2}=1\pm
\sqrt{1-\frac{3\lambda\nu}{16\tau_{3}^{3/2}}}$, whose solution is
given by
\begin{equation}
\langle\tau_{3}\rangle=\frac{1}{g_{s}}\left(\frac{\xi}
{2\alpha\gamma}\right)^{2/3}\sim \frac{1}{g_{s}}. \label{x}
\end{equation}
On the contrary, from (\ref{due}) we work out
\begin{equation}
\langle\mathcal{V}\rangle=\frac{3(\alpha\gamma)^{2/3} W_{0}}{4
a_{3}\sqrt{g_{s}}}\left(\frac{\xi}{2}\right)^{1/3}
e^{\frac{a_{3}}{g_{s}}\left(\frac{\xi}
{2\alpha\gamma}\right)^{2/3}}\sim \frac{W_{0}}{
a_{3}\sqrt{g_{s}}}e^{\frac{a_{3}}{g_{s}}}. \label{duet}
\end{equation}
There is therefore an exponentially large volume minimum. Setting
$\alpha=\gamma=1$, $\xi=2$, $g_{s}=0.1$, $a_{3}=\pi$ and
$W_{0}=1$, we finally obtain $\langle\tau_{3}\rangle=10$ and
$\langle\mathcal{V}\rangle=3.324\cdot10^{13}$. However there is
still the presence of an exactly flat direction which can be
better appreciated after the following change of coordinates:
\begin{equation}
(\tau_{1},\tau_{2})\text{ \ \ }\longrightarrow\text{ \ \ }(
\mathcal{V},\Omega):\text{\ \ \ \ \ \ \ \ \ \ \ \ } \left\{
\begin{array}{l}
\mathcal{V}\simeq
\alpha\left[\sqrt{\tau_{1}}\left(\tau_{2}-\beta\tau_{1}\right)\right] \\
\Omega=\alpha\left[\sqrt{\tau_{1}}\left(\tau
_{2}+\beta\tau_{1}\right)\right]
\end{array}
\right.  \label{43}
\end{equation}

\begin{figure}[ht]
\begin{center}
\epsfig{file=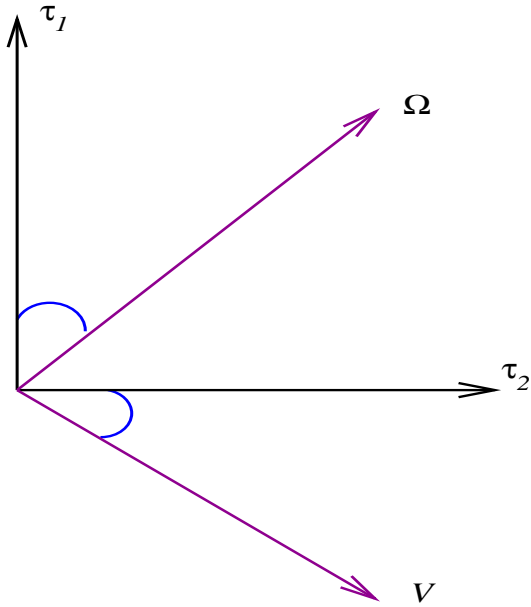, height=80mm,width=70mm} \caption{Change of
coordinates of the K\"{a}hler moduli space.}
\end{center}
\end{figure}
From (\ref{duet}) and (\ref{43}) we see that the stabilisation of
$\mathcal{V}$ and $\tau_{3}$ does not depend on $\Omega$ at all,
implying that $\Omega$ is a flat direction. We plot below in
Figure 2 the behaviour of this scalar potential where the flat
direction is manifest: $\tau_{3}$ has been already fixed as
$\langle\tau_{3}\rangle=10$, $\mathcal{V}$ is plotted on the
\textit{x}-axis and $\Omega$ on the \textit{y}-axis.

\begin{figure}[ht]
\begin{center}
\epsfig{file=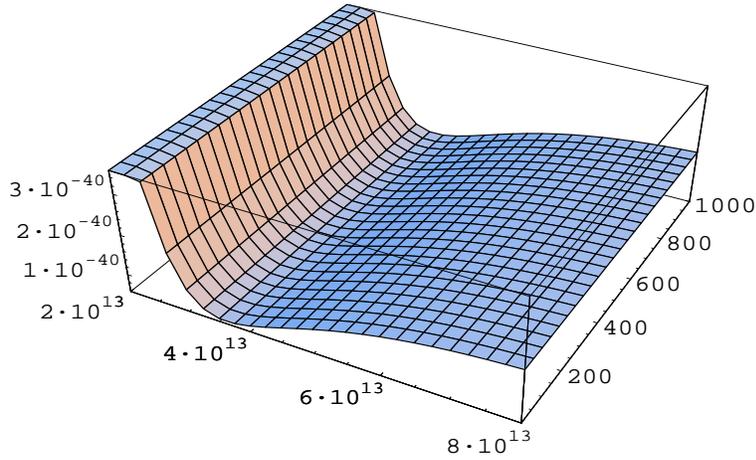, height=70mm,width=100mm} \caption{"Sofa"
potential with the presence of a flat direction.}
\end{center}
\end{figure}

\bigskip

\textbf{2) $\tau_{1}$ small}

In this case the large volume limit is taken keeping $\tau_{1}$
small and the scalar potential takes the general form
(\ref{ygfdo}). The minimisation equation with respect to
$\tau_{1}$ reads
\begin{equation}
\frac{\partial V}{\partial \tau_{1}}=0\text{\ \ \ \
}\Leftrightarrow \text{ \ \ }\frac{4}{\mathcal{V}^{2}}W_{0}
a_{1}e^{-a_{1}\tau_{1}}\left(a_{1}\tau _{1}-1\right)=0, \label{II}
\end{equation}
which implies $a_{1}\tau_{1}=1$ and so we cannot neglect higher
instanton corrections.

There is therefore no trustable minimum for the $\tau_1$ field. We
may however think about a situation in which the system still has
an exponentially large internal volume, with $\tau_{3}$ and
$\mathcal{V}$ sitting at their minimum
$\langle\mathcal{V}\rangle\sim e^{a_{3}\langle\tau_{3}\rangle}$,
while $\tau_{1}$ plays the r\^{o}le of a quintessence field
rolling in a region at large $\tau_{1} \gg 1$ away from
$a_{1}\tau_{1}=1$. The quintessence scale would be set by the
$e^{-a_1 \tau_1}$ exponent. Setting $a_{1}\tau_{1}\gg 1$ it is
easy to see that this is possible. However, the values of $a_1$
and $\tau_1$ need to be tuned to get a realistically small mass
for $\tau_1$ and even if this is done the fifth force problems of
quintessence fields would seem to be unavoidable.

 Finally, let us summarize in the table
below the results found without string loop corrections to $K$.
\begin{gather*}
1)\text{\ \ \ }\tau _{1}\rightarrow\infty\text{\ \ \ \ \ \
}\left\{
\begin{array}{l}
W_{0}\text{\ \ small \ \ \ \ \ No LVS,} \\
W_{0}\sim\mathcal{O}(1)\text{\ \ \ \ LVS + exactly flat direction
}\bot \mathcal{V},
\end{array}
\right. \\
2)\text{\ }\tau _{1}\text{ \ small \ }\left\{
\begin{array}{l}
W_{0}\text{\ \ small \ \ \ \ \ No LVS,} \\
W_{0}\sim\mathcal{O}(1)\text{\ and}\left\{
\begin{array}{l}
a_{1}\tau_{1}\gtrsim a_{3}\langle\tau_{3}\rangle\text{\ \ LVS +
almost
flat direction }\bot \text{\ }\mathcal{V}\text{ (quintessence),} \\
a_{1}\tau_{1}<a_{3}\langle\tau_{3}\rangle\text{\ \ No LVS.}
\end{array}
\right.
\end{array}
\right.
\end{gather*}

\section{Inclusion of the String Loop Corrections}
\label{5}

One of the purposes
 of this paper is to study the effect of string loop corrections
on moduli stabilisation. On a Calabi-Yau it is of course not
possible to explicitly determine the full functional form of loop
corrections. Nonetheless, it may still be possible to extract the
leading scaling behaviour with the moduli of the loop corrections,
even if the detailed form of the prefactors cannot be determined.
As the K\"{a}hler moduli are not stabilised at tree-level, even
this leading scaling behaviour is very significant.

This is the philosophy that was pursued in \cite{07040737, cicq},
where the expected parametric behaviour of the string loop
corrections to the K\"{a}hler potential was studied, even if the
detailed prefactors cannot be computed. On the torus, the explicit
string computation of \cite{bhk} does exist. For the $T^{6}/(
\mathbb{Z}_{2}\times \mathbb{Z}_{2})$ case the loop corrections
take the form
\begin{equation}
\delta K_{(g_{s})}=\delta K_{(g_{s})}^{KK}+\delta K_{(g_{s})}^{W},
\end{equation}
where $\delta K_{(g_{s})}^{KK}$ comes from the exchange between D7
and D3-branes of closed strings which carry Kaluza-Klein momentum,
and reads (for vanishing open string scalars)
\begin{equation}
\delta K_{(g_{s})}^{KK}= -\frac{1}{128\pi
^{4}}\sum\limits_{i=1}^{3}
\frac{\mathcal{E}_{i}^{KK}(U,\bar{U})}{\hbox{Re}\left( S\right)
\tau _{i}}. \label{KK}
\end{equation}
In the previous expression we assumed the all the three 4-cycles
of the torus are wrapped by D7-branes and $\tau_{i}$ denotes the
volume of the 4-cycle wrapped by the $i$-th D7-brane. The other
correction $\delta K_{(g_{s})}^{W}$ is interpreted in the closed
string channel as due to exchange of winding strings between
intersecting stacks of D7-branes. It takes the form
\begin{equation}
\delta K_{(g_{s})}^{W}=-\frac{1}{128\pi
^{4}}\sum\limits_{i=1}^{3}\left.
\frac{\mathcal{E}_{i}^{W}(U,\bar{U})}{\tau _{j}\tau
_{k}}\right\vert _{j\neq k\neq i},  \label{W}
\end{equation}
where $\tau_{i}$ and $\tau_{j}$ denote the volume of the 4-cycles
wrapped by the $i$-th and the $j$-th intersecting D7-branes. Note
in both cases there is a very
 complicated dependence of the corrections on the $U$ moduli, encoded
 in the functions $\mathcal{E}_{i}(U,\bar{U})$, but a
 very simple
dependence on the $T$ moduli.

These formulae were generalised by \cite{07040737}
 for the behaviour of loop corrections on
general Calabi-Yau three-folds. Given that these corrections can
be interpreted as the tree-level propagation of a closed KK string
and a Weyl rescaling is always necessary to convert the string
computation to Einstein frame, they proposed
\begin{equation}
\delta K_{(g_{s})}^{KK}\sim \sum\limits_{i=1}^{h_{1,1}}
\frac{\mathcal{C}_{i}^{KK}(U,\bar{U})m_{KK}^{-2} }{\hbox{Re}
\left( S\right) \mathcal{V}} \sim \sum\limits_{i=1}^{h_{1,1}}
\frac{\mathcal{C}_{i}^{KK}(U,\bar{U})\left( a_{il}t^{l}\right)
}{\hbox{Re} \left( S\right) \mathcal{V}},  \label{UUU}
\end{equation}
where $a_{il}t^{l}$ is a linear combination of the basis 2-cycle
volumes $t_{l}$ that is transverse to the 4-cycle wrapped by the
$i$-th D7-brane. A similar line of argument for the winding
corrections gives
\begin{equation}
\delta K_{(g_{s})}^{W}\sim \sum\limits_{i}\frac{\mathcal{C}
_{i}^{W}(U,\bar{U})m_{W}^{-2}}{\mathcal{V}}\sim
\sum\limits_{i}\frac{\mathcal{C} _{i}^{W}(U,\bar{U})}{\left(
a_{il}t^{l}\right) \mathcal{V}}, \label{UUUU}
\end{equation}
with $a_{il}t^{l}$ the 2-cycle where the two D7-branes intersect.
$\mathcal{C}^{KK}_i $ and $\mathcal{C}^W_i$ are unknown functions
of the complex structure moduli; however, as the complex structure
moduli are flux-stabilised these reduce to an (unknown) constant.
This approach is therefore useful to fix the leading order
dependence on K\"{a}hler moduli.

In \cite{cicq} the current authors
 showed that the proposed form of these correction is consistent
with what would be expected based on the form of the
Coleman-Weinberg potential in supergravity \cite{cw, fkz},
\begin{equation}
\delta V_{1-loop}\sim \Lambda ^{4} STr\left( M^{0}\right)
+2\Lambda ^{2} STr\left( M^{2}\right) + STr \left( M^{4}\ln \left(
\frac{M^{2}}{\Lambda ^{2}}\right) \right), \label{Coleman}
\end{equation}
where $\Lambda$ denotes the cut-off scale and $STr
\left(M^{n}\right)$ is the supertrace. The cut-off scale is taken
to be the scale at which the higher supersymmetry of the
ten-dimensional theory becomes apparent, effectively giving an
extended supersymmetry within the loops.

In terms of the corrections to the K\"{a}hler potential, the
leading contribution of these corrections to the scalar potential
is always vanishing, giving an "extended no-scale structure". This
result holds in general as long as the corrections are homogeneous
functions of degree $-2$ in the 2-cycle volumes.

In \cite{cicq} a general formula was also worked out for the first
non-vanishing contribution to the effective scalar potential of
$\delta K_{(g_{s})}$ as given by the conjectures
(\ref{UUU})-(\ref{UUUU}). It turns out to be relatively simple and
expressible in terms of the tree-level K\"{a}hler metric
$K_{0}=-2\ln \left( \mathcal{V}\right)$ and the winding correction
to the K\"{a}hler potential:
\begin{equation}
\delta V_{\left( g_{s}\right)
}^{1-loop}=\sum\limits_{i=1}^{h_{1,1}}\left( \frac{\left(
\mathcal{C} _{i}^{KK}\right)
^{2}}{\hbox{Re}(S)^{2}}K_{ii}^{0}-2\delta K_{(g_{s}),\tau
_{i}}^{W}\right) \frac{W_{0}^{2}}{\mathcal{V}^{2}}.  \label{V at
1-loop}
\end{equation}

The low energy interpretation of the 1-loop scalar potential
(\ref{V at 1-loop}) comes from matching it with the
Coleman-Weinberg potential (\ref{Coleman}). The leading term in
both cases vanishes, and so the comparison should involve the
leading non-zero terms in both cases. These match precisely for
the various cases studied in \cite{cicq}.

This low-energy interpretation can easily be illustrated in the
simple one-modulus case where the volume takes the form
$\mathcal{V}=\tau ^{3/2}$. The formula for the 1-loop scalar
potential (\ref{V at 1-loop}) applied to this case produces
(dropping the dilaton dependence since $S$ is fixed at tree level)
\begin{equation}
\delta V_{\left( g_{s}\right) ,1-loop}^{KK} = \left( 0\cdot
\frac{-3\mathcal{C}^{KK}}{\mathcal{V}^{8/3}}+\frac{3\alpha
_{2}\left( \mathcal{C}^{KK}\right)
^{2}}{\mathcal{V}^{10/3}}-\frac{6\alpha _{3}\left(
\mathcal{C}^{KK}\right) ^{3}}{\mathcal{V}^{4}}+\mathcal{O}\left(
\frac{1}{\mathcal{V}^{14/3}}\right) \right) W_{0}^{2}.
\label{torus}
\end{equation}
To compare with (\ref{Coleman}) we recall that in supergravity the
supertrace is proportional to the gravitino mass:
\begin{equation}
STr\left( M^{2}\right) \simeq m_{3/2}^{2}=e^{K}W_{0}^{2}\simeq
\frac{1}{\mathcal{V}^{2}}.
\end{equation}
The cut-off $\Lambda$ is identified with the compactification
scale given by
\begin{equation}
\Lambda =m_{KK}\simeq \frac{M_{s}}{R}=\frac{M_{s}}{\tau
^{1/4}}=\frac{1}{ \tau
^{1/4}}\frac{M_{P}}{\sqrt{\mathcal{V}}}=\frac{M_{P}}{\mathcal{V}^{2/3}}.
\end{equation}
Therefore in units of the Planck mass, (\ref{Coleman}) scales, in
agreement with (\ref{torus}), as
\begin{equation}
\label{fqx} \delta V_{1-loop} \simeq 0\cdot
\frac{1}{\mathcal{V}^{8/3}}+\frac{1}{\mathcal{V}^{10/3}}+
\frac{1}{\mathcal{V}^{4}}.
\end{equation}
We now use these results for the study of K\"{a}hler moduli
stabilisation.

\bigskip
\subsection{LARGE Volume and String Loop Corrections}

The results reviewed in the previous section are very important
for K\"{a}hler moduli stabilisation. The general picture for LVS
which we presented in the previous sections, was neglecting the
effect of string loop corrections to the scalar potential. However
just looking at the K\"{a}hler potential we have seen that, in
terms of powers of the K\"{a}hler moduli, the leading order
$\alpha'$ correction (\ref{eq}) scales as $\delta
K_{(\alpha')}\sim \frac{1}{\mathcal{V}}$, whereas from
(\ref{UUU}), the scaling behaviour of the Kaluza-Klein loop
correction is $\delta K_{(g_{s})}^{KK}\sim
\frac{\sqrt{\tau}}{\mathcal{V}}$. Naively it seems incorrect to
neglect $\delta K_{(g_{s})}^{KK}$ while including the effects of
$\delta K_{(\alpha')}$. However, as discussed in \cite{bhk,
07040737, cicq} due to the extended no-scale structure at the
level of the scalar potential the $\alpha'$ corrections dominate
over the $g_s$ corrections. This allows loop corrections to be
neglected compared to $\alpha'$ corrections for the stabilisation
of the volume.

However in our general analysis earlier, we saw that for fibration
models the inclusion of $\alpha'$ corrections still left almost
flat directions corresponding to non blow-up moduli orthogonal to
the overall volume. Loop corrections to the scalar potential are
much more important than non-perturbative superpotential
corrections, and we realise that they can play a crucial role in
stabilising these non blow-up moduli transverse to the overall
volume.

Thus we conclude that the extended no-scale structure renders the
LVS  robust not only because it allows $\delta V_{(g_{s})}$ to be
neglected when stabilising the volume, but also because it ensures
that when $\delta V_{(g_{s})}$ is introduced to lift the remaining
flat directions, even though it will reintroduce a dependence in
$V$ on $\mathcal{V}$ and blow-up moduli, it will not destroy the
minimum already found but will give just a small perturbation
around it.

The general picture is that all corrections - $\alpha'$, loop and
non-perturbative - play a r\^{o}le in a generic Calabi-Yau
compactification. We can summarise our general analysis for the
LVS as:

\begin{enumerate}

\item{}
In order to stabilise all the K\"{a}hler moduli at exponentially
large volume one needs at least one 4-cycle which is a blow-up
mode resolving point-like singularities.

\item{}
 This 4-cycle,
together with other blow-up modes possibly present, are fixed
small by the interplay of non-perturbative and $\alpha'$
corrections, which stabilise also the overall volume mode.

\item{}
 The
$g_{s}$ corrections are subleading and so can be safely neglected.

\item{}
All the other 4-cycles, as those corresponding to fibrations, even
though they have non-perturbative effects, cannot be stabilised
small. Thus they are sent large so making their non-perturbative
corrections negligible.

\item{}
These moduli, which are large and transverse to the overall
volume,
 can then be frozen by $g_{s}$ corrections,
which dominate over the (tiny) non-perturbative ones.

\end{enumerate}

In general $\delta V_{(g_{s})}$ only lifts the flat directions
associated to non blow-up moduli transverse to the overall volume.
One could wonder whether they indeed yield a real minimum for such
moduli as opposed to a runaway direction. We do not address this
problem in general terms here and so in principle this looks like
a model-dependent issue. However, as the overall volume is
stabilised, the internal moduli space is compact. Therefore these
non blow-up moduli cannot run-away to infinity and so we expect
that loop corrections will induce a minimum for the potential. In
fact, one example in the next section will illustrate this idea
explicitly.

\section{Moduli Stabilisation via String Loop Corrections}
\label{6}

We will now see in detail how the inclusion of string loop
corrections can affect the results found in the previous examples
which, neglecting $g_{s}$ corrections, can be summarised as:
\begin{enumerate}
\item  $\mathbb{C}P_{[1,1,1,6,9]}^{4}\rightarrow$
LVS without flat directions.

\item  3-parameter K3 Fibration with
$\tau_{1}$ ``small'' and $a_{1}\tau_{1}>a_{3}\langle\tau_{3}\rangle
\rightarrow$ LVS with an almost flat direction.

\item  3-parameter K3 Fibration with $\tau_{3}\ll\tau_{1}<\tau_{2}
\rightarrow$ LVS with one flat direction.

\item  $\mathbb{C}P_{[1,3,3,3,5]}^{4}\rightarrow$ LVS with a tachyonic
direction.

\item  $\mathbb{C}P_{[1,1,2,2,6]}^{4}$ and 3-parameter K3 Fibration with
$\tau_{1}$ "small" and $a_{1}\tau_{1}<a_{3}\tau_{3}\rightarrow$ No
LVS.
\end{enumerate}
We shall find that the inclusion of loop corrections modifies the
previous picture as follows:
\begin{enumerate}
\item  $\mathbb{C}P_{[1,1,1,6,9]}^{4}\rightarrow$
Not affected by $\delta V_{(g_{s})}$.

\item  3-parameter K3 Fibration with
$\tau_{1}$ ``small'' and
$a_{1}\tau_{1}>a_{3}\langle\tau_{3}\rangle \rightarrow \delta
V_{(g_{s})}$ ruins the almost flat direction $\Longrightarrow$ No
LVS \footnote{Notice that this case is the same as case 3 below but  in a
  different region of moduli space. The fact that there is no LVS
  realised here only means that for this model the LARGE Volume is
  realised as in the conditions of case 3 but not in those
  for case 2. In particular it requires both fibration moduli to be large.}.

\item  3-parameter K3 Fibration with $\tau _{3}\ll\tau _{1}<\tau
_{2}\rightarrow \delta V_{(g_{s})}$ lifts the flat direction
$\Longrightarrow$ LVS without flat directions.

\item  $\mathbb{C}P_{[1,3,3,3,5]}^{4}\rightarrow \delta V_{(g_{s})}$
stabilises the tachyonic direction $\Longrightarrow$ LVS without
flat directions.

\item  $\mathbb{C}P_{[1,1,2,2,6]}^{4}$ and 3-parameter K3 Fibration with
$\tau_{1}$ ``small'' and $a_{1}\tau_{1}<a_{3}\tau_{3}\rightarrow$
Not affected by $\delta V_{(g_{s})}$ - still no LVS.
\end{enumerate}

The $\mathbb{C}P_{[1,1,2,2,6]}^{4}$ case can never give large
volume due to the fibration 4-cycle $\tau_{1}$ which is impossible
to stabilise small. However in the example of the 3-parameter K3
fibration, LARGE Volume can be achieved by including a third
K\"{a}hler modulus which is a local blow-up and then sending
$\tau_{1}$ large. We shall use the expression (\ref{V at 1-loop})
for the form of string loop corrections to the scalar potential.

\subsection{The single-hole Swiss cheese:
$\mathbb{C}P_{[1,1,1,6,9]}^{4}$}

The influence of the $g_{s}$ corrections in the
$\mathbb{C}P_{[1,1,1,6,9]}^{4}$ case has been studied in detail in
\cite{07040737}. The authors showed that the loop corrections are
subleading and so can be neglected, as we claimed above. The loop
corrected K\"{a}hler potential looks like
\begin{eqnarray}
K &=&K_{tree}+\delta K_{(\alpha ^{\prime })}+\delta K_{(g_{s},\tau
_{5})}^{KK}+\delta K_{(g_{s},\tau _{4})}^{KK}  \nonumber \\
&=&-2\ln \mathcal{V}-\frac{\xi }{\mathcal{V}g_{s}^{3/2}}+\frac{
g_{s}C_{5}^{KK}\sqrt{\tau
_{5}}}{\mathcal{V}}+\frac{g_{s}C_{4}^{KK}\sqrt{\tau
_{4}}}{\mathcal{V}}, \label{Gret}
\end{eqnarray}
but due to the ``extended no scale structure'', we obtain for the
scalar potential
\begin{eqnarray}
V &=&V_{np}+V_{(\alpha ^{\prime })}+V_{(g_{s},\tau
_{5})}^{KK}+V_{(g_{s},\tau _{4})}^{KK}  \nonumber \\
&=&\frac{\lambda _{1}\sqrt{\tau _{4}}e^{-2a_{4}\tau
_{4}}}{\mathcal{V}}- \frac{\lambda _{2}W_{0}\tau _{4}e^{-a_{4}\tau
_{4}}}{\mathcal{V}^{2}}+\frac{ 3\xi
W_{0}^{2}}{4\mathcal{V}^{3}g_{s}^{3/2}}+\frac{g_{s}^{2}(C_{5}^{KK})^{2}
}{\mathcal{V}^{3}\sqrt{\tau
_{5}}}+\frac{g_{s}^{2}(C_{4}^{KK})^{2}}{\mathcal{V}^{3}\sqrt{\tau
_{4}}}. \label{Grett}
\end{eqnarray}
Without taking the loop corrections into account, we have found a
minimum located at $\mathcal{V}\sim
e^{a_{4}\tau_{4}}\Leftrightarrow a_{4}\tau_{4}\sim
\ln\mathcal{V}$. Therefore the various terms in (\ref{Grett})
scale as
\begin{eqnarray}
V &=&V_{np}+V_{(\alpha ^{\prime })}+V_{(g_{s},\tau
_{5})}^{KK}+V_{(g_{s},\tau _{4})}^{KK}  \nonumber \\
&\sim&\frac{\sqrt{\ln\mathcal{V}}}{\mathcal{V}^{3}}-
\frac{\ln\mathcal{V}}{\mathcal{V}^{3}}+\frac{1}{\mathcal{V}^{3}}
+\frac{1}{\mathcal{V}^{10/3}}+\frac{1}{\mathcal{V}^{3}\sqrt{\ln\mathcal{V}}},
\label{Gretta}
\end{eqnarray}
and it is straightforward to realise that at exponentially large
volume the last two terms in (\ref{Gretta}) are suppressed with
respect to the first three ones.

\subsection{The multiple-hole Swiss cheese:
$\mathbb{C}P_{[1,3,3,3,5]}^{4}$} \label{SM}

In Section \ref{Swisscheese2} we have seen that if the
non-perturbative corrections in the SM cycle $\tau_{SM}$ are
absent, the F-term scalar potential (\ref{mio2}) for the
$\mathbb{C}P_{[1,3,3,3,5]}^{4}$ Calabi-Yau does not present a LVS
with all the K\"{a}hler moduli stabilised. Following the same
procedure as in \cite{hepth0602233}, we shall now illustrate how
the $g_{s}$ corrections can turn the maximum in the $\tau_{SM}$
direction into a minimum without destroying the exponentially
large volume minimum
$\mathcal{V}\sim\sqrt{\tau_{E3}}e^{2\pi\tau_{E3}}$.

To derive the conjectured scaling behaviour of the loop
corrections, we use the formula (\ref{V at 1-loop}) setting
$\mathcal{C}_{i}^{KK}=\hbox{Re}(S)$ $\forall i$ and $W_{0}=1$. Two
stacks of D7-branes wrap the $\tau_{SM}$ and $\tau_{c}$ cycle
respectively and both will give rise to Kaluza-Klein $g_{s}$
corrections. From (\ref{V at 1-loop}), we estimate the first kind
of corrections by writing the overall volume (\ref{NuovoVolume})
in the $(\tau_{a},\tau_{SM},\tau_{c})$ basis and computing the
relevant elements of the direct K\"{a}hler metric. We find:
\begin{equation}
\mathcal{V}=\sqrt{\frac{2}{45}}\left(\tau_{a}^{3/2}-\frac{1}{3}\left(3\tau_{SM}+2\tau_{c}\right)^{3/2}
-\frac{\sqrt{5}}{3}\tau_{c}^{3/2}\right), \label{NuovoVolume2}
\end{equation}
along with
\begin{equation}
\frac{\partial^{2}K_{tree}}{\partial\tau_{SM}^{2}}\simeq
\frac{3}{\sqrt{10}}\frac{1}{\mathcal{V}\sqrt{3\tau_{SM}+2\tau_{c}}},
\end{equation}
and
\begin{equation}
\frac{\partial^{2}K_{tree}}{\partial\tau_{c}^{2}}\simeq
\frac{2\sqrt{2}}{3\sqrt{5}}\left(\frac{\sqrt{5}}{4\sqrt{\tau_{c}}}
+\frac{1}{\sqrt{3\tau_{SM}+2\tau_{c}}}\right)\frac{1}{\mathcal{V}},
\label{az2}
\end{equation}
where in the large volume limit we have approximated the volume as
$\mathcal{V}\simeq\sqrt{\frac{2}{45}}\tau_{a}^{3/2}$. Thus the
Kaluza-Klein loop corrections to (\ref{mio2}) look like
\begin{equation}
\delta V_{(g_{s})}^{KK}\simeq \left(\frac{5}{\sqrt{\tau_{c}}}
+\frac{13\sqrt{5}}{\sqrt{3\tau_{SM}+2\tau_{c}}}\right)
\frac{1}{15\sqrt{2}\mathcal{V}^{3}}. \label{azs2}
\end{equation}
Writing (\ref{azs2}) back in terms of $\tau_{SM}$ and
$\tau_{E3}=\tau_{c}+\tau_{SM}$, we obtain
\begin{equation}
\delta V_{(g_{s})}^{KK}\simeq
\left(\frac{5}{\sqrt{\tau_{E3}-\tau_{SM}}}
+\frac{13\sqrt{5}}{\sqrt{2\tau_{E3}+\tau_{SM}}}\right)
\frac{1}{15\sqrt{2}\mathcal{V}^{3}}. \label{imp1}
\end{equation}
Due to the particulary simple form of the volume
(\ref{NuovoVolume2}), it is very sensible to expect that the
winding corrections will scale like the Kaluza-Klein ones
(\ref{imp1}). Therefore adding (\ref{imp1}) to (\ref{mio2}) we end
up with
\begin{eqnarray}
V+\delta V_{(g_{s})} &=&\frac{\lambda _{1}\left( \sqrt{5\left(
2\tau _{E3}+\tau _{SM}\right) }+\sqrt{\tau _{E3}-\tau
_{SM}}\right) e^{-4\pi \tau _{E3}}}{\mathcal{V}}-\frac{3\lambda
_{2}\tau _{E3}e^{-2\pi \tau _{E3}}}{
\mathcal{V}^{2}}  \notag \\
&&+\frac{\lambda _{3}}{\mathcal{V}^{3}}+\left( \frac{\lambda
_{4}}{\sqrt{ \tau _{E3}-\tau _{SM}}}+\frac{\lambda
_{5}}{\sqrt{2\tau _{E3}+\tau _{SM}}} \right)
\frac{1}{\mathcal{V}^{3}}.  \label{mio20}
\end{eqnarray}
We notice that the string loop corrections are suppressed with
respect to the $\alpha'$ ones by a factor of $1/\sqrt{\tau_{E3}}$
and so do not affect the large volume minimum
$\mathcal{V}\sim\sqrt{\tau_{E3}}e^{2\pi\tau_{E3}}$ given that we
require $\tau_{E3}\gg 1$ to neglect higher order instanton
contributions. On the contrary $\delta V_{(g_{s})}$ can become
important to fix the SM direction when $\tau_{SM}$ gets small. In
fact, the maximum in that direction is now accompanied by a
minimum, as illustrated in Figure 3.
\begin{figure}[ht]
\begin{center}
\epsfig{file=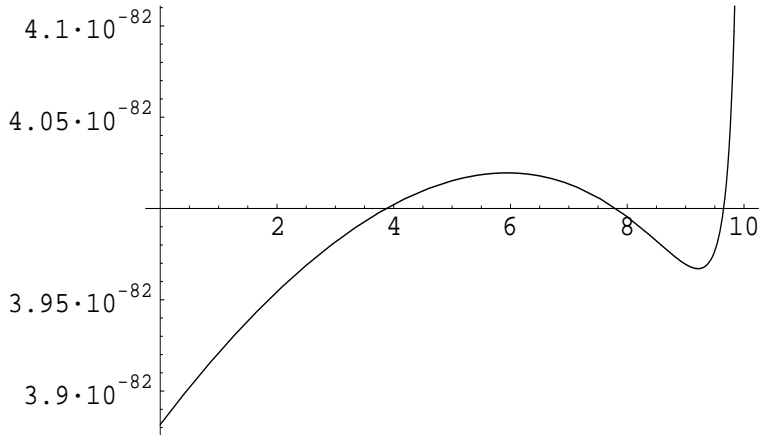, height=70mm,width=100mm}
\caption{$\tau_{SM}$ fixed by string loop corrections. The
numerical values used are $\lambda_{1}=\lambda_{2}=1$,
$\lambda_{3}=50$, $\lambda_{4}=\lambda_{5}=5$ and then we have
fixed $\tau_{E3}=10$ and $\mathcal{V}=\sqrt{10}e^{20\pi}$. }
\end{center}
\end{figure}

Thus we have shown that $g_{s}$ corrections can indeed freeze the
SM direction so giving rise to a LVS without any tachyonic
direction. The physics of this stabilisation is simply that if a
D7-brane wraps a 4-cycle, then loop corrections induced by the
brane will become large as the cycle size collapses. This repels
the modulus from collapsing and induces a minimum of the the
potential.

This example is illustrative in nature and shows how a cycle,
which is required to be small and which does not admit
nonperturbative effects, can potentially be stabilised by loop
corrections. In a fully realistic model, the D-term contribution
to the potential should also be included and the combined F- and
D-term potential studied. Usually the D-term will include, besides
the Fayet-Iliopoulos term depending on the moduli, also the
charged matter fields. Minimising the D-term will generically fix
one of the Standard Model singlets to essentially cancel the
Fayet-Iliopoulos term. Thus we can foresee a scenario in which the
Standard Model cycle is fixed by string loop corrections whereas
the D-term fixes not the size of that cycle but instead the VEV of
a Standard Model singlet as a function of the moduli. In this way
we address the challenge of \cite{07113389}. The form of the
D-term however depends on the model and in particular on the
details of the charged matter content and whether or not they
acquire VEVs. We therefore do not try and specify this, but note
that it will be necessary to include it in a realistic model.

\subsection{2-Parameter K3 Fibration:
$\mathbb{C}P_{[1,1,2,2,6]}^{4}$}

One could wonder whether including the string loop corrections in
the case of the K3 Fibration with two K\"{a}hler moduli treated in
Section \ref{2modK3noLoop}, could generate an exponentially large
volume minimum which was absent when only non-perturbative and the
$\alpha'$ corrections are included. In reality, the answer is
negative as these further perturbative corrections produce a
contribution $V_{(g_{s})}^{KK}+$ $V_{(g_{s})}^{W}$ to the scalar
potential (\ref{oooo}), which is subdominant and cannot help to
stabilise the moduli. In fact, in the large volume limit
(\ref{ooooo}) and for $W_{0}\sim \mathcal{O}(1)$, the full
corrected scalar potential, now takes the form
\begin{eqnarray}
V &=&V_{np}+V_{(\alpha \prime )}+V_{(g_{s},\tau_{1})}^{KK}
+V_{(g_{s},\tau_{2})}^{KK}+V_{(g_{s},\tau_{1})}^{W}
+V_{(g_{s},\tau_{2})}^{W}\simeq \nonumber \\
&\simeq &-\frac{4}{\mathcal{V}^{2}}W_{0}a_{1}\tau
_{1}e^{-a_{1}\tau _{1}}+ \frac{3}{4}\frac{\xi \hbox{Re}\left(
S\right)^{3/2}}{\mathcal{V}^{3}}W_{0}^{2}\nonumber \\
&&+\frac{W_{0}^{2}}{\mathcal{V}^{2}}\left( \frac{\left(\mathcal{C}
_{1}^{KK}\right) ^{2}}{\hbox{Re}\left( S\right)
^{2}}\frac{1}{\tau_{1}^{2}}+ \frac{\left(
\mathcal{C}_{2}^{KK}\right) ^{2}}{\hbox{Re}\left( S\right)
^{2}}\frac{1}{2}\frac{\tau
_{1}}{\mathcal{V}^{2}}-2\mathcal{C}_{1}^{W}\frac{\tau
_{1}}{\mathcal{V}^{2}}-\frac{2\mathcal{C}_{2}^{W}}{\mathcal{V}\sqrt{\tau_{1}
}}\right) \nonumber \\
&\simeq &-\frac{4}{\mathcal{V}^{2}}W_{0}a_{1}\tau
_{1}e^{-a_{1}\tau _{1}}+ \frac{3}{4}\frac{\xi\hbox{Re}\left(
S\right)^{3/2}}{\mathcal{V}^{3}}W_{0}^{2}+\frac{W_{0}^{2}}{\mathcal{V}
^{2}}\left( \frac{\left( \mathcal{C}_{1}^{KK}\right)
^{2}}{\hbox{Re}\left( S\right) ^{2}}\frac{1}{\tau
_{1}^{2}}-\frac{2\mathcal{C}_{2}^{W}}{\mathcal{V}\sqrt{\tau
_{1}}}\right). \label{freg}
\end{eqnarray}
First of all we have to check that the minimum in the volume is
exponentially large. Therefore let us take the derivative
\begin{equation}
\frac{4\mathcal{V}^{4}}{W_{0}^{2}}\frac{\partial V}{\partial
\mathcal{V}} =\left( \frac{32}{W_{0}}a_{1}\tau _{1}e^{-a_{1}\tau
_{1}}-\frac{\left( \mathcal{C}_{1}^{KK}\right)
^{2}}{\hbox{Re}\left( S\right) ^{2}}\frac{8}{ \tau
_{1}^{2}}\right) \mathcal{V}+\left(
\frac{24\mathcal{C}_{2}^{W}}{\sqrt{\tau _{1}}}-9\xi
\hbox{Re}\left( S\right) ^{3/2}\right) =0,
\end{equation}
whose solution is
\begin{equation}
\left\langle \mathcal{V}\right\rangle
=\frac{3}{8}\frac{\hbox{Re}\left( S\right) ^{2}\left\langle \tau
_{1}\right\rangle ^{3/2}W_{0}\left( 8\mathcal{
C}_{2}^{W}-3\sqrt{\left\langle \tau _{1}\right\rangle
}\xi\hbox{Re}\left( S\right) ^{3/2} \right) }{\left( \left(
\mathcal{C}_{1}^{KK}\right) ^{2}W_{0}e^{a_{1}\left\langle \tau
_{1}\right\rangle }-4a_{1}\hbox{Re}\left( S\right)
^{2}\left\langle \tau _{1}\right\rangle ^{3}\right)
}e^{a_{1}\left\langle \tau _{1}\right\rangle }. \label{ttt}
\end{equation}
From (\ref{ttt}) we realise that in order to have an exponentially
large volume, we need to fine tune $\left(
\mathcal{C}_{1}^{KK}\right) ^{2}\sim e^{-a_{1}\tau_{1}}\ll 1$. We
assume that this is possible and so the denominator of (\ref{ttt})
scales as
\begin{equation}
W_{0}-4a_{1}\hbox{Re}(S)^{2}\langle\tau_{1}\rangle^{3}\simeq
-4a_{1}\hbox{Re}(S)^{2}\langle\tau_{1}\rangle^{3},
\end{equation}
given that we are working in a regime where
$W_{0}\sim\mathcal{O}(1)$, $\hbox{Re}(S)\simeq 10$ and
$a_{1}\tau_{1}\gg 1$. Finally the VEV of the volume reads
\begin{equation}
\left\langle \mathcal{V}\right\rangle \simeq
\frac{3}{8}\frac{W_{0}\left( 8
\mathcal{C}_{2}^{W}-3\sqrt{\left\langle \tau _{1}\right\rangle
}\xi \hbox{Re}(S)^{3/2}\right) }{-4a_{1}\left\langle \tau
_{1}\right\rangle ^{3/2}}e^{a_{1}\left\langle \tau
_{1}\right\rangle },  \label{t3}
\end{equation}
with $\mathcal{C}_{2}^{W}$ chosen such that
\begin{equation}
\left( 1-\frac{3\sqrt{\left\langle \tau _{1}\right\rangle
}\xi\hbox{Re}(S)^{3/2}}{8\mathcal{C} _{2}^{W}}\right) <0,
\label{t2}
\end{equation}
to have a positive result. Now we neglect the
$\left(\mathcal{C}_{1}^{KK}\right) ^{2}$ term in $V$ (\ref{freg})
when we perform the derivative with respect to $\tau_{1}$ and we
obtain
\begin{equation}
\frac{\mathcal{V}^{2}}{W_{0}}\frac{\partial V}{\partial \tau _{1}}
=4a_{1}e^{-a_{1}\left\langle \tau _{1}\right\rangle }\left(
a_{1}\left\langle \tau _{1}\right\rangle -1\right)
+\frac{W_{0}\mathcal{C}_{2}^{W}}{\left\langle
\mathcal{V}\right\rangle \left\langle \tau _{1}\right\rangle
^{3/2}}=0.  \label{t4}
\end{equation}
Substituting back (\ref{t3}), (\ref{t4}) becomes
\begin{equation}
a_{1}\left\langle \tau _{1}\right\rangle =1+\frac{1}{3\left(
1-\frac{3\sqrt{ \left\langle \tau _{1}\right\rangle
}\xi\hbox{Re}(S)^{3/2}}{8\mathcal{C}_{2}^{W}}\right) },
\end{equation}
but (\ref{t2}) forces us to get $a_{1}\left\langle \tau
_{1}\right\rangle <1$, clearly in disagreement with our starting
point when we ignored higher order instanton corrections. Hence we
conclude that the inclusion of the string loop corrections does
not help to stabilise the moduli at exponentially large volume
since they render this attempt even worse.

\subsection{3-Parameter K3 Fibration}
\label{bene}

The results of the study of the K3 Fibration with three K\"{a}hler
moduli are summarised in the table at the end of Section
\ref{3modK3noLoopCalc}. We will now try to address the problem
left unsolved in that section. Without loop corrections it was
possible to find an exponentially large volume in this class of
models but there was still a flat direction left, which we named
$\Omega$. Let us see now how this direction is lifted. We shall
work in the regime $W_{0}\sim\mathcal{O}(1)$ where the
perturbative corrections are important. We start off wrapping
stacks of D7 branes around all the 4-cycle $\tau_{1}$, $\tau_{2}$
and $\tau_{3}$. We immediately notice that the Kaluza-Klein loop
correction to $V$ in $\tau_{3}$ takes the form
\begin{equation}
\delta V^{KK}_{(g_{s}),\tau _{3}}=
\frac{g_{s}^{2}(\mathcal{C}_{3}^{KK})^{2}}
{\sqrt{\tau_{3}}\mathcal{V}^{3}}, \label{eq3}
\end{equation}
and so does not depend on $\Omega$ and is subdominant to the
$\alpha'$ correction. Thus we will confidently neglect it. More
precisely, it could modify the exact locus of the minimum but not
the main feature of the model, that is the presence of an
exponentially large volume. Let us now focus on the region:
$\tau_{3}\ll\tau_{1}<\tau_{2}$. We recall the form of the scalar
potential and the K\"{a}hler potential without loop corrections:
\begin{equation}
V =\frac{16 a_{3}^{2}}{3\mathcal{V}}\sqrt{\tau_{3}}e^{-2a_{3}\tau
_{3}}-\frac{4}{\mathcal{V}^{2}}a_{3}\tau_{3}e^{-a_{3}\tau
_{3}}+\frac{3}{2g_{s}^{3/2}\mathcal{V}^{3}}, \label{41new}
\end{equation}
\begin{equation}
K=K_{tree}+\delta K_{(\alpha')}\underset{\mathcal{V}\gg 1}{\simeq
}-2\ln \mathcal{V}-\frac{2}{g_{s}^{3/2}\mathcal{V}}.
\end{equation}
We study now the possible corrections to $V$ coming from
$\tau_{1}$ and $\tau_{2}$ according to the general 1-loop formula
(\ref{V at 1-loop}). We realise that the form of the volume
(\ref{hhh}) implies that in this base of the K\"{a}hler cone, the
blow-up mode $\tau_{3}$ has only its triple self-intersection
number non-vanishing and so it does not intersect with any other
cycle. This is a typical feature of a blow-up mode which resolves
a point-like singularity: due to the fact that this exceptional
divisor is a \textit{local} effect, it is always possible to find
a suitable basis where it does not intersect with any other
divisor. Now we have seen that some string loop corrections come
from the exchange of closed winding strings at the intersection of
stacks of D7 branes. Hence the topological absence of these
intersections, implies an absence of these corrections. At the
end, the only relevant loop corrections are
\begin{equation}
\delta V_{(g_{s})}=\delta V^{KK}_{(g_{s}), \tau_{1}}+\delta
V^{KK}_{(g_{s}), \tau_{2}}+\delta V^{W}_{(g_{s}),
\tau_{1}\tau_{2}},
\end{equation}
which look like
\begin{eqnarray}
\delta V_{(g_{s}),\tau _{1}}^{KK} &=&g_{s}^{2}\left(
C_{1}^{KK}\right) ^{2}\left( \frac{1}{\tau _{1}^{2}}+\frac{2\beta
^{2}}{P}\right) \frac{
W_{0}^{2}}{\mathcal{V}^{2}}, \notag \\
\delta V_{(g_{s}),\tau _{2}}^{KK} &=&g_{s}^{2}\left(
C_{2}^{KK}\right) ^{2}
\frac{2}{P}\frac{W_{0}^{2}}{\mathcal{V}^{2}}, \label{LOOP} \\
\delta V_{(g_{s}),\tau _{1}\tau _{2}}^{W}
&=&-2C_{12}^{W}\frac{W_{0}^{2}}{ \mathcal{V}^{3}t_{\ast }}, \notag
\end{eqnarray}
where the 2-cycle $t_{*}$ is the intersection locus of the two
4-cycles whose volume is given by $\tau_{1}$ and $\tau_{2}$. In
order to work out the form of $t_{*}$, we need to write down the
volume of the K3 Fibration (\ref{hhh}) in terms of 2-cycle moduli:
\begin{equation}
\mathcal{V}=(\lambda_{1}t_{1}+\lambda_{2}t_{2})t_{2}^{2}+\lambda_{3}t_{3}^{3}.
\label{volumeet}
\end{equation}
Then
\begin{equation}
\left\{
\begin{array}{c}
\tau _{1}=\frac{\partial \mathcal{V}}{\partial t_{1}}=t_{2}\left(
\lambda
_{1}t_{2}\right) , \\
\tau _{2}=\frac{\partial \mathcal{V}}{\partial t_{2}}=t_{2}\left(
2\lambda _{1}t_{1}+3\lambda _{2}t_{2}\right),
\end{array}
\right. \label{taus}
\end{equation}
and so $t_{*}=t_{2}=\sqrt{\frac{\tau_{1}}{\lambda_{1}}}$.
Therefore the $g_{s}$ corrections to the scalar potential
(\ref{LOOP}) take the general form:
\begin{equation}
\delta V_{(g_{s})}= \left(\frac{A}{\tau_{1}^{2}}
+\frac{B}{\mathcal{V}\sqrt{\tau_{1}}} +\frac{C\tau_{1}}
{\mathcal{V}^{2}}\right)\frac{W_{0}^{2}}{\mathcal{V}^{2}},
\label{74}
\end{equation}
where
\begin{equation}
\left\{
\begin{array}{c}
A=g_{s}^{2}\left( C_{1}^{KK}\right) ^{2}>0, \\
B=-2C_{12}^{W}\sqrt{\lambda _{1}}\equiv -\frac{C_{12}^{W}}{\alpha }, \\
C=2\alpha ^{2}g_{s}^{2}\left[ \left( C_{1}^{KK}\right) ^{2}\beta
^{2}+\left( C_{2}^{KK}\right) ^{2}\right] >0.
\end{array}
\right.
\end{equation}
Notice that due to the ``extended no-scale structure'' which
causes the vanishing of the leading Kaluza-Klein correction to
$V$, we know the sign of the coefficients $A$ and $C$ because the
parameters are squared (see (\ref{V at 1-loop})) but we do not
have any control over the sign of $B$. It is now convenient to
take advantage of the field redefinition (\ref{43}) and recast the
loop corrections (\ref{LOOP}) in terms of $\mathcal{V}$ and
$\Omega$. Inverting the relation (\ref{43}), we get
\begin{equation}
\tau_{1}=\left(\frac{\Omega-\mathcal{V}}{2\alpha\beta}\right)^{2/3},\text{
\ \ \
}\tau_{2}=\left(\frac{\beta}{4\alpha^{2}}\right)^{1/3}\frac{\left(\Omega+\mathcal{V}\right)}
{\left(\Omega-\mathcal{V}\right)^{1/3}}. \label{CambiaCoord}
\end{equation}
Substituting these results back in (\ref{LOOP}) we can find the
relevant dependence of the scalar potential on $\Omega$:
\begin{equation}
\delta
V_{(g_{s})}=\frac{d_{1}\Omega^{2}+d_{2}\Omega\mathcal{V}+d_{3}\mathcal{V}^{2}}
{\left(\Omega-\mathcal{V}\right)^{4/3}\mathcal{V}^{4}}, \label{mo}
\end{equation}
where
\begin{eqnarray}
d_{1} &=&g_{s}^{2}\left( \frac{2\alpha ^{4}}{\beta ^{2}}\right)
^{1/3}\left[ \left( C_{1}^{KK}\right) ^{2}\beta ^{2}+\left(
C_{2}^{KK}\right) ^{2}\right]
W_{0}^{2}, \label{1} \\
d_{2} &=&-\left( \frac{2}{\alpha ^{2}\beta ^{2}}\right)
^{1/3}\left\{ \beta C_{12}^{W}+2\alpha ^{2}g_{s}^{2}\left[ \left(
C_{1}^{KK}\right) ^{2}\beta
^{2}+\left( C_{2}^{KK}\right) ^{2}\right] \right\} W_{0}^{2}, \label{2} \\
d_{3} &=&\left( \frac{2}{\alpha ^{2}\beta ^{2}}\right)
^{1/3}\left\{ \beta C_{12}^{W}+\alpha ^{2}g_{s}^{2}\left[ 3\left(
C_{1}^{KK}\right) ^{2}\beta ^{2}+\left( C_{2}^{KK}\right)
^{2}\right] \right\} W_{0}^{2}. \label{3}
\end{eqnarray}
For generic values of $d_{1}$, $d_{2}$ and $d_{3}$ we expect to
lift the flat direction $\Omega$. Consistency requirements imply
that any meaningful minimum must lie within the K\"{a}hler cone so
that no 2-cycle or 4-cycle shrinks to zero and the overall volume
is always positive. Let us work out the boundaries of the
K\"{a}hler moduli space in terms of $\mathcal{V}$ and $\Omega$ and
then look for a minimum in the $\Omega$ direction. Given that we
are sending both $\tau_{1}$ and $\tau_{2}$ large while keeping
$\tau_{3}$ small we can approximate the volume
(\ref{hhh})-(\ref{volumeet}) as follows
\begin{equation}
\mathcal{V}\simeq\alpha\sqrt{\tau _{1}}(\tau _{2}-\beta \tau
_{1})=(\lambda_{1}t_{1}+\lambda_{2}t_{2})t_{2}^{2},
\label{volumes}
\end{equation}
where $\lambda_{1}=\frac{1}{4\alpha^{2}}>0$ and
$\lambda_{2}=\frac{\beta}{4\alpha^{2}}>0$. Then looking at
(\ref{taus}) and (\ref{volumes}) it is clear that when $t_{1}$ and
$t_{2}$ are positive then also $\mathcal{V}>0$ and $\tau_{i}>0$
$\forall i=1,2$. Hence the boundaries of the K\"{a}hler cone are
where one of the 2-cycle moduli $t_{1,2} \to 0$. The expression of
the 2-cycles in terms of $\mathcal{V}$ and $\Omega$ reads
\begin{equation}
t_{1}=\left(\frac{2\mathcal{V}-\Omega}{\lambda_{1}}\right)\left(\frac{\lambda_{2}}
{\Omega-\mathcal{V}}\right)^{2/3},\text{ \ \ \
}t_{2}=\left(\frac{\Omega-\mathcal{V}}{\lambda_{2}}\right)^{1/3},
\label{ts}
\end{equation}
and so we realise that the K\"{a}hler cone is given by
$\mathcal{V}<\Omega<2\mathcal{V}$. In fact, looking at
(\ref{CambiaCoord}) and (\ref{ts}) we obtain
\begin{equation*}
\left\{
\begin{array}{c}
\Omega \rightarrow \mathcal{V}^{+}\Longleftrightarrow \tau
_{1}\rightarrow 0\Longleftrightarrow \tau _{2}\rightarrow \infty
\Longleftrightarrow
t_{1}\rightarrow \infty \Longleftrightarrow t_{2}\rightarrow 0, \\
\Omega \rightarrow \left( 2\mathcal{V}\right)
^{-}\Longleftrightarrow \tau _{1}\rightarrow \lambda _{1}\left(
\frac{\mathcal{V}}{\lambda _{2}}\right) ^{2/3}\Longleftrightarrow
\tau _{2}\rightarrow 3\lambda _{2}^{1/3}\mathcal{V}
^{2/3}\Longleftrightarrow t_{1}\rightarrow 0\Longleftrightarrow
t_{2}\rightarrow \left( \frac{\mathcal{V}}{\lambda _{2}}\right)
^{1/3}.
\end{array}
\right.
\end{equation*}
We look now for possible minima along the $\Omega$ direction
considering the volume already fixed:
\begin{equation}
\frac{\partial(\delta V_{(g_{s})})}{\partial\Omega}=\frac{2 d_{1}
\Omega
\left(\Omega-3\mathcal{V}\right)-\mathcal{V}\left(d_{2}\left(\Omega+3\mathcal{V}\right)+4
d_{3}\mathcal{V} \right)
}{3\left(\Omega-\mathcal{V}\right)^{7/3}\mathcal{V}^{4}}=0.
\label{der}
\end{equation}
Equation (\ref{der}) admits a solution of the form
$\langle\Omega\rangle=\kappa\langle\mathcal{V}\rangle$ where
\begin{equation}
\kappa=\frac{6 d_1 + d_2 + \sqrt{36 d_1^2+36 d_1 d_2 + d_2^2 + 32
d_1 d_3 }}{4 d_{1}}. \label{kappa}
\end{equation}
A consistent minimum within the walls of the K\"{a}hler cone
requires a choice of $d_{1}$, $d_{2}$ and $d_{3}$ such that
$1<\kappa<2$. In Section \ref{3modK3noLoopCalc} we have set the
parameters $\alpha=\gamma=1$, $\xi=2$, $g_{s}=0.1$, $a_{3}=\pi$
and $W_{0}=1$, and then obtained $\langle\tau_{3}\rangle=10$ and
$\langle\mathcal{V}\rangle\simeq 3.324\cdot10^{13}$ from
(\ref{x})-(\ref{duet}). We now keep the same choice of parameters
and set also $\beta=1/2$, $C_{1}^{KK}=1$,
$C_{2}^{KK}=C_{12}^{W}=10$. It follows that $d_{1}=2.005$,
$d_2=-14.01$ and $d_3=12.015$ which gives $\kappa\simeq 1.004$
correctly in the required regime. Then the minimum for $\Omega$
shown in Figure 4, is located at $\langle\Omega\rangle=\kappa\cdot
3.324\cdot 10^{13}\simeq 3.337\cdot 10^{13}$. We stress that we
have stabilised $\Omega$ without fine tuning any parameter.
\begin{figure}[ht]
\begin{center}
\epsfig{file=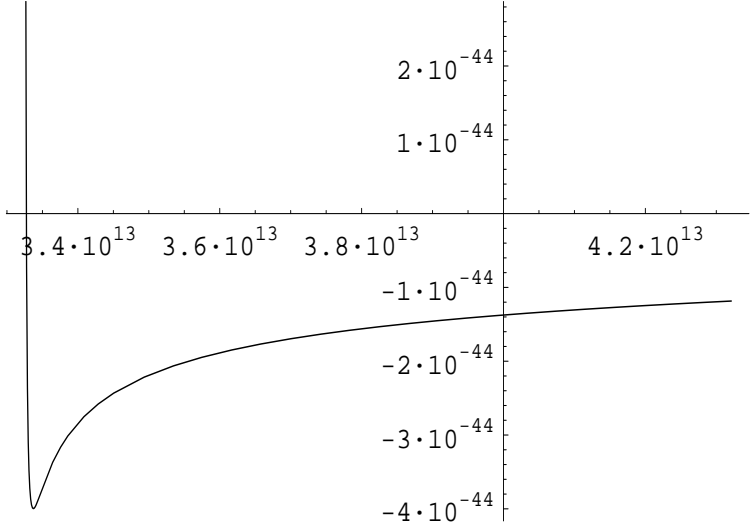, height=70mm,width=100mm} \caption{$V$
versus $\Omega$ at $\mathcal{V}$ and $\tau_{3}$ fixed.}
\end{center}
\end{figure}

We could also contemplate the case where we do not have D7 branes
wrapping one of the 4-cycles $\tau_{1}$ and $\tau_{2}$. In this
case there is no correction coming from the exchange of winding
strings because we have just one stack of D7 branes with no
intersection. Only the Kaluza-Klein corrections would survive.

\begin{enumerate}
\item no D7 brane wrapping the $\tau _{1}$ cycle

In this case the 1-loop correction looks like
\begin{equation}
\delta V_{(g_{s})}=\delta V^{KK}_{(g_{s}),
\tau_{2}}=2\alpha^{2}g_{s}^{2}\left( C_{2}^{KK}\right) ^{2}
\frac{W_{0}^{2}\tau_{1}}{\mathcal{V}^{4}}=
d\frac{\left(\Omega-\mathcal{V}\right)^{2/3}}{\mathcal{V}^{4}},
\label{xy}
\end{equation}
where $d=\left(\alpha^{2}\sqrt{2}/\beta\right)^{2/3}
g_{s}^{2}\left(C_{2}^{KK}\right)^{2}W_{0}^{2}$. However (\ref{xy})
has no minimum in $\Omega$ regardless of the value of $d$.

\item no D7 brane wrapping the $\tau _{2}$ cycle

\begin{equation}
\delta V_{(g_{s})}=\delta V^{KK}_{(g_{s}),
\tau_{1}}=g_{s}^{2}\left( C_{1}^{KK}\right) ^{2}\left(
\frac{1}{\tau
_{1}^{2}}+\frac{2(\alpha\beta)^{2}\tau_{1}}{\mathcal{V}^{2}}\right)
\frac{ W_{0}^{2}}{\mathcal{V}^{2}}=\frac{\mu_{1}\Omega^{2}
+\mu_{2}\Omega\mathcal{V}+\mu_{3}\mathcal{V}^{2}}
{\left(\Omega-\mathcal{V}\right)^{4/3}\mathcal{V}^{4}}, \label{mm}
\end{equation}
where
\begin{equation}
\mu _{1}=\delta ,\text{ \ \ \ \ }\mu _{2}=-2\delta ,\text{ \ \ \ \
}\mu _{3}=3\delta ,\text{ \ \ \ \ \ }\delta =2^{1/3}\left( \alpha
\beta \right) ^{4/3}g_{s}^{2}\left( C_{1}^{KK}\right)
^{2}W_{0}^{2}.
\end{equation}
Since the potential (\ref{mm}) has the same form of (\ref{mo}), we
can follow the same line of reasoning as above and conclude that
this case admits a minimum located at
$\langle\Omega\rangle=\kappa\langle\mathcal{V}\rangle$ if
$1<\kappa<2$, where $\kappa$ now is given by (\ref{kappa}) with
the replacement $d_i\leftrightarrow \mu_i$ $\forall i=1,2,3,$.
\end{enumerate}

We consider now the matrix of second derivatives in the
$(\tau_1,\tau_2)$ space $M_{ij}=V_{ij}$ and using the known
expression for the K\"{a}hler metric $K_{ij}$ we construct the
matrix $K^{-1}_{ik}M_{kj}$. The two eigenvalues of this matrix
correspond to the mass-squared of the canonically normalised
particles corresponding to the volume modulus and the originally
flat direction $\Omega$:
$$m^2_{\mathcal{V}}\ \sim\ 1/\mathcal{V}^3, \ \ \ m^2_{\Omega}\ \sim\ 1/\mathcal{V}^{10/3}.$$

Finally we have looked at minima in which $\tau_2\gg \tau_1$. This
would be an interesting configuration because of the following
observation: since $t_2=\sqrt{\tau_1}$ and
$t_1=(\tau_2-2\tau_1)/2\sqrt{\tau_1}$, we can see that $\tau_2\gg
\tau_1$ would imply that $t_1\gg t_2$ and we would effectively
have a very anisotropic compactification with the 2-cycle much
bigger than its dual 4-cycle. This could then lead to a
realisation of the supersymmetric 2 large extra dimensions
scenario \cite{add,cliff} in which the extra dimension could be as
large as a fraction of a millimetre. This would correspond to
looking for solutions $\langle\Omega\rangle
=(1+\epsilon)\langle\mathcal{V}\rangle$ with $\epsilon\to 0$.

However, we have shown that in the case $\tau_{1}$ ``small'', the
inclusion of the string loop corrections ruins the presence of the
almost flat direction which we had found before, without being
able to produce an actual minimum at exponentially large volume
with all moduli stabilised in a reliable region of moduli space.

\section{Conclusions and Potential Applications}
\label{7}

This paper has studied the general conditions needed to find
exponentially large volume in type IIB compactifications. The
necessary and sufficient conditions are simple to state: negative
Euler number, more than one K\"{a}hler moduli with at least one of
them being a blow-up mode resolving a point-like singularity.

We have also uncovered the important r\^{o}le played by $g_{s}$
corrections in moduli stabilisation.
 This has allowed us to find new
classes of LVS with a fibration structure in which not only the
volume but the fibre moduli  are exponentially large whereas the
blow-up modes are stabilised at the usual small values. Therefore
in general all of $\alpha'$, non-perturbative and $g_s$
corrections, may be important to stabilise the different classes
of K\"{a}hler moduli.

Here we briefly discuss some of the applications. First, our
results do not appear to change significantly the standard
phenomenology of LVS explored in \cite{hepth0610129, 07043403},
where we imagine the Standard Model localised on D7 branes
wrapping a small 4-cycle. The reason is that the volume modulus is
still the main source of supersymmetry breaking leading to an
approximate no-scale structure, which can be argued in general
terms \cite{joemirror}. As the Standard Model is localised around
a blow-up cycle, the effects from other exponentially large moduli
will be suppressed. However, it may be interesting to explore the
potential implications of hidden sectors localised on those
cycles. Also, in the multiple-hole Swiss cheese case where the
Standard Model cycle is stabilised by perturbative rather than
non-perturbative effects, the general structure of soft terms will
not change significantly, again since the main source of
supersymmetry breaking is the volume modulus. The only difference
could be the absence of the small hierarchy between the scale of
the soft terms $M_{soft}$ and the gravitino mass $m_{3/2}$, since
if the SM cycle is not stabilised non-perturbatively, then the
suppression of $M_{soft}$ with respect to $m_{3/2}$ by
$\ln(M_P/m_{3/2})$ \cite{suppression} is probably not present, but
it would be interesting to study this case in further detail.

A potentially more interesting application is to cosmology. The
cosmological implications of LVS have been explored in
\cite{hepth0509012, hepth0612197, 08043653, aalok} only for Swiss
cheese compactifications.
 Small moduli were found to be good candidates for
inflatons as long as $h_{11}>2$ without the need to fine tune.
However a difficulty with this is that loop corrections are
expected to modify this result if there is a D7 brane wrapping the
inflaton cycle, while if there is no such brane then it is
difficult to reheat the Standard Model brane since there is no
direct coupling of the inflaton to Standard Model fields.

The volume modulus is not suitable for inflation as $m_{\mc{V}}
\sim H$ and so it suffers directly from the $\eta$ problem.
However for fibration models such as
$\mathbb{C}P^{4}_{[1,1,2,2,6]}$, there is the transverse field
$\Omega$ which is stabilised by the loop corrections. As the loop
corrections are parametrically weaker than the $\alpha'$
corrections which stabilise the volume, $\Omega$ is parametrically
lighter than the volume modulus and thus the Hubble scale. In fact
 $m_{\Omega}\sim
\mathcal{V}^{-(3/2+\alpha)}$, with $\alpha=1/6$. It follows that
the slow-roll $\eta$ parameter is
\begin{equation}
\eta\sim M_P^2\,  \frac{m_{\Omega}^{2}}{H^2}\sim
\frac{1}{\mathcal{V}^{1/3}}\ll 1.
\end{equation}
Therefore such fibration models seem
 promising for string theory realisations of modular inflation, as at large
volume the mass
 scale induced by loop corrections is parametrically smaller than the Hubble
scale.
  A detailed study of the potential for large values of the field, away from
the
 minimum, will be required in order to see if this is a viable model
 of inflation, including the value of density perturbations and the
 potential for reheating.

The fact that the spectrum of moduli fields includes further
candidate light fields, besides the volume modulus, is a
 new source for the cosmological moduli problem.
 In the LARGE volume scenario this problem is
already present as long as the string scale is smaller than
$10^{13}$ GeV since the volume modulus would be lighter than $10$
TeV, and coupling with gravitational strength interactions it
would overclose the universe or decay so late to ruin
nucleosynthesis. Given a solution to this problem - such as a late
period of inflation - the corresponding modulus becomes a dark
matter candidate. With an intermediate string scale and TeV
supersymmetry, the volume modulus has a mass $m \sim 1
\hbox{MeV}$. The additional light moduli fields are also potential
dark matter candidates and
 have masses $m \sim 10 \hbox{keV}$.
Furthermore, they can decay into photons with a clean
monochromatic line similar to the volume modulus. A proper
analysis of their couplings to photons along the lines of
\cite{conlonquev} should be made in order to see if this effect
could be eventually detected.

It is worth pointing that the multiple hole Swiss cheese example
provides an explicit example of K\"{a}hler moduli inflation, in
which at least three K\"{a}hler moduli were needed (but no
explicit example was provided in \cite{hepth0509012}). Also this
is a good example to explore the issue about stabilisation of the
Standard Model cycle that has to be small (and then a blow-up
mode) but without the presence of a non-perturbative
superpotential which is not desired if the corresponding axion is
the QCD axion \cite{hepth0602233} and if D-terms could induce a
breaking of the Standard Model group \cite{07113389}. Our results
indicate that it is actually possible to achieve this.

We would finally like to emphasise that this is only a first
attempt to investigate the relevance of loop corrections in the
LVS and much work remains to be done. In particular, although we
have used a well motivated volume dependence of the leading
quantum corrections to the K\"{a}hler potential, explicit
calculations are still lacking. While we believe that given the
general importance of loop corrections, it is important to study
their effects even with incomplete knowledge of their form,
further information about these corrections for general
Calabi-Yaus is very desirable.

{\bf Acknowledgements}

We thank M. Berg, C.P. Burgess, P. Candelas, X. de la Ossa, F.
Denef, M. Haack, M. Kreuzer,
 A. Maharana and L. McAllister for useful conversations. MC is partially funded
by St John's College, EPSRC and CET. JC is funded by Trinity
College, Cambridge and also thanks the University of Texas at
Austin for hospitality while some of this work was carried out. FQ
is partially funded by STFC and a Royal Society Wolfson merit
award. He also thanks the Mitchell family for hospitality at the
Cambridge-Texas A\& M Cook's Branch meeting 2008. This material is
based upon work supported by the National Science Foundation under
Grant No. PHY-0455649.

\appendix
\section{Proof of the LARGE Volume Claim}
\label{Appendix A}

Let us now present a comprehensive argument in favour of the LARGE
Volume Claim which establishes the existence of LARGE Volumes in
IIB string compactifications.

\subsection{Proof for $N_{small}=1$}
\label{Appendix A1}

\begin{proof}(LARGE Volume Claim for $N_{small}=1$)
Let us start from the scalar potential (\ref{scalar}) which we now
rewrite as
\begin{equation}
V=V_{np1}+V_{np2}+V_{\left( \alpha' \right) },
\end{equation}
and perform the large volume limit as described in (\ref{limit})
with $N_{small}=1$ corresponding to $\tau_{1}$. In this limit
$V_{\left( \alpha'\right) }$ behaves as
\begin{equation}
V_{\left( \alpha' \right) }\underset{\mathcal{V}\rightarrow \infty
}{ \longrightarrow }+\frac{3\hat{\xi}
}{4\mathcal{V}^{3}}e^{K_{cs}}\left\vert W\right\vert
^{2}+\mathcal{O}\left( \frac{1}{\mathcal{V}^{4}}\right) .
\label{Alfa}
\end{equation}
We also point out that
\begin{equation}
e^{K}\underset{\mathcal{V}\rightarrow \infty }{\longrightarrow
}\frac{ e^{K_{cs}}}{\mathcal{V}^{2}}+\mathcal{O}\left(
\frac{1}{\mathcal{V}^{3}} \right) .  \label{Eallak}
\end{equation}
Let us now study $V_{np1}$ which reduces to
\begin{equation}
V_{np1}\underset{\mathcal{V}\rightarrow \infty }{\longrightarrow}
e^{K}K_{11}^{-1}a_{1}^{2}\left\vert
A_{1}\right\vert^{2}e^{-a_{1}\left(T_{1}+\bar{T}_{1}\right)
}=\frac{K_{11}^{-1}}{\mathcal{V}^{2}}a_{1}^{2}\left\vert
A_{1}\right\vert^{2}e^{-2 a_{1}\tau_{1}}. \label{Imp}
\end{equation}
Switching to the study of $V_{np2}$, we find that
\begin{equation}
V_{np2}\underset{\mathcal{V}\rightarrow \infty }{\longrightarrow }
-e^{K}\sum\limits_{k=1}^{h_{1,1}}K_{1k}^{-1} \left[ \left(
a_{1}A_{1}e^{-a_{1}\tau _{1}}e^{-ia_{1}b_{1}}\bar{W}\partial _{
\bar{T}_{k}}K\right) +\left( a_{1}\bar{A}_{1}e^{-a_{1}\tau
_{1}}e^{+ia_{1}b_{1}}W\partial _{T_{k}}K\right) \right] ,
\label{Jjj}
\end{equation}
where we have used the fact that $K_{1k}^{-1}=K_{k1}^{-1}$.
Equation (\ref{Jjj}) can be rewritten as
\begin{eqnarray}
&&V_{np2}\underset{\mathcal{V}\rightarrow \infty
}{\longrightarrow}-e^{K}\sum\limits_{k=1}^{h_{1,1}}K_{1k}^{-1}\left(
\partial _{T_{k}}K\right)a_{1} e^{-a_{1}\tau _{1}}\left[ \left(
A_{1}\bar{W}e^{-ia_{1}b_{1}}\right) +\left(
\bar{A}_{1}We^{+ia_{1}b_{1}}\right) \right]  \notag \\
&=&\left(
X_{1}e^{+ia_{1}b_{1}}+\bar{X}_{1}e^{-ia_{1}b_{1}}\right),
\end{eqnarray}
where
\begin{equation}
X_{1}\equiv -e^{K}K_{1k}^{-1}\left( \partial _{T_{k}}K\right)
a_{1}\bar{A}_{1}W e^{-a_{1}\tau _{1}}.  \label{Def1}
\end{equation}
We note that for a general Calabi-Yau, the following relation
holds:
\begin{equation}
K_{1k}^{-1}\left( \partial _{T_{k}}K\right) =-2\tau _{1},
\end{equation}
and thus the definition (\ref{Def1}) can be simplified to
\begin{equation}
X_{1}\equiv 2e^{K}a_{1}\tau _{1}\left\vert A_{1}\right\vert e^{-i
\vartheta_{1}}\left\vert W \right\vert e^{i \vartheta_{W}}
e^{-a_{1}\tau _{1}}=\left\vert X_{1}\right\vert e^{i(\vartheta
_{W}-\vartheta_{1})}. \label{DDEF1}
\end{equation}
Therefore
\begin{equation}
V_{np2}\underset{\mathcal{V}\rightarrow \infty }{\longrightarrow }
\left\vert X_{1}\right\vert \left( e^{+i\left(
\vartheta_{W}-\vartheta _{1}+a_{1}b_{1}\right) }+e^{-i\left(
\vartheta_{W}-\vartheta _{1}+a_{1}b_{1}\right) }\right)
=2\left\vert X_{1}\right\vert \cos \left( \vartheta_{W}-\vartheta
_{1}+a_{1}b_{1}\right).
\end{equation}
$V_{np2}$ is a scalar function of the axion $b_{1}$ whereas
$\vartheta_{1}$ and $\vartheta_{W}$ are to be considered just as
parameters. In order to find a minimum for $V_{np2}$ let us set
its first derivative to zero:
\begin{equation}
\frac{\partial V_{np2}}{\partial b_{1}}=-2a_{1} \left\vert
X_{1}\right\vert \sin (\vartheta_{W}-\vartheta _{1}+a_{1}b_{1})=0.
\label{Ggg}
\end{equation}
The solution of (\ref{Ggg}) is given by
\begin{equation}
a_{1}b_{1}=p_{1}\pi +\vartheta _{1}-\vartheta_{W} ,\text{ \
}p_{1}\in \mathbb{Z}. \label{VVEV}
\end{equation}
We have still to check the sign of the second derivative evaluated
at $b_{1}$ as given in (\ref{VVEV}) and require it to be positive:
\begin{equation}
\frac{\partial^{2} V_{np2}}{\partial b_{1}^{2}} =-2 a_{1}^{2}
\left\vert X_{1}\right\vert \cos (\vartheta_{W}-\vartheta
_{1}+a_{1}b_{1})>0  \Longleftrightarrow  p_{1}\in 2\mathbb{Z}+1.
\label{secder}
\end{equation}
Thus we realise that at the minimum
\begin{equation}
V_{np2} =-2 \left\vert X_{1}\right\vert=-2 \left\vert W\right\vert
\left\vert A_{1}\right\vert a_{1}\tau _{1}
\frac{e^{-a_{1}\tau_{1}}}{\mathcal{V}^{2}}.  \label{Fff}
\end{equation}
We notice that the phases of $W$ and $A_{1}$ do not enter into
$V_{np2}$ once the axion has been properly minimised and so,
without loss of generality, we can consider $W$ and $A_{1}\in
\mathbb{R} ^{+}$ from now on.

We may now study the full potential by combining equations
(\ref{Alfa}), (\ref{Imp}) and (\ref{Fff})
\begin{equation}
V \simeq \frac{K^{-1}_{11}}{\mathcal{V}^{2}} A_{1}^{2}a_{1}^{2}
e^{-2a_{1}\tau_{1}}-\frac{W_{0}}{\mathcal{V}^{2}}A_{1}a_{1}\tau
_{1}e^{-a_{1}\tau_{1}}+\frac{\hat{\xi}
}{\mathcal{V}^{3}}W_{0}^{2}, \label{final potenzial}
\end{equation}
where we have substituted $W$ with its tree-level expectation
value $W_{0}$ because the non-perturbative corrections are always
subleading by a power of $\mathcal{V}$. Moreover, we have dropped
all the factors since they are superfluous for our reasoning.

We would like to emphasize that we know that the first term in
(\ref{final potenzial}) is indeed positive. In fact it comes from
\begin{equation}
K_{11}^{-1}(\partial _{1}W)(\partial _{1}\bar{W}),
\end{equation}
and we know that the K\"{a}hler matrix is positive definite since
it gives rise to the kinetic terms. Moreover, as we have just
seen, the second term in (\ref{final potenzial}) comes from the
axion minimisation as so is definitely negative. Only the sign of
$V_{\left( \alpha'\right) }$ is in principle unknown, but the
condition $h_{2,1}(X)>h_{1,1}(X)$ ensures that it is positive.
This condition will turn out to be crucial in showing that the
volume direction has indeed a minimum at exponentially large
volume.

We need now to study the form of $K_{11}^{-1}$. For a general
Calabi-Yau, the inverse K\"{a}hler matrix with $\alpha'$
corrections included, reads \cite{bobkov}:
\begin{equation}
K_{ij}^{-1}=-\frac{2}{9}\left(2\mathcal{V}+\hat{\xi}\right)k_{ijk}t^{k}+
\frac{4\mathcal{V}-\hat{\xi}}{\mathcal{V}-\hat{\xi}}\tau _{i}\tau
_{j}, \label{inversaalfa}
\end{equation}
which at large volume becomes
\begin{equation}
K_{ij}^{-1}=-\frac{4}{9}\mathcal{V}k_{ijk}t^{k}+4\tau _{i}\tau
_{j}+\left( \text{terms subleading in }\mathcal{V}\right).
\label{inversa}
\end{equation}
Hence we can classify the behaviour of $K_{11}^{-1}$ depending on
the volume dependence of the quantity $k_{11j}t^{j}$ and find 4
different cases:

\begin{enumerate}
\item $k_{11j}t^{j}=0\text{ \ or \ }k_{11j}t^{j}\simeq\frac
{\tau _{1}^{1/2+3\alpha/2}}{\mathcal{V}^{\alpha }},$ $\alpha \geq
1$ $\Longrightarrow K_{11}^{-1}\simeq \tau _{1}^{2},$

\item $k_{11j}t^{j}=\frac{\tau _{1}^{1/2+3\alpha/2}}{\mathcal{V}^{\alpha }},$
$0<\alpha<1$ $\Longrightarrow K_{11}^{-1}\simeq\mathcal{V}^{\alpha
}\tau _{1}^{2-3\alpha /2},$ $0<\alpha <1,$

\item $k_{11j}t^{j}\simeq\sqrt{\tau _{1}}$ $\Longrightarrow
K_{11}^{-1}\simeq \mathcal{V}\sqrt{\tau _{1}},$

\item $k_{11j}t^{j}\simeq\mathcal{V}^{\alpha }\tau _{1}^{1/2-3\alpha/2},$
$\alpha >0$ $\Longrightarrow K_{11}^{-1}\simeq\mathcal{V}^{\alpha
}\tau _{1}^{2-3\alpha /2},$ $\alpha >1.$
\end{enumerate}

One could wonder why we are setting the conditions of the Theorem
on the elements of the inverse K\"{a}hler matrix and not on the
intersection numbers or the form of the overall volume of the
Calabi-Yau from which it would be easier to understand their
topological meaning. The reason is that it is the inverse
K\"{a}hler matrix which enters directly with the superpotential
into the form of the scalar potential which is the one that
determines the physics.

Moreover, the Claim applies if the superpotential has the
expression (\ref{explicit}), but in this case we can still make
linear field redefinitions that will not change $W$, corresponding
to proper changes of basis, of the form
\begin{equation}
\left\{
\begin{array}{c}
\tau _{j}\longrightarrow \tau _{j}'=\tau _{j},\text{ }\forall
j=1,...,N_{small}, \\
\tau _{j}\longrightarrow \tau
_{j}'=\tau_{j}+g_{1}(\tau_{i}),\text{ }\forall
j=N_{small}+1,...,h_{1,1}(X),
\end{array}
\right. \label{ChangeBasis}
\end{equation}
where $g_{1}(\tau_{i})$, $i=1,...,N_{small},$ is an homogeneous
function of degree 1. This means that the small 4-cycles will stay
small and the large ones will just be perturbed by the small ones.
We are therefore in the same situation and the physics should not
change. We conclude that the inverse K\"{a}hler matrix should not
change but both the intersection numbers and the form of the
volume can indeed vary. In fact, for an arbitrary Calabi-Yau, the
elements of the inverse K\"{a}hler matrix are given by:
\begin{equation}
K_{ij}^{-1}=-\frac{4}{9}\mathcal{V}k_{ijk}t^{k}+4\tau _{i}\tau
_{j}, \label{inversaAlfa}
\end{equation}
and so we see that in order to keep the form of $K^{-1}_{ij}$
unaltered, the quantity $(k_{ijk}t^{k})$ has not to vary, but the
intersection numbers $k_{ijk}$ can indeed change. This is the main
reason why we need to put our conditions on the $K^{-1}_{ij}$.

Let us illustrate this statement in the explicit example of the
orientifold of the Calabi-Yau threefold
$\mathbb{C}P^{4}_{[1,1,1,6,9]}$ whose volume in terms of 2-cycle
volumes is given by
\begin{equation}
\mathcal{V}=6\left(t_{5}^{3}+t_{4}^{3}\right).
\end{equation}
The corresponding 4-cycle volumes look like
\begin{equation}
\left\{
\begin{array}{c}
\tau _{4}=\frac{\partial \mathcal{V}}{\partial t_{4}}=18t_{4}^{2}, \\
\tau _{5}=\frac{\partial \mathcal{V}}{\partial t_{5}}=18t_{5}^{2},
\end{array}
\right. \text{ \ }\Longleftrightarrow \text{ \ }\left\{
\begin{array}{c}
t_{4}=-\frac{\sqrt{\tau _{4}}}{3\sqrt{2}}, \\
t_{5}=+\frac{\sqrt{\tau _{5}}}{3\sqrt{2}},
\end{array}
\right.
\end{equation}
and the volume in terms of the 4-cycles is
\begin{equation}
\mathcal{V}=\frac{1}{9
\sqrt{2}}\left(\tau_{5}^{3/2}-\tau_{4}^{3/2}\right).
\label{initial form}
\end{equation}
Finally the superpotential reads
\begin{equation}
W=W_{0}+A_{4}e^{-a_{4}T_{4}}+A_{5}e^{-a_{5}T_{5}}.
\end{equation}
It exists a well defined large volume limit when the 4-cycle
$\tau_{4}$ is kept small and $\tau_{5}$ is sent to infinity. In
this case the superpotential can be approximated as
\begin{equation}
W\simeq W_{0}+A_{4}e^{-a_{4}T_{4}}. \label{form of W}
\end{equation}
We can now perform the following field redefinition
\begin{equation}
\left\{
\begin{array}{c}
\tau _{4}\longrightarrow \tau _{4}'=\tau _{4}, \\
\tau _{5}\longrightarrow \tau _{5}'=\tau _{5}+\tau_{4},
\end{array}
\right. \label{redef}
\end{equation}
which will not change the form of $W$ (\ref{form of W}). However
now the volume reads
\begin{equation}
\mathcal{V}'=\frac{1}{9
\sqrt{2}}\left((\tau_{5}'-\tau_{4}')^{3/2}-\tau_{4}'^{3/2}\right)\simeq
\frac{1}{9
\sqrt{2}}\left(\tau_{5}'^{3/2}-\tau_{4}'\sqrt{\tau_{5}'}-\tau_{4}'^{3/2}\right),
\label{volumeprime}
\end{equation}
which is clearly different from the initial form (\ref{initial
form}). This means that also the intersection numbers are
different. However the elements of the inverse K\"{a}hler matrix
do not change. In particular we are interested in
$K^{-1}_{44}\simeq \mathcal{V}\sqrt{\tau_{4}}$ in this case as
$\tau_{4}$ is the small cycle. Its form stays unchanged since
$K'^{-1}_{44}\simeq \mathcal{V'}\sqrt{\tau_{4}'}$. From
(\ref{inversaAlfa}), this implies that
\begin{equation}
\sqrt{\tau_{4}'}=(k_{44k}'t'^{k})=k_{444}'t'_{4}+k_{445}'t'_{5},
\end{equation}
and one would tend to say that $k_{445}'$ has to be zero but we
know from (\ref{volumeprime}) that this is definitely not the
case. This means that the field redefinition (\ref{redef}) will
have the corresponding redefinition of the 2-cycle volumes which
will produce $t_{4}'$ and $t_{5}'$ that are both large 2-cycles
but such that the combination
$\left(k_{444}'t'_{4}+k_{445}'t'_{5}\right)$ stays small. This is
the reason why the form of the inverse K\"{a}hler matrix is left
invariant while the intersection numbers do vary. This can be
rephrased by saying that if $\tau_{j}$ is a small 4-cycle, in
general the corresponding $t_{j}$ has not to be a small 2-cycle
and viceversa. This is clear without the need to perform any field
redefinition in the case of the Calabi-Yau K3 fibration described
by the degree 12 hypersurface in $\mathbb{C}P^{4}_{[1,1,2,2,6]}$
whose overall volume in terms of 2-cycle volumes is
\begin{equation}
\mathcal{V}=t_{1}t_{2}^{2}+\frac{2}{3}t_{2}^{3},
\end{equation}
giving relations between the 2- and 4-cycle volumes, \bea
\label{tay} \tau _{1}=t_{2}^{2}, & \qquad &
\tau_{2}=2t_{2}\left(t_{1}+t_{2}\right),
\nonumber \\
t_{2}=\sqrt{\tau _{1}}, & \qquad & t_{1}=\frac{\tau _{2}-2\tau
_{1}}{2\sqrt{\tau _{1}}}, \label{viceversa} \eea that allow us to
write
\begin{equation}
\mathcal{V}=\frac{1}{2}\sqrt{\tau _{1}}\left( \tau
_{2}-\frac{2}{3}\tau _{1}\right) .  \label{volu11226}
\end{equation}
Looking at (\ref{volu11226}) we see that the large volume limit
can be performed keeping $\tau_{1}$ small and taking $\tau_{2}$
large. Nonetheless, as it is clear from (\ref{viceversa}), $t_{1}$
is big whereas $t_{2}$ is small. Therefore it is impossible to
impose that the quantity $k_{jji}t^{i}$ does not introduce any
volume dependence by requiring that some intersection numbers have
to vanish.

Going back to the proof of the LARGE Volume Claim for
$N_{small}=1$, let us assume that we are in case (3), so that
(\ref{final potenzial}) becomes
\begin{equation}
V \simeq \frac{\sqrt{\tau_{1}}}{\mathcal{V}} A_{1}^{2}a_{1}^{2}
e^{-2a_{1}\tau_{1}} -\frac{W_{0}}{\mathcal{V}^{2}}A_{1}a_{1}\tau
_{1} e^{-a_{1}\tau_{1}} +\frac{\hat{\xi}
}{\mathcal{V}^{3}}W_{0}^{2}, \label{Finale potenzial}
\end{equation}
and when we take the decompactification limit given by
\begin{equation}
\mathcal{V}\rightarrow \infty \ \ \text{with}\ \ e^{a_{1}\tau
_{1}}=\frac{\mathcal{V}}{W_{0}}, \label{VEry}
\end{equation}
all the terms in (\ref{Finale potenzial}) have the same volume
dependence
\begin{equation}
V \simeq \frac{W_{0}^{2}}{\mathcal{V}^{3}}\left[ \left(A_{1}a_{1}
-\sqrt{\tau _{1}}\right)A_{1}a_{1}\sqrt{\tau _{1}}+\hat{\xi}
\right]. \label{Jkj}
\end{equation}
We can finally express the scaling behaviour of (\ref{Jkj}) as
\begin{equation}
V\simeq \frac{W_{0}^{2}}{\mathcal{V}^{3}}\left( C_{1} \sqrt{\ln
\mathcal{V}}-C_{2} \ln \mathcal{V}+\hat{\xi} \right),
\label{Chiave}
\end{equation}
where $C_{1}$ and $C_{2}$ are positive constants of order 1 for
natural values of the parameter $A_{1}\simeq 1$. We conclude that
at large volume, the dominant term in (\ref{Chiave}) is the second
one and the scalar potential approaches zero from below. It is now
straightforward to argue that there must exist an exponentially
large volume AdS minimum.

In fact, at smaller volumes the dominant term in the potential
(\ref{Chiave}) is either the first or the third term, depending on
the exact value of the constants. Both are positive as we have
explained above. Thus at smaller volumes the potential is
positive, and so since it must go to zero at infinity from below,
there must exist a local AdS minimum along the direction in
K\"{a}hler moduli space where the volume changes.

One could argue that if at smaller volumes the dominant term in
(\ref{Chiave}) is the first one, then there is no need to require
$h_{2,1}(X)>h_{1,1}(X)$. In reality this is wrong, because $\xi
<0$ could still ruin the presence of the large volume minimum. In
fact we can rewrite the full scalar potential (\ref{Finale
potenzial}) as
\begin{equation}
V=\frac{\lambda }{\mathcal{V}}\sqrt{\tau _{1}}e^{-2a_{1}\tau
_{1}}-\frac{\mu }{\mathcal{V}^{2}}\tau _{1}e^{-a_{1}\tau
_{1}}+\frac{\hat{\xi} W_{0}^{2}}{\mathcal{V}^{3}},  \label{Asfk}
\end{equation}
where $\lambda $, $\mu $ and $\nu $ are positive constants
depending on the exact details of this model. We can integrate out
$\tau _{1}$, so ending up with just a potential for $\mathcal{V}$.
Under the requirement $a_{1}\tau _{1}\gg 1$, $\partial V /
\partial \tau _{1}=0$ gives
\begin{equation}
e^{-a_{1}\tau _{1}}=\frac{\mu }{2}\frac{\sqrt{\tau _{1}}}{\lambda
\mathcal{V}},
\end{equation}
which substituted back in (\ref{Asfk}) yields
\begin{equation}
V=-\frac{1}{2}\frac{\mu ^{2}}{\lambda }\frac{\tau
_{1}^{3/2}}{\mathcal{V}^{3}}+\frac{\hat{\xi}
W_{0}^{2}}{\mathcal{V}^{3}}\sim \frac{-\left( \ln \mathcal{V}
\right) ^{3/2}+\hat{\xi} W_{0}^{2}}{\mathcal{V}^{3}}
\end{equation}
and is straightforward to see that we need $\hat{\xi} >0$ even
though the dominant term at small volumes in (\ref{Asfk}) is the
first one.

It remains to show that the scalar potential has also a minimum in
the other direction of the moduli space. In order to do that, let
us fix the Calabi-Yau volume and see what happens if we vary the
small K\"{a}hler modulus along that surface. Then as one
approaches the walls of the K\"{a}hler cone the positive first
term in (\ref{Finale potenzial}) dominates since it has the fewest
powers of volume in the denominator and the exponential
contributions of the modulus that is becoming small cannot be
neglected. Thus at large overall volume, we expect the potential
to grow in the positive direction towards the walls of the
K\"{a}hler cone.

On the other hand, when the small K\"{a}hler modulus becomes
bigger then the dominant term in (\ref{Finale potenzial}) is the
positive $V_{(\alpha')}$ due to the exponential suppressions in
the other two terms. Given that the potential is negative along
the special direction in the moduli space that we have identified
and eventually raises to be positive or to vanish in the other
direction, we are sure to have an AdS exponentially large volume
minimum.

Since $V\sim \mathcal{O}(1/\mathcal{V}^{3})$ at the minimum, while
$-3e^{K}\left\vert W\right\vert ^{2}\sim
\mathcal{O}(1/\mathcal{V}^{2})$, it is clear that this minimum is
non-supersymmetric. We can heuristically see why the minimum we
are arguing for can be at exponentially large volume. The naive
measure of its location is the value of the volume at which the
negative term in (\ref{Chiave}) becomes dominant. As this occurs
only when $(\ln \mathcal{V})$ is large, we expect to find the
vacuum at large values of $(\ln \mathcal{V})$.

In reality the way in which we have taken the limit (\ref{VEry}),
tells us how the volume will scale, even though this can very well
not be the correct location of the minimum
\begin{equation}
\mathcal{V}\sim W_{0} e^{a_{1}\tau _{1}}. \label{Hyu}
\end{equation}
Looking at (\ref{Hyu}) we realise that $W_{0}$ cannot be too
small, otherwise we would get a small volume minimum merging with
the KKLT one and our derivation would not make sense anymore.
However $W_{0}$ is multiplying an exponential, which means that in
order to destroy the large volume minimum $W_{0}$ has to be really
small.

Furthermore, we stress that there is no need to require
$h_{2,1}(X) \gg h_{1,1}(X)$ instead of just
$h_{2,1}(X)>h_{1,1}(X)$. In fact, in this proof we have used
$\hat{\xi}$ instead of $\xi$, so obscuring the presence of any
factors of $g_{s}$ but, as it is written explicitly in
(\ref{explicit}), in Einstein frame $\hat{\xi}$ is equivalent to
$\xi / g_{s}^{3/2}$. Therefore if we just have
$h_{2,1}(X)>h_{1,1}(X)$ then we can still adjust $g_{s}$ to make
sure that the AdS minimum is indeed at large volume.

We are now able to understand what happens if $K^{-1}_{11}$ is not
in case (3). For example, when it is in case (4) then the first
term in (\ref{final potenzial}) beats all the other ones and along
the direction (\ref{VEry}) the scalar potential either presents a
runaway or has no minimum at large volume depending on the exact
value of $\alpha$.

Moreover if $K^{-1}_{11}$ is in case (1) or (2) then the first
term in (\ref{final potenzial}) is subleading with respect to the
other two and at leading order in the volume, the scalar potential
looks like
\begin{equation}
V \simeq -\frac{W_{0}}{\mathcal{V}^{2}}A_{1}a_{1}\tau
_{1}e^{-a_{1}\tau_{1}}+\frac{\hat{\xi}
}{\mathcal{V}^{3}}W_{0}^{2}. \label{POt}
\end{equation}
The minimisation equation for $\tau_{1}$, $\partial V/\partial
\tau_{1}=0$, admits the only possible solution $a_{1}\tau_{1}=1$,
that has to be discarded since we need $a_{1}\tau_{1}\gg 1$ in
order to avoid higher instanton corrections.

Finally, let us argue in favour of the last statement of Claim 1.
At the end of all our derivation we realised that the small
K\"{a}hler modulus $\tau_{1}$ plus a particular combination which
is the overall volume are stabilised. Therefore we have in general
$(N_{small}+1)$ fixed K\"{a}hler moduli and is straightforward to
see that if we have just one big K\"{a}hler modulus then it will
be fixed, whereas if we have more than one big K\"{a}hler moduli,
only one of them will be fixed and the others will give rise to
exactly $(h_{1,1}(X)-N_{small}-1)$ flat directions. This is
because they do not appear in the non-perturbative corrections to
the superpotential due to the limit (\ref{limit}). This terminates
our proof of the LARGE Volume Claim for $N_{small}$=1.
\end{proof}

\subsection{Proof for $N_{small}>1$}
\label{Appendix A2}

\begin{proof} (LARGE Volume Claim for $N_{small}>1$)
When $N_{small}>1$ the situation is more involved due to the
presence of cross terms. However $V_{\left( \alpha'\right) }$ has
still the form (\ref{Alfa}). Without loss of generality, we shall
focus on the case with $N_{small}=2$ K\"{a}hler moduli, which we
will call $\tau_{1}$ and $\tau_{2}$. $V_{np1}$ generalises to
\begin{eqnarray}
&&V_{np1}\underset{\mathcal{V}\rightarrow \infty
}{\longrightarrow}
e^{K}\sum\limits_{j,k=1}^{2}K_{jk}^{-1}a_{j}A_{j}a_{k}\bar{A}
_{k}e^{-\left( a_{j}T_{j}+a_{k}\bar{T}_{k}\right) } \label{imp} \\
&=&e^{K}\left\{
\sum\limits_{j=1}^{2}K_{jj}^{-1}a_{j}^{2}\left\vert
A_{j}\right\vert ^{2}e^{-2a_{j}\tau
_{j}}+K_{12}^{-1}a_{1}A_{1}a_{2}\bar{A}_{2}e^{-\left( a_{1}\tau
_{1}+a_{2}\tau _{2}\right) }e^{i\left(
a_{2}b_{2}-a_{1}b_{1}\right) }\right\}.  \nonumber
\end{eqnarray}
In order to consider separately the axion-dependent part of
$V_{np1}$, we write
\begin{equation}
V_{np1}=V_{np1}^{real}+V_{np1}^{AX}.
\end{equation}
Switching to the study of $V_{np2}$, we find that
\begin{equation}
V_{np2}\underset{\mathcal{V}\rightarrow \infty }{\longrightarrow }
-e^{K}\sum\limits_{k=1}^{h_{1,1}}\sum\limits_{j=1}^{2}K_{jk}^{-1}
\left[ \left( a_{j}A_{j}e^{-a_{j}\tau
_{j}}e^{-ia_{j}b_{j}}\bar{W}\partial _{ \bar{T}_{k}}K\right)
+\left( a_{j}\bar{A}_{j}e^{-a_{j}\tau
_{j}}e^{+ia_{j}b_{j}}W\partial _{T_{k}}K\right) \right] ,
\label{JJjh}
\end{equation}
where we have used the fact that $K_{jk}^{-1}=K_{kj}^{-1}$.
Equation (\ref{JJjh}) can be rewritten as
\begin{eqnarray}
&&V_{np2}\underset{\mathcal{V}\rightarrow \infty
}{\longrightarrow}-e^{K}\sum\limits_{k=1}^{h_{1,1}}\sum
\limits_{j=1}^{2}K_{jk}^{-1}\left( \partial _{T_{k}}K\right)a_{j}
e^{-a_{j}\tau _{j}}\left[ \left(
A_{j}\bar{W}e^{-ia_{j}b_{j}}\right) +\left(
\bar{A}_{j}We^{+ia_{j}b_{j}}\right) \right]  \notag \\
&=&\sum\limits_{j=1}^{2}\left(
X_{j}e^{+ia_{j}b_{j}}+\bar{X}_{j}e^{-ia_{j}b_{j}}\right) ,
\end{eqnarray}
where
\begin{equation}
X_{j}\equiv -e^{K}K_{jk}^{-1}\left( \partial _{T_{k}}K\right)
a_{j}\bar{A}_{j}W e^{-a_{j}\tau _{j}}.  \label{defin1}
\end{equation}
We note that for a general Calabi-Yau, the following relation
holds:
\begin{equation}
K_{jk}^{-1}\left( \partial _{T_{k}}K\right) =-2\tau _{j},
\end{equation}
and thus the definition (\ref{defin1}) can be simplified to
\begin{equation}
X_{j}\equiv 2e^{K}a_{j}\tau _{j}\left\vert A_{j}\right\vert e^{-i
\vartheta_{j}}\left\vert W \right\vert e^{i \vartheta_{W}}
e^{-a_{j}\tau _{j}}=\left\vert X_{j}\right\vert e^{i(\vartheta
_{W}-\vartheta_{j})}. \label{DEFin1}
\end{equation}
Therefore
\begin{equation}
V_{np2}\underset{\mathcal{V}\rightarrow \infty }{\longrightarrow }
\sum\limits_{j=1}^{2}\left\vert X_{j}\right\vert \left(
e^{+i\left( \vartheta_{W}-\vartheta _{j}+a_{j}b_{j}\right)
}+e^{-i\left( \vartheta_{W}-\vartheta _{j}+a_{j}b_{j}\right)
}\right).
\end{equation}
Let us now reconsider $V_{np1}^{AX}$, which we had set aside for a
moment. It can be rewritten as
\begin{equation}
V_{np1}^{AX}=e^{K} K_{12}^{-1}a_{1}a_{2}e^{-\left( a_{1}\tau
_{1}+a_{2}\tau _{2}\right) }\left( A_{1}\bar{A}_{2}e^{i\left(
a_{2}b_{2}-a_{1}b_{1}\right) }+A_{2}\bar{A}_{1}e^{-i\left(
a_{2}b_{2}-a_{1}b_{1}\right) }\right),
\end{equation}
and finally as
\begin{equation}
V_{np1}^{AX}=Y_{12}e^{i\left( a_{2}b_{2}-a_{1}b_{1}\right)
}+\bar{Y}_{12}e^{-i\left( a_{2}b_{2}-a_{1}b_{1}\right)},
\end{equation}
where
\begin{equation}
Y_{12}\equiv e^{K}K_{12}^{-1}a_{1}a_{2}A_{1}\bar{A}_{2}e^{-\left(
a_{1}\tau _{1}+a_{2}\tau _{2}\right) }=\left\vert
Y_{12}\right\vert e^{i(\vartheta _{1}-\vartheta_{2})}.
\label{DEFin2}
\end{equation}
Therefore
\begin{equation}
V_{np1}^{AX}=\left\vert Y_{12}\right\vert \left( e^{i\left(
\vartheta _{1}-\vartheta_{2}+a_{2}b_{2}-a_{1}b_{1}\right)
}+e^{-i\left( \vartheta
_{1}-\vartheta_{2}+a_{2}b_{2}-a_{1}b_{1}\right)}\right).
\end{equation}
Thus, the full axion-dependent part of the scalar potential
$V_{AX}$ looks like
\begin{equation}
V_{AX}=V_{np2}+V_{np1}^{AX}=2\sum\limits_{j=1}^{2}\left\vert
X_{j}\right\vert \cos \left( \vartheta_{W}-\vartheta
_{j}+a_{j}b_{j}\right) +2\left\vert Y_{12}\right\vert \cos \left(
\vartheta _{1}-\vartheta_{2}+a_{2}b_{2}-a_{1}b_{1}\right) .
\end{equation}

\subsubsection{Axion stabilisation}

$V_{AX}$ is a scalar function of the axions $b_{1}$ and $b_{2}$
whereas $\vartheta_{1}$, $\vartheta_{2}$ and $\vartheta_{W}$ are
to be considered just as parameters. In order to find a minimum
for $V_{AX}$ let us set its gradient to zero
\begin{equation}
\left\{
\begin{array}{c}
\partial V_{AX}/\partial b_{1}=0\text{ \
}\Longleftrightarrow \text{ \ }\left\vert X_{1}\right\vert \sin
(\vartheta _{W}-\vartheta _{1}+a_{1}b_{1})=+\left\vert
Y_{12}\right\vert \sin \left( \vartheta
_{1}-\vartheta _{2}+a_{2}b_{2}-a_{1}b_{1}\right) ,\text{ } \\
\partial V_{AX}/\partial b_{2}=0\text{ \ \
}\Longleftrightarrow \text{ \ }\left\vert X_{2}\right\vert \sin
(\vartheta _{W}-\vartheta _{2}+a_{2}b_{2})=-\left\vert
Y_{12}\right\vert \sin \left( \vartheta _{1}-\vartheta
_{2}+a_{2}b_{2}-a_{1}b_{1}\right) ,\text{ \ }
\end{array}
\right.   \label{GGgui}
\end{equation}
The solution of (\ref{GGgui}) is given by
\begin{equation}
\left\{
\begin{array}{c}
\psi _{1}\equiv \left( \vartheta _{W}-\vartheta
_{1}+a_{1}b_{1}\right)
=p_{1}\pi ,\text{ \ }p_{1}\in \mathbb{Z}, \\
\psi _{2}\equiv \left( \vartheta _{W}-\vartheta
_{2}+a_{2}b_{2}\right) =p_{2}\pi ,\text{ \ }p_{2}\in \mathbb{Z},
\end{array}
\right.   \label{VEVov}
\end{equation}
and
\begin{equation}
\psi_{12}\equiv \left( \vartheta
_{1}-\vartheta_{2}+a_{2}b_{2}-a_{1}b_{1}\right) =p_{12}\pi ,\text{
\ } p_{12}\in \mathbb{Z}. \label{ADjust}
\end{equation}
From (\ref{VEVov}) equation (\ref{ADjust}) requires
$p_{12}=p_{2}-p_{1}$. Let us summarise the points where the
gradient of the axion potential is zero in the following table
\begin{equation}
\begin{tabular}{|c|c|c|c|c|}
\hline & (a) & (b) & (c) & (d) \\ \hline $\cos \psi _{1}$ & +1 &
-1 & +1 & -1 \\ \hline $\cos \psi _{2}$ & +1 & -1 & -1 & +1 \\
\hline $\cos \psi _{12}$ & +1 & +1 & -1 & -1 \\ \hline
\end{tabular}
\label{TABLE}
\end{equation}
We notice that the phases of $W$, $A_{1}$ and $A_{2}$ will not
enter into $V_{np2}$ once the axions have been properly minimised
and so, without loss of generality, we can consider $W$, $A_{1}$
and $A_{2}\in \mathbb{R} ^{+}$ from now on.

We have still to check the Hessian matrix evaluated at $b_{1}$ and
$b_{2}$ as given in (\ref{VEVov}) and require it to be positive
definite. Its diagonal elements are given by
\begin{equation}
\left\{
\begin{array}{c}
\partial ^{2}V_{AX}/\partial b_{1}^{2}=-2a_{1}^{2}\left(
\left\vert X_{1}\right\vert \cos \psi _{1}+\left\vert
Y_{12}\right\vert \cos \psi _{12}
\right) , \\
\partial ^{2}V_{AX}/\partial b_{2}^{2}=-2a_{2}^{2}\left(
\left\vert X_{2}\right\vert \cos \psi _{2}+\left\vert
Y_{12}\right\vert \cos \psi _{12} \right),
\end{array}
\right.   \label{HEssian1}
\end{equation}
whereas the non-diagonal ones read
\begin{equation}
\frac{\partial^{2} V_{AX}}{\partial b_{2} \partial b_{1}}
=\frac{\partial^{2} V_{AX}}{\partial b_{1} \partial b_{2}}=2 a_{1}
a_{2} \left\vert Y_{12}\right\vert \cos \psi_{12}.
\label{HEssian2}
\end{equation}
We can diagonalise the Hessian $\mathcal{H}$ to the identity by
decomposing it $a$ $la$ Choleski into
$\mathcal{H}=\mathcal{U}^{T}\mathbb{I}\mathcal{U}$, where the
elements of the upper triangular matrix $\mathcal{U}$ are given by
the following recursive relations:
\begin{equation}
\mathcal{U}_{11}^{2}=-2a_{1}^{2}\left(\left\vert X_{1}\right\vert
\cos\psi_{1}+\left\vert Y_{12}\right\vert \cos \psi_{12}\right),
\label{UU11}
\end{equation}
\begin{equation}
\mathcal{U}_{12}=\frac{2a_{1}a_{2}\left\vert Y_{12}\right\vert
\cos\psi_{12}}{\mathcal{U}_{11}}, \label{UU12}
\end{equation}
\begin{equation}
\mathcal{U}_{22}^{2}=-\mathcal{U}_{12}^{2}-2a_{2}^{2} \left(
\left\vert X_{2}\right\vert \cos\psi_{2} +\left\vert
Y_{12}\right\vert \cos\psi_{12}\right), \label{UU22}
\end{equation}
with $\mathcal{U}_{21}=0$. Determining if the Hessian is positive
definite is equal to checking that $\mathcal{U}$ is a real matrix.
Looking at (\ref{UU12}) we realise that $\mathcal{U}_{12}$ is
automatically real if $\mathcal{U}_{11}$ is real. Hence we have to
make sure just that both $\mathcal{U}_{11}^{2}>0$ and
$\mathcal{U}_{22}^{2}>0$. When we analyse the cases listed in
table (\ref{TABLE}) we realise that:

\begin{enumerate}
\item[(a)] can never be a minimum since $\textit{ }\mathcal{U}_{11}^{2}<0$;
in reality it turns out to be always a maximum,

\item[(b)] is a minimum only if $\left\vert X_{1}\right\vert >\left\vert
Y_{12}\right\vert $ and $\left\vert X_{1}\right\vert \left\vert
X_{2}\right\vert >\left\vert Y_{12}\right\vert \left( \left\vert
X_{1}\right\vert +\left\vert X_{2}\right\vert \right) ,$

\item[(c)] is a minimum only if $\left\vert Y_{12}\right\vert >\left\vert
X_{1}\right\vert $ and $\left\vert X_{2}\right\vert \left\vert
Y_{12}\right\vert >\left\vert X_{1}\right\vert \left( \left\vert
X_{2}\right\vert +\left\vert Y_{12}\right\vert \right) ,$

\item[(d)] is a minimum only if  $\left\vert Y_{12}\right\vert \left\vert
X_{1}\right\vert >\left\vert X_{2}\right\vert \left( \left\vert
X_{1}\right\vert +\left\vert Y_{12}\right\vert \right) ,$
\end{enumerate}
where, according to the definitions (\ref{DEFin1}) and
(\ref{DEFin2}), we have
\begin{equation}
\left\{
\begin{array}{c}
\left\vert X_{1}\right\vert =2\left\vert A_{1}\right\vert
a_{1}\tau _{1}\left\vert W \right\vert \frac{e^{-a_{1}\tau
_{1}}}{\mathcal{V}^{2}}, \\
\left\vert X_{2}\right\vert =2\left\vert A_{2}\right\vert
a_{2}\tau _{2}\left\vert W \right\vert \frac{e^{-a_{2}\tau
_{2}}}{\mathcal{V}^{2}}, \\
\left\vert Y_{12}\right\vert =K_{12}^{-1}\left\vert
A_{1}\right\vert a_{1}\left\vert A_{2}\right\vert a_{2}\frac{
e^{-a_{1}\tau _{1}}e^{-a_{2}\tau _{2}}}{\mathcal{V}^{2}}.
\end{array}
\right. \label{ineq}
\end{equation}
In order to study the cases (b), (c) and (d), it is therefore
crucial to know the order of magnitude of the two exponentials
$e^{-a_{1}\tau _{1}}$ and $e^{-a_{2}\tau _{2}}$ given by their
scaling behaviour in the volume. This depends on the direction we
are looking at to find the minimum in the large volume limit which
can be performed in three different ways:
\begin{eqnarray}
I)\text{ }\mathcal{V} &\sim &e^{\gamma a_{1}\tau _{1}}\sim
e^{\gamma a_{2}\tau _{2}},\text{ \ }\gamma \in \mathbb{R}^{+}
\notag \\
II)\text{ }\mathcal{V} &\sim &e^{\beta a_{1}\tau _{1}}\sim
e^{\gamma a_{2}\tau _{2}},\text{ \ }\beta <\gamma,
\text{ \ }\gamma,\beta \in \mathbb{R}^{+},  \label{DIrections} \\
III)\mathcal{V} &\sim &e^{\beta a_{1}\tau _{1}}\sim e^{\gamma
a_{2}\tau _{2}},\text{ \ }\beta >\gamma\text{ \ }\gamma,\beta \in
\mathbb{R}^{+}.  \notag
\end{eqnarray}
Finally we need also to know the form of $K^{-1}_{12}$. We can
classify its behaviour according to the volume dependence of the
quantity $k_{12j}t^{j}$ and find 4 different cases:

\begin{enumerate}
\item $k_{12j}t^{j}=0\text{ \ or \ }k_{12j}t^{j}=\frac{f
(\tau _{1},\tau_{2})}{\mathcal{V}^{\alpha }},$ $\alpha \geq 1$
$\Longrightarrow K_{12}^{-1}=\tau _{1}\tau _{2};$

\item $k_{12j}t^{j}=\frac{g_{\gamma }(\tau _{1},\tau_{2})}{\mathcal{V}^{\alpha
}},$ $0<\alpha<1$, $g$ homogeneous function of degree $\gamma
=\frac{1+3\alpha}{2}$ $\Longrightarrow
K_{12}^{-1}=\mathcal{V}^{\alpha }g_{2-3\alpha /2}(\tau _{1},\tau
_{2}),$ \ $0<\alpha <1;$

\item $k_{12j}t^{j}=f_{1/2}(\tau _{1},\tau_{2}),$ $f$
homogeneous function of degree $1/2$ $\Longrightarrow
K_{12}^{-1}=\mathcal{V}f_{1/2}(\tau _{1},\tau _{2});$

\item $k_{12j}t^{j}=\mathcal{V}^{\alpha }h_{\beta }(\tau _{1},\tau_{2}),$
$\alpha >0$, $h$ homogeneous function of degree $\beta =
\frac{1-3\alpha}{2}$ $\Longrightarrow
K_{12}^{-1}=\mathcal{V}^{\alpha }h_{2-3\alpha /2}(\tau _{1},\tau
_{2}),$ \ $\alpha >1.$
\end{enumerate}
Let us now focus on the axion minimisation by analysing each of
these 4 cases in full detail. For each case we will have to study
if the inequalities (b), (c) and (d) admit a solution for any of
the three possible ways to take the large volume limit as
expressed in (\ref{DIrections}). We will always consider natural
values of the parameters $\left\vert A_{1} \right\vert \simeq
\left\vert A_{2} \right\vert \simeq \left\vert W \right\vert\simeq
1$.

From (I) of (\ref{DIrections}), we can immediately realise that,
regardless of the form of $K^{-1}_{12}$, at large volume
$\left\vert X_{1}\right\vert$ and $\left\vert X_{2}\right\vert$
have the same scaling with the volume and so we can denote both of
them as $\left\vert X \right\vert$. It is then straightforward to
see that both the second (c)-condition and the (d)-condition can
never be satisfied. In fact they take the form
\begin{equation}
\left\vert X \right\vert \left\vert Y_{12} \right\vert >
\left\vert X \right\vert \left\vert Y_{12} \right\vert +
\left\vert X \right\vert^{2},
\end{equation}
which is manifestly an absurd. This implies that neither (c) nor
(d) can be a minimum along the direction (I) for any value of
$K^{-1}_{12}$. A further analysis reveals that the points (c) and
(d) can never be maxima so since we proved that they cannot be
minima, they are forced to be saddle points. We do not present the
details of this analysis here since it is not important for our
reasoning. On the other hand, the first (b)-condition is
automatically satisfied if the second one is true since it reduces
to
\begin{equation}
\left\vert X \right\vert^{2} > 2 \left\vert X \right\vert
\left\vert Y_{12} \right\vert\textit{ \
}\Longleftrightarrow\textit{ \ }\left\vert X \right\vert > 2
\left\vert Y_{12} \right\vert. \label{disug}
\end{equation}

From (II) of (\ref{DIrections}), we also notice that, regardless
of the form of $K^{-1}_{12}$, at large volume $\left\vert
X_{1}\right\vert<\left\vert X_{2}\right\vert$ since
$\beta<\gamma$. It is then straightforward to see that in this
situation the (d)-condition can never be satisfied. This implies
that (d) is always a saddle point along the direction (II) for any
value of $K^{-1}_{12}$.

Furthermore (III) of (\ref{DIrections}) implies that, regardless
of the form of $K^{-1}_{12}$, at large volume $\left\vert
X_{1}\right\vert>\left\vert X_{2}\right\vert$ as $\beta>\gamma$.
Then we immediately see that the second (c)-condition can never be
satisfied. Therefore (c) is always a saddle point along the
direction (III) for any value of $K^{-1}_{12}$.

\bigskip

\textbf{Case (1):} $K^{-1}_{12}\simeq \tau_{1}\tau_{2}$

\begin{enumerate}
\item[$\bullet$] direction (I)

The volume dependence of the parameters (\ref{ineq}) is
\begin{equation}
\left\{
\begin{array}{l}
\left\vert X_{1} \right\vert \simeq \left\vert X_{2} \right\vert
\simeq \mathcal{V}^{-(2+1/\gamma)}\ln\mathcal{V}, \\
\left\vert Y_{12} \right\vert \simeq
\mathcal{V}^{-(2+2/\gamma)}(\ln\mathcal{V})^{2}.
\end{array}
\right.  \label{ABsol}
\end{equation}
Looking at (\ref{ABsol}), we realise that at large volume
$\left\vert X_{1}\right\vert > 2\left\vert Y_{12} \right\vert$.
Therefore the second (b)-condition (\ref{disug}) is satisfied and
(b) is a minimum of the axion potential.

\item[$\bullet$] direction (II)

The parameters (\ref{ineq}) now read
\begin{equation}
\left\{
\begin{array}{l}
\left\vert X_{1} \right\vert \simeq
\mathcal{V}^{-(2+1/\beta)}\ln\mathcal{V}, \\
\left\vert X_{2} \right\vert \simeq
\mathcal{V}^{-(2+1/\gamma)}\ln\mathcal{V}, \\
\left\vert Y_{12} \right\vert \simeq
\mathcal{V}^{-(2+1/\beta+1/\gamma)}(\ln\mathcal{V})^{2}.
\end{array}
\right.  \label{dabs}
\end{equation}
Looking at (\ref{dabs}), we realise that at large volume
$\left\vert X_{1}\right\vert > \left\vert Y_{12} \right\vert$,
which implies that (c) is a saddle point. Thus the first
(b)-condition is satisfied and the second becomes
\begin{equation}
\frac{(\ln \mathcal{V})^{2}}{\mathcal{V}^{4+1/\beta +1/\gamma
}}>\frac{(\ln \mathcal{V})^{2}}{\mathcal{V}^{2+1/\beta +1/\gamma
}}\left( \frac{\ln \mathcal{V}}{\mathcal{V}^{2+1/\beta
}}+\frac{\ln \mathcal{V}}{\mathcal{V}^{2+1/\gamma}}\right)
\underset{1/\beta >1/\gamma }{\simeq }\frac{(\ln
\mathcal{V})^{3}}{\mathcal{V}^{4+1/\beta +2/\gamma}},
\end{equation}
which is true at large volume for values of $\gamma>\beta>0$ not
extremely big. Thus (b) is a minimum of the axion potential.

\item[$\bullet$] direction (III)

The parameters (\ref{ineq}) take the same form as (\ref{dabs}) but
now with $\beta>\gamma$. We have still $\left\vert
X_{1}\right\vert > \left\vert Y_{12} \right\vert$, which implies
that the first (b)-condition is satisfied. The second looks like
\begin{equation}
\frac{(\ln \mathcal{V})^{2}}{\mathcal{V}^{4+1/\beta +1/\gamma
}}>\frac{(\ln \mathcal{V})^{2}}{\mathcal{V}^{2+1/\beta +1/\gamma
}}\left( \frac{\ln \mathcal{V}}{\mathcal{V}^{2+1/\beta
}}+\frac{\ln \mathcal{V}}{\mathcal{V}^{2+1/\gamma}}\right)
\underset{1/\beta<1/\gamma }{\simeq }\frac{(\ln
\mathcal{V})^{3}}{\mathcal{V}^{4+2/\beta +1/\gamma}},
\end{equation}
which is true at large volume. Thus (b) is a minimum of the axion
potential. On the contrary the simplified (d)-condition reads
\begin{equation}
\frac{\ln\mathcal{V}}{\mathcal{V}^{1/\beta}}>1
+\frac{\ln\mathcal{V}}{\mathcal{V}^{1/\gamma}}
\end{equation}
which at large volume is clearly false for values of
$\beta>\gamma>0$ not extremely big. It follows that in this case
(d) is a saddle point.
\end{enumerate}

Let us summarise the results found in case (1) in the following
table
\begin{center}
\begin{tabular}{|c|c|c|c|}
\hline & (I) & (II) & (III) \\ \hline (a) & max & max & max \\
\hline (b) & min & min & min \\ \hline (c) & saddle & saddle &
saddle \\ \hline (d) & saddle & saddle & saddle \\ \hline
\end{tabular}
\end{center}

\bigskip

\textbf{Case (2):} $K_{12}^{-1}=\mathcal{V}^{\alpha }g_{2-3\alpha
/2}(\tau _{1},\tau _{2}),$ \ $0<\alpha <1$

\begin{enumerate}
\item[$\bullet$] direction (I)

The volume dependence of the parameters (\ref{ineq}) now reads
\begin{equation}
\left\{
\begin{array}{l}
\left\vert X_{1} \right\vert \simeq \left\vert X_{2} \right\vert
\simeq \mathcal{V}^{-(2+1/\gamma)}\ln\mathcal{V}, \\
\left\vert Y_{12} \right\vert \simeq
\mathcal{V}^{-(2+2/\gamma-\alpha)}(\ln\mathcal{V})^{2-3\alpha/2}.
\end{array}
\right.  \label{tABsol}
\end{equation}
Substituting the expressions (\ref{tABsol}) in (\ref{disug}) we
find
\begin{equation}
1>2
\frac{(\ln\mathcal{V})^{1-3\alpha/2}}{\mathcal{V}^{1/\gamma-\alpha}},
\label{IO}
\end{equation}
which at large volume is true if $\alpha<(1/\gamma)$, false if
$\alpha< (1/\gamma)$ or $\alpha=(1/\gamma)\leq 2/3$. On the
contrary, for $2/3<\alpha=(1/\gamma)<1$ the minimum is present.
Thus (b) can be a minimum of the axion potential.

\item[$\bullet$] direction (II)

The parameters (\ref{ineq}) now read
\begin{equation}
\left\{
\begin{array}{l}
\left\vert X_{1} \right\vert \simeq
\mathcal{V}^{-(2+1/\beta)}\ln\mathcal{V}, \\
\left\vert X_{2} \right\vert \simeq
\mathcal{V}^{-(2+1/\gamma)}\ln\mathcal{V}, \\
\left\vert Y_{12} \right\vert \simeq
\mathcal{V}^{-(2+1/\beta+1/\gamma-\alpha)}(\ln\mathcal{V})^{2-3\alpha/2}.
\end{array}
\right.  \label{ddabso}
\end{equation}
The first (b)-condition becomes
\begin{equation}
1>\frac{(\ln
\mathcal{V})^{1-3\alpha/2}}{\mathcal{V}^{1/\gamma-\alpha}},
\label{question}
\end{equation}
that is satisfied if either $\alpha<(1/\gamma)$ or
$2/3<\alpha=(1/\gamma)<1$. Otherwise (\ref{question}) is false
unless $\alpha=(1/\gamma)=2/3$ in which case we cannot conclude
anything just looking at the volume dependence. However the second
(b)-condition reads
\begin{equation}
1>(\ln{\mathcal{V}})^{1-3\alpha/2}\left(\frac{1}
{\mathcal{V}^{1/\beta-\alpha}}+\frac{1}{\mathcal{V}^{1/\gamma-\alpha}}\right),
\label{SECONDB}
\end{equation}
which is definitely true at large volume if $\alpha<(1/\gamma)$ or
$2/3<\alpha=(1/\gamma)<1$. On the contrary, in the case
$\alpha=(1/\gamma)=2/3\Leftrightarrow (1/\beta-2/3)>0$,
(\ref{SECONDB}) becomes
\begin{equation}
1>1+\frac{1} {\mathcal{V}^{1/\beta-2/3}},
\end{equation}
which is clearly impossible. Thus (b) can be a minimum of the
axion potential. We need now just to study the case (c) for which
the first inequality is
\begin{equation}
1<\frac{(\ln
\mathcal{V})^{1-3\alpha/2}}{\mathcal{V}^{1/\gamma-\alpha}},
\label{Question}
\end{equation}
that is satisfied if either $(1/\gamma)<\alpha<1$ or
$\alpha=(1/\gamma)<2/3$. Otherwise (\ref{Question}) is false
unless $\alpha=(1/\gamma)=2/3$ in which case we cannot conclude
anything just looking at the volume dependence. However the second
(c)-condition can be simplified to give
\begin{equation}
\frac{(\ln
\mathcal{V})^{1-3\alpha/2}}{\mathcal{V}^{1/\gamma-\alpha}}>
1+\frac{(\ln
\mathcal{V})^{1-3\alpha/2}}{\mathcal{V}^{1/\beta-\alpha}},
\label{llook}
\end{equation}
which is clearly satisfied at large volume if either
$(1/\gamma)<\alpha<1$ or $\alpha=(1/\gamma)<2/3$. On the contrary
when $\alpha=(1/\gamma)=2/3\Leftrightarrow (1/\beta-2/3)>0$,
(\ref{llook}) becomes
\begin{equation}
1> 1+\frac{1}{\mathcal{V}^{1/\beta-2/3}},
\end{equation}
which is clearly false.

\item[$\bullet$] direction (III)

The parameters (\ref{ineq}) assume the same form as (\ref{ddabso})
but now with $\beta>\gamma$. Following lines of reasoning similar
to those used for direction (II), we get the results summarised in
the following table along with all the others for case (2).
\end{enumerate}

\begin{center}
\begin{tabular}{|c|c|c|c|}
\hline & (I) & (II) & (III) \\ \hline (a) & max & max & max \\
\hline (b) & $\left\{
\begin{array}{c}
\alpha <1/\gamma \text{ min,}\\
2/3<\alpha =1/\gamma<1 \text{ min,}\\
\alpha =1/\gamma\leq 2/3\text{ saddle,}\\
\alpha >1/\gamma \text{ saddle.}
\end{array}
\right.$ & $\left\{
\begin{array}{c}
\alpha <1/\gamma \text{ min,}\\
2/3<\alpha =1/\gamma<1 \text{ min,}\\
\alpha =1/\gamma\leq 2/3\text{ saddle,}\\
\alpha >1/\gamma \text{ saddle.}
\end{array}
\right.$ & $\left\{
\begin{array}{c}
\alpha <1/\beta \text{ min,}\\
2/3< \alpha=1/\beta <1 \text{ min,}\\
\alpha=1/\beta\leq 2/3, \text{ saddle,}\\
\alpha>1/\beta \text{ saddle.}
\end{array}
\right.$ \\ \hline (c) & saddle & $\left\{
\begin{array}{c}
\alpha <1/\gamma \text{ saddle,}\\
2/3 \leq\alpha =1/\gamma<1 \text{ saddle,}\\
\alpha =1/\gamma< 2/3\text{ min,}\\
1/\gamma<\alpha<1 \text{ min.}
\end{array}
\right.$ & saddle \\ \hline (d) & saddle & saddle & $\left\{
\begin{array}{c}
\alpha <1/\beta \text{ saddle,}\\
2/3\leq \alpha=1/\beta <1 \text{ saddle,}\\
\alpha=1/\beta< 2/3, \text{ min,}\\
\alpha>1/\beta \text{ min.}
\end{array}
\right.$ \\ \hline
\end{tabular}
\end{center}

\textbf{Case (3):} $K_{12}^{-1}=\mathcal{V}f_{1/2}(\tau _{1},\tau
_{2})$

\begin{enumerate}
\item[$\bullet$] direction (I)

The volume dependence of the parameters (\ref{ineq}) now looks
like
\begin{equation}
\left\{
\begin{array}{l}
\left\vert X_{1} \right\vert \simeq \left\vert X_{2} \right\vert
\simeq \mathcal{V}^{-(2+1/\gamma)}\ln\mathcal{V}, \\
\left\vert Y_{12} \right\vert \simeq
\mathcal{V}^{-(1+2/\gamma)}\sqrt{\ln\mathcal{V}}.
\end{array}
\right.  \label{ABSOl}
\end{equation}
Substituting the expressions (\ref{ABSOl}) in (\ref{disug}) we
find
\begin{equation}
\frac{\ln\mathcal{V}}{\mathcal{V}^{2+1/\gamma}}>2
\frac{\sqrt{\ln\mathcal{V}}}{\mathcal{V}^{1+2/\gamma}},
\end{equation}
which at large volume is true if $\gamma\leq 1$, false if
$\gamma>1$. Thus (b) can be a minimum of the axion potential.

\item[$\bullet$] direction (II)

The parameters (\ref{ineq}) now read
\begin{equation}
\left\{
\begin{array}{l}
\left\vert X_{1} \right\vert \simeq
\mathcal{V}^{-(2+1/\beta)}\ln\mathcal{V}, \\
\left\vert X_{2} \right\vert \simeq
\mathcal{V}^{-(2+1/\gamma)}\ln\mathcal{V}, \\
\left\vert Y_{12} \right\vert \simeq
\mathcal{V}^{-(1+1/\beta+1/\gamma)}\sqrt{\ln\mathcal{V}}.
\end{array}
\right.  \label{dabS}
\end{equation}
Looking at (\ref{dabS}), we realise that the first (b)-condition
becomes
\begin{equation}
\frac{\sqrt{\ln
\mathcal{V}}}{\mathcal{V}}>\frac{1}{\mathcal{V}^{1/\gamma}},
\label{firstB}
\end{equation}
which is satisfied only if $\gamma\leq 1$. Viceversa the first
(c)-condition is satisfied only for $\gamma>1$. Let us check now
the validity of the second (b)-condition which reads
\begin{equation}
\frac{\sqrt{\ln
\mathcal{V}}}{\mathcal{V}}>\frac{1}{\mathcal{V}^{1/\beta}}
+\frac{1}{\mathcal{V}^{1/\gamma}}, \label{secondB}
\end{equation}
which, at large volume, is automatically true if $\gamma\leq 1$.
The second (c)-condition is also correctly satisfied for
$\gamma>1$ since it reads
\begin{equation}
\frac{1}{\mathcal{V}^{1/\gamma}}>\frac{\sqrt{\ln
\mathcal{V}}}{\mathcal{V}}+\frac{1}{\mathcal{V}^{1/\beta}}.
\label{segno}
\end{equation}
Thus both (b) and (c) can be a minimum of the axion potential.

\item[$\bullet$] direction (III)

The parameters (\ref{ineq}) assume the same form as (\ref{dabS})
but now with $\beta>\gamma$. The inequality corresponding to the
(d)-condition reads
\begin{equation}
\frac{1}{\mathcal{V}^{1/\beta}}>\frac{1}{\mathcal{V}^{1/\gamma}}+
\frac{\sqrt{\ln\mathcal{V}}}{\mathcal{V}},
\end{equation}
and becomes true if $\beta>1$. Moreover, the first (b)-condition
(\ref{firstB}) is again satisfied for $\gamma\leq 1$. On the other
hand, the second looks like (\ref{secondB}) and now, at large
volume, is true only if $\beta\leq 1$, which implies correctly
$\gamma\leq 1$ since in this case $\gamma<\beta$. It follows that
both (b) and (d) can be a minimum.
\end{enumerate}

Let us summarise the results found in case (3) in the following
table
\begin{center}
\begin{tabular}{|c|c|c|c|}
\hline & (I) & (II) & (III) \\ \hline (a) & max & max & max \\
\hline (b) & $\left\{
\begin{array}{c}
0<\gamma \leq 1\text{ min,} \\
\gamma >1\text{ saddle.}
\end{array}
\right. $ & $\left\{
\begin{array}{c}
0<\gamma \leq 1\text{ min,} \\
\gamma >1\text{ saddle.}
\end{array}
\right. $ & $\left\{
\begin{array}{c}
0<\gamma <\beta \leq 1\text{ min,} \\
\beta>1 \text{ saddle.} \\
\end{array}
\right. $ \\ \hline (c) & saddle & $\left\{
\begin{array}{c}
0<\gamma \leq 1\text{ saddle,} \\
\gamma >1\text{ min.}
\end{array}
\right. $ & saddle \\ \hline (d) & saddle & saddle & $\left\{
\begin{array}{c}
0<\gamma<\beta \leq 1\text{ saddle,} \\
\beta>1\text{ min.}
\end{array}
\right. $ \\ \hline
\end{tabular}
\end{center}

\bigskip

\textbf{Case (4):} $K_{12}^{-1}=\mathcal{V}^{\alpha }h_{2-3\alpha
/2}(\tau _{1},\tau _{2}),$ \ $\alpha >1$

\begin{enumerate}
\item[$\bullet$] direction (I)

The volume dependence of the parameters (\ref{ineq}) is given
again by (\ref{tABsol}) and (\ref{disug}) takes the same form as
the inequality (\ref{IO}) which at large volume is true if
$\alpha<1/\gamma$, false if $\alpha>1/\gamma$. The situation
$\alpha=1/\gamma$ is more involved and (\ref{IO}) simplifies to
\begin{equation}
1>2(\ln\mathcal{V})^{1-3\alpha/2},
\end{equation}
which gives a positive result if $\alpha>2/3$. This is definitely
true in our case where $\alpha>1$. Thus (b) can be a minimum of
the axion potential.

\item[$\bullet$] direction (II)

The parameters (\ref{ineq}) now take the same form given in
(\ref{ddabso}). It follows then that the first (b)-condition
$\left\vert X_{1}\right\vert > \left\vert Y_{12} \right\vert$
looks like (\ref{question}) and is verified for $1/\gamma\geq
\alpha$. The second (b)-condition looks like (\ref{SECONDB}) which
at large volume is correctly true for $1/\gamma\geq \alpha$. Thus
(b) is a minimum of the axion potential. On the contrary the first
(c)-condition implies $1/\gamma<\alpha$, whereas the second is
similar to the inequality (\ref{llook}) which is again clearly
true for $1/\gamma<\alpha$. It follows that in this case (c) can
also be a minimum.

\item[$\bullet$] direction (III)

The parameters (\ref{ineq}) assume the same form as (\ref{ddabso})
but now with $\beta>\gamma$. The (d)-condition looks like
\begin{equation}
\frac{(\ln{\mathcal{V}})^{1-3\alpha/2}}{\mathcal{V}^{1/\beta-\alpha}}>1
+\frac{(\ln{\mathcal{V}})^{1-3\alpha/2}}{\mathcal{V}^{1/\gamma-\alpha}},
\end{equation}
and is verified only if $1/\beta<\alpha$. On the other hand, the
first (b)-condition is again given by (\ref{question}) and so is
still solved for $1/\gamma\geq \alpha$. The second (b)-condition
looks like (\ref{SECONDB}) but now at large volume it is satisfied
for $1/\beta\geq\alpha$. It follows that in this case both (b) and
(d) can be a minimum of the axion potential.
\end{enumerate}

Let us summarise the results found in case (4) in the following
table
\begin{center}
\begin{tabular}{|c|c|c|c|}
\hline & (I) & (II) & (III) \\ \hline (a) & max & max & max \\
\hline (b) & $\left\{
\begin{array}{c}
1<\alpha \leq 1/\gamma \text{ min,} \\
\alpha >1/\gamma \text{ saddle.}
\end{array}
\right. $ & $\left\{
\begin{array}{c}
1<\alpha \leq 1/\gamma \text{ min,} \\
\alpha >1/\gamma \text{ saddle.}
\end{array}
\right. $ & $\left\{
\begin{array}{c}
1<\alpha \leq 1/\beta <1/\gamma \text{ min,} \\
1/\beta <\alpha \text{ saddle.} \\
\end{array}
\right. $ \\ \hline (c) & saddle & $\left\{
\begin{array}{c}
1<\alpha \leq 1/\gamma \text{ saddle,} \\
\alpha >1/\gamma \text{ min.}
\end{array}
\right. $ & saddle \\ \hline (d) & saddle & saddle & $\left\{
\begin{array}{c}
\alpha \leq 1/\beta \text{ saddle,} \\
\alpha > 1/\beta \text{ min.}
\end{array}
\right. $ \\ \hline
\end{tabular}
\end{center}
\bigskip

\subsubsection{K\"{a}hler moduli stabilisation}

After this long analysis of the axion minimisation, let us now
focus again step by step on the four cases according to the
different possible values of $K^{-1}_{12}$. In each case, we shall
fix the axions at their possible VEVs and then study the
K\"{a}hler moduli stabilisation depending on the particular form
of $K^{-1}_{11}$ and $K^{-1}_{22}$.

However, before focusing on each particular case, let us point out
some general features. At the axion minimum we will have:
\begin{equation}
\left\langle V_{AX}\right\rangle =2\left(\pm\left\vert
Y_{12}\right\vert \pm\left\vert X_{1}\right\vert \pm\left\vert
X_{2}\right\vert\right), \label{assione}
\end{equation}
where the "$\pm$" signs depend on the specific locus of the
minimum, that is (a) or (b) or (c), as specified in (\ref{TABLE}).
Now to write (\ref{assione}) explicitly, recall (\ref{ineq}) and
get:
\begin{equation}
V_{np2}+V_{np1}^{AX}=\frac{2}{\mathcal{V}^{2}}\left\{
W\sum\limits_{j=1}^{2}\left(\pm 2 a_{j}\tau
_{j}e^{-a_{j}\tau_{j}}\right) \pm K^{-1}_{12}
a_{1}a_{2}e^{-(a_{1}\tau_{1}+a_{2}\tau_{2})}
 \right\}, \label{fffor}
\end{equation}
where we have set $A_{1}=A_{2}=1$. We may now study the full
potential by combining equations (\ref{Alfa}), (\ref{imp}) and
(\ref{fffor})
\begin{eqnarray}
V &\sim &\frac{1}{\mathcal{V}^{2}}\left[
\sum\limits_{j=1}^{2}K^{-1}_{jj} a_{j}^{2} e^{-2a_{j}\tau_{j}}\pm
2K^{-1}_{12}a_{1}a_{2}e^{-(a_{1}\tau_{1}+a_{2}\tau_{2})}\right]  \notag \\
&&+4\frac{W_{0}}{\mathcal{V}^{2}}\sum\limits_{j=1}^{2}\left(\pm
a_{j}\tau _{j}e^{-a_{j}\tau_{j}}\right)+\frac{3}{4}\frac{\hat{\xi}
}{\mathcal{V}^{3}}W_{0}^{2}, \label{finalPotential}
\end{eqnarray}
where we have substituted $W$ with its tree-level expectation
value $W_{0}$ because the non-perturbative corrections are always
subleading by a power of $\mathcal{V}$.

When we take the generic large volume limit $\mathcal{V}\sim
e^{\beta a_{1}\tau_{1}}\sim e^{\gamma a_{2}\tau_{2}}$, with
$\beta$ and $\gamma \in \mathbb{R}$ without any particular
relation among them to take into account all the possible limits
(\ref{DIrections}), (\ref{finalPotential}) has the following
volume scaling
\begin{equation}
V \sim \frac{K^{-1}_{11}}{\mathcal{V}^{2+2/\beta}}
+\frac{K^{-1}_{22}}{\mathcal{V}^{2+2/\gamma}}
\pm\frac{K^{-1}_{12}}{\mathcal{V}^{2+1/\beta+1/\gamma}}
\pm\frac{\tau _{1}}{\mathcal{V}^{2+1/\beta}}\pm\frac{\tau
_{2}}{\mathcal{V}^{2+1/\gamma}}+\frac{1}{\mathcal{V}^{3}}.
\label{semPlifica}
\end{equation}
Now given that we are already aware of the volume scaling of
$K^{-1}_{12}$, which was our starting point to stabilise the
axions, in order to understand what are the leading terms in
(\ref{semPlifica}), we need to know only the form of
$K^{-1}_{jj}$, $j=1,2$. We can classify its behaviour according to
the volume dependence of the quantity $k_{jjk}t^{k}$ and find 4
different cases as we did for $K^{-1}_{12}$:

\begin{enumerate}
\item $k_{jjk}t^{k}=0\text{ \ or \ }k_{jjk}t^{k}=\frac{f
(\tau _{1},\tau_{2})}{\mathcal{V}^{\alpha }},$ $\alpha \geq 1$
$\Longrightarrow K_{jj}^{-1}=\tau _{j}^{2};$

\item $k_{jjk}t^{k}=\frac{g_{\gamma }(\tau _{1},\tau_{2})}{\mathcal{V}^{\alpha
}},$ $0<\alpha<1$, $g$ homogeneous function of degree $\gamma
=\frac{1+3\alpha}{2}$ $\Longrightarrow
K_{jj}^{-1}=\mathcal{V}^{\alpha }g_{2-3\alpha /2}(\tau _{1},\tau
_{2}),$ \ $0<\alpha <1;$

\item $k_{jjk}t^{k}=f_{1/2}(\tau _{1},\tau_{2}),$ $f$
homogeneous function of degree $1/2$ $\Longrightarrow
K_{jj}^{-1}=\mathcal{V}f_{1/2}(\tau _{1},\tau _{2});$

\item $k_{jjk}t^{k}=\mathcal{V}^{\alpha }h_{\beta }(\tau _{1},\tau_{2}),$
$\alpha >0$, $h$ homogeneous function of degree $\beta =
\frac{1-3\alpha}{2}$ $\Longrightarrow
K_{jj}^{-1}=\mathcal{V}^{\alpha }h_{2-3\alpha /2}(\tau _{1},\tau
_{2}),$ \ $\alpha >1.$
\end{enumerate}
Before focusing on the K\"{a}hler moduli minimisation by analysing
all these 4 cases in full detail for each direction
(\ref{DIrections}), we stress that we can already show in general
that some situations do not lead to any LARGE Volume minimum.

For example, let us assume that the elements of $K^{-1}$ are such
that the dominant terms in (\ref{semPlifica}) are
\begin{equation}
V \sim \frac{K^{-1}_{11}}{\mathcal{V}^{2+2/\beta}}-\frac{\tau
_{1}}{\mathcal{V}^{2+1/\beta}}-\frac{\tau
_{2}}{\mathcal{V}^{2+1/\gamma}}+\frac{1}{\mathcal{V}^{3}},
\label{SemPlifica}
\end{equation}
with $\beta=\gamma=1$ and
$K^{-1}_{11}=\mathcal{V}f_{1/2}(\tau_{1},\tau_{2})$. Therefore the
potential (\ref{finalPotential}) with $W_{0}=1$ looks like
\begin{equation}
V \sim \frac{f_{1/2}(\tau_{1},\tau_{2}) a_{1}^{2}
e^{-2a_{1}\tau_{1}}}{\mathcal{V}} -\frac{4
a_{1}\tau_{1}e^{-a_{1}\tau_{1}}}{\mathcal{V}^{2}} -\frac{4
a_{2}\tau_{2}e^{-a_{2}\tau_{2}}}{\mathcal{V}^{2}}
+\frac{3}{4}\frac{\hat{\xi}}{\mathcal{V}^{3}}.
\label{FinalPotential}
\end{equation}
Thus from $\frac{\partial V}{\partial \tau_{1}}=0$ we find
\begin{equation}
\mathcal{V}=\frac{4\tau_{1}}{\left(2a_{1}f_{1/2}-\frac{\partial
f_{1/2}}{\partial \tau_{1}}\right)}e^{a_{1}\tau_{1}},
\label{Etichetta1}
\end{equation}
whereas $\frac{\partial V}{\partial \tau_{2}}=0$ gives
\begin{equation}
\mathcal{V}=- 4
\left(\frac{a_{2}}{a_{1}}\right)^{2}\frac{\tau_{2}}{\frac{\partial
f_{1/2}}{\partial \tau_{2}}}\frac{e^{2
a_{1}\tau_{1}}}{e^{a_{2}\tau_{2}}}. \label{Etichetta2}
\end{equation}
Now since we have $\beta=\gamma=1$, from the form of the large
volume limit (I) of (\ref{DIrections}), we infer that the minimum
should be located at $a_{1}\tau_{1}\simeq a_{2}\tau_{2}$. Making
this substitution and combining (\ref{Etichetta1}) with
(\ref{Etichetta2}), we end up with the following equation
\begin{equation}
\frac{\partial f_{1/2}}{\partial
\tau_{2}}=\frac{a_{2}}{a_{1}}\frac{\partial f_{1/2}}{\partial
\tau_{1}}-2a_{2}f_{1/2}. \label{eor}
\end{equation}
Now using the homogeneity property of $f_{1/2}$, that is
$\tau_{1}\frac{\partial f_{1/2}}{\partial
\tau_{1}}+\tau_{2}\frac{\partial f_{1/2}}{\partial
\tau_{2}}=\frac{1}{2}f_{1/2}$, (\ref{eor}) takes the form
\begin{equation}
\frac{\partial f_{1/2}}{\partial
\tau_{2}}=a_{2}\left(\frac{1}{4a_{2}\tau_{2}}-1\right)f_{1/2}.
\label{Eor}
\end{equation}
We can solve the previous differential equation getting
$f_{1/2}=\tau_{2}^{1/4}e^{-a_{2}\tau_{2}}$, which is not an
homogeneous function of degree $1/2$. Thus we deduce that this
case gives no LVS.

Another case in which we can show explicitly that no LARGE Volume
minimum is present, is the one where the dominant terms in
(\ref{semPlifica}) read
\begin{equation}
V \sim \frac{K^{-1}_{11}}{\mathcal{V}^{2+2/\beta}}
+\frac{K^{-1}_{22}}{\mathcal{V}^{2+2/\gamma}}+\frac{1}{\mathcal{V}^{3}}.
\label{sSEMPLIl}
\end{equation}
with $\beta\leq\gamma$. The fact that all the three terms in
(\ref{sSEMPLIl}) are strictly positive leads us to conclude that
there would definitely be no LARGE Volume minimum in the volume
direction once we integrate out the small moduli. In fact,
(\ref{sSEMPLIl}) would take the generic form
\begin{equation}
V\sim\frac{a(\ln{\mathcal{V}})^{b}+c}{\mathcal{V}^{3}},\text{ \
}c>0,
\end{equation}
which can be easily seen to have a minimum only if $a<0$ and
$b>0$.

We illustrate now a further case in which it is possible to show
explicitly that no LVS is present. The leading terms in
(\ref{semPlifica}) read
\begin{equation}
V \sim \frac{K^{-1}_{11}}{\mathcal{V}^{2+2/\beta}}
+\frac{K^{-1}_{22}}{\mathcal{V}^{2+2/\gamma}}-\frac{\tau_{2}}{\mathcal{V}^{2+1/\gamma}},
\label{Se}
\end{equation}
with $\beta<\gamma$ and $\gamma>1$ to be able to neglect the
$\alpha'$ corrections that scale as $\mathcal{V}^{-3}$. The
necessary but not sufficient condition to fix the small K\"{a}hler
moduli is $K^{-1}_{11}=\mathcal{V}^{\delta}\tau_{1}^{2-3\delta/2}$
and
$K^{-1}_{22}=\mathcal{V}^{\eta}f_{2-3\eta/2}(\tau_{1},\tau_{2})$
with $\delta=2/\beta-1/\gamma$ and $\eta=1/\gamma$. We can now
prove that it is never possible to stabilise $a_{1}\tau_{1}\gg 1$.
In fact, the relevant part of the scalar potential
(\ref{finalPotential}) would read
\begin{equation}
V \simeq \frac{a_{1}^{2}\tau_{1}^{2-3\delta/2}
e^{-2a_{1}\tau_{1}}}{\mathcal{V}^{2-\delta}}+\frac{a_{2}^{2}
f_{2-3\eta/2}(\tau_{1},\tau_{2})
 e^{-2a_{2}\tau_{2}}}{\mathcal{V}^{2-\eta}}-\frac{4
a_{2}\tau_{2}e^{-a_{2}\tau_{2}}}{\mathcal{V}^{2}}. \label{Flo}
\end{equation}
Now the equation $\frac{\partial V}{\partial \tau_{1}}=0$, admits
a solution of the form
\begin{equation}
\mathcal{V}^{\eta-\delta}=\frac{2a_{1}^{3}
\tau_{1}^{2-3\delta/2}}{a_{2}^{2}\frac{\partial
f_{2-3\eta/2}}{\partial
\tau_{1}}}e^{2(a_{2}\tau_{2}-a_{1}\tau_{1})}, \label{Ppp}
\end{equation}
whereas $\frac{\partial V}{\partial \tau_{2}}=0$ gives
\begin{equation}
\mathcal{V}^{\eta}=\frac{2
\tau_{2}}{a_{2}f_{2-3\eta/2}}e^{a_{2}\tau_{2}}. \label{Pp}
\end{equation}
The third minimisation equation $\frac{\partial V}{\partial
\mathcal{V}}=0$ looks like
\begin{equation}
(\eta-2)a_{2}^{2}f_{2-3\eta/2}\mathcal{V}^{\eta}e^{-2a_{2}\tau_{2}}
+(\delta-2)a_{1}^{2}\tau_{1}^{2-3\delta/2}\mathcal{V}^{\delta}e^{-2a_{1}\tau_{1}}
+8a_{2}\tau_{2}e^{-a_{2}\tau_{2}}=0, \label{PPp}
\end{equation}
and substituting the results (\ref{Pp}) and (\ref{Ppp}), we obtain
\begin{equation}
2(\eta-2)+\frac{(\delta-2)}{a_{1}f_{2-3\eta/2}}\frac{\partial
f_{2-3\eta/2}}{\partial \tau_{1}}+8=0. \label{PPP}
\end{equation}
Solving the differential equation (\ref{PPP}), we realise that
$f_{2-3\eta/2}$ has an exponential behaviour in $\tau_{1}$ which
is in clear contrast with the requirement that it has to be
homogeneous. Following arguments very similar to this one it can
be seen that, as in the case with just one small modulus, the
presence of the $\alpha'$ corrections is crucial to find a LARGE
Volume minimum. In fact if we omit them, either it is impossible
to fix the small moduli large enough to ignore higher instanton
corrections or, once we integrate them out, we are left with a
run-away in the volume direction.

Lastly, we describe the final case in which it is possible to
prove the absence of a LARGE Volume vacuum. The leading terms in
(\ref{semPlifica}) are given by
\begin{equation}
V \sim \frac{K^{-1}_{11}}{\mathcal{V}^{2+2/\beta}}
+\frac{K^{-1}_{22}}{\mathcal{V}^{4}}-\frac{\tau_{2}}
{\mathcal{V}^{3}}+\frac{1}{\mathcal{V}^{3}}, \label{USe}
\end{equation}
with $\beta<1$ and the axion minimum along the direction
$\mathcal{V}\sim e^{\beta a_{1}\tau_{1}}\sim e^{a_{2}\tau_{2}}$.
The necessary but not sufficient condition to fix the small
K\"{a}hler moduli is
$K^{-1}_{11}\simeq\mathcal{V}^{\delta}\tau_{1}^{2-3\delta/2}$ with
$\delta=2/\beta-1$, and
$K^{-1}_{22}=\mathcal{V}f_{1/2}(\tau_{1},\tau_{2})$. The relevant
part of the scalar potential (\ref{finalPotential}) takes the form
\begin{equation}
V \simeq \frac{a_{1}^{2}\tau_{1}^{2-3\delta/2}
e^{-2a_{1}\tau_{1}}}{\mathcal{V}^{2-\delta}}+\frac{a_{2}^{2}
f_{1/2}(\tau_{1},\tau_{2})
 e^{-2a_{2}\tau_{2}}}{\mathcal{V}}-\frac{4
a_{2}\tau_{2}e^{-a_{2}\tau_{2}}}{\mathcal{V}^{2}}
+\frac{3}{4}\frac{\hat{\xi}}{\mathcal{V}^{3}}. \label{UFlo}
\end{equation}
Now the equation $\frac{\partial V}{\partial \tau_{1}}=0$, admits
a solution of the form
\begin{equation}
a_{1}^{2}\tau_{1}^{2-3\delta/2}\mathcal{V}^{\delta}e^{-2a_{1}\tau_{1}}=
\frac{a_{2}^{2}}{2a_{1}}\frac{\partial
f_{1/2}}{\partial\tau_{1}}\mathcal{V}e^{-2a_{2}\tau_{2}},
\label{UPp}
\end{equation}
whereas $\frac{\partial V}{\partial \tau_{2}}=0$ gives
\begin{equation}
a_{2}\tau_{2}e^{-a_{2}\tau_{2}}=\frac{a_{2}^{2}}{2}f_{1/2}\mathcal{V}e^{-2a_{2}\tau_{2}}.
\label{UPpp}
\end{equation}
The third minimisation equation $\frac{\partial V}{\partial
\mathcal{V}}=0$ corresponds to
\begin{equation}
(\delta-2)a_{1}^{2}\tau_{1}^{2-3\delta/2}\mathcal{V}^{\delta}e^{-2a_{1}\tau_{1}}
-a_{2}^{2}f_{1/2}\mathcal{V}e^{-2a_{2}\tau_{2}}-8a_{2}\tau_{2}e^{-a_{2}\tau_{2}}=
\frac{9}{4}\frac{\hat{\xi}}{\mathcal{V}}, \label{UPPp}
\end{equation}
and substituting the results (\ref{UPp}) and (\ref{UPpp}), we
obtain
\begin{equation}
4a_{2}^2\mathcal{V}^{2}\left[\frac{(\delta-2)}{2
a_{1}}\frac{\partial f_{1/2}}{\partial\tau_{1}} -5 f_{1/2}\right]=
9\hat{\xi}e^{2a_{2}\tau_{2}}. \label{UPPPl}
\end{equation}
Now writing $f_{1/2}(\tau_{1},\tau_{2})=F(\frac{\beta
a_{1}}{a_{2}})\sqrt{\tau_{1}}$ for appropriate function $F$,
(\ref{UPPPl}) becomes
\begin{equation}
\frac{a_{2}^2}{a_{1}\sqrt{\tau_{1}}}\mathcal{V}^{2}F\left(\frac{\beta
a_{1}}{a_{2}}\right)\left(\delta-2-20 a_{1}\tau_{1}\right)=
9\hat{\xi}e^{2a_{2}\tau_{2}}. \label{UQmuU}
\end{equation}
Given that a trustable minimum requires $a_{1}\tau_{1}\gg 1$, the
LHS of (\ref{UQmuU}) is negative while the RHS is definitively
positive and so this case does not allow us to find any LVS.

The general path that we shall follow to derive the conditions
which guarantee that we have enough terms with the correct volume
scaling to stabilise all the moduli at exponentially large volume,
is the following one. We learnt from the proof of Claim 1 for the
case with just one small modulus $\tau_{s}$, that we need to have
two terms in the scalar potential with the same volume scaling
that depend on $\tau_{s}$ so that it can be stabilised rather
large in order to be able to neglect higher instanton corrections.
Then if we integrate out $\tau_{s}$, we have to be left with at
least two terms that depend on the overall volume and have the
same volume scaling. Lastly in order to find the exponentially
large volume minimum, the leading term at large volume has to be
negative. As we have seen before, the same arguments apply here.
Thus we shall first work out the conditions to be able to fix both
$a_{1}\tau_{1}\gg 1$ and $a_{2}\tau_{2}\gg 1$ by having at least
two terms in the potential with a dependence on these moduli and
the same volume scaling. Then, we shall imagine to integrate out
these moduli, and derive the conditions to be left with at least
two terms dependent on $\mathcal{V}$ with the leading one which is
negative.

\bigskip

\textbf{Case (1):} $K^{-1}_{12}\simeq \tau_{1}\tau_{2}$

\bigskip
The previous analysis tells us that, regardless of the particular
direction considered, the axion minimum is always in the case (b).
Thus we realise that at the minimum
\begin{equation}
\left\langle V_{AX}\right\rangle =2\left(\left\vert
Y_{12}\right\vert -\left\vert X_{1}\right\vert-\left\vert
X_{2}\right\vert\right). \label{axione}
\end{equation}
Now recalling that in case (1) $K^{-1}_{12}\simeq
\tau_{1}\tau_{2}$, (\ref{finalPotential}) takes the form:
\begin{eqnarray}
V &\sim &\frac{1}{\mathcal{V}^{2}}\left[
\sum\limits_{j=1}^{2}K^{-1}_{jj} a_{j}^{2}
e^{-2a_{j}\tau_{j}}+2\tau_{1}\tau_{2}
a_{1}a_{2}e^{-(a_{1}\tau_{1}+a_{2}\tau_{2})}\right]  \notag \\
&&-4\frac{W_{0}}{\mathcal{V}^{2}}\sum\limits_{j=1}^{2}a_{j}\tau
_{j}e^{-a_{j}\tau_{j}}+\frac{3}{4}\frac{\hat{\xi}
}{\mathcal{V}^{3}}W_{0}^{2}. \label{finalpotential}
\end{eqnarray}
We shall now study the behaviour of (\ref{finalpotential}) by
taking the large volume limit along each direction
(\ref{DIrections}) and then considering all the possible forms of
$K^{-1}_{jj}$, $j=1,2$. When we take the large volume limit (I) of
(\ref{DIrections}), (\ref{finalpotential}) has the following
volume scaling
\begin{equation}
V \sim \frac{K^{-1}_{11}}{\mathcal{V}^{2+2/\gamma}}
+\frac{K^{-1}_{22}}{\mathcal{V}^{2+2/\gamma}}
+\frac{\tau_{1}\tau_{2}}{\mathcal{V}^{2+2/\gamma}} -\frac{\tau
_{1}}{\mathcal{V}^{2+1/\gamma}}-\frac{\tau
_{2}}{\mathcal{V}^{2+1/\gamma}}+\frac{1}{\mathcal{V}^{3}}.
\label{semplifica}
\end{equation}
The third term in (\ref{semplifica}) is subleading with respect to
the fourth and the fifth. Thus it can be neglected
\begin{equation}
V \sim \frac{K^{-1}_{11}}{\mathcal{V}^{2+2/\gamma}}
+\frac{K^{-1}_{22}}{\mathcal{V}^{2+2/\gamma}}-\frac{\tau
_{1}}{\mathcal{V}^{2+1/\gamma}}-\frac{\tau
_{2}}{\mathcal{V}^{2+1/\gamma}}+\frac{1}{\mathcal{V}^{3}}.
\label{Semplifica}
\end{equation}
We have seen that the presence of the $\alpha'$ corrections is
crucial to find the exponentially large volume minimum. Therefore
the fact that the third and the fourth terms in (\ref{Semplifica})
have to scale as $\mathcal{V}^{-3}$ tells us that $\gamma=1$.
\begin{equation}
V \sim \frac{K^{-1}_{11}}{\mathcal{V}^{4}}
+\frac{K^{-1}_{22}}{\mathcal{V}^{4}}-\frac{\tau
_{1}}{\mathcal{V}^{3}}-\frac{\tau
_{2}}{\mathcal{V}^{3}}+\frac{1}{\mathcal{V}^{3}}.
\label{SEmplifica}
\end{equation}
At this point it is straightforward to realise that if either
$K^{-1}_{11}$ or $K^{-1}_{22}$ were in case (4), then we would
have a run-away behaviour of the volume direction. Similarly the
situation with $K^{-1}_{11}$ and $K^{-1}_{22}$ either in case (1)
or (2) is not giving a LARGE Volume minimum since the first two
terms in (\ref{SEmplifica}) should be neglected without then the
possibility to stabilise $\tau_{1}$ and $\tau_{2}$ large. What
happens if either $K^{-1}_{11}$ or $K^{-1}_{22}$ is in case (3)
and the other one is either in case (1) or (2)? We do not find any
minimum. In fact, let us say that $K^{-1}_{11}$ is in case (3) and
$K^{-1}_{22}$ in case (1) or (2): then the second term in
(\ref{SEmplifica}) can be neglected. If we want to have still some
hope to stabilise $\tau_{2}$ large, then $K^{-1}_{11}$ should
better depend also on $\tau_{2}$: $K^{-1}_{11}\simeq
f_{1/2}(\tau_{1},\tau_{2})\mathcal{V}$. However this case has been
studied explicitly to show that it leads to an absurd. Thus only
if both $K^{-1}_{11}$ and $K^{-1}_{22}$ is in case (3) we can have
a LARGE Volume minimum.

On the contrary, if we took either the large volume limit (II) or
(III) in (\ref{DIrections}), (\ref{finalpotential}) would scale as
\begin{equation}
V \sim \frac{K^{-1}_{11}}{\mathcal{V}^{2+2/\beta}}
+\frac{K^{-1}_{22}}{\mathcal{V}^{2+2/\gamma}}
+\frac{\tau_{1}\tau_{2}}{\mathcal{V}^{2+1/\beta+1/\gamma}}
-\frac{\tau _{1}}{\mathcal{V}^{2+1/\beta}}-\frac{\tau
_{2}}{\mathcal{V}^{2+1/\gamma}}+\frac{1}{\mathcal{V}^{3}}.
\label{semplific}
\end{equation}
Let us focus on the direction (II) where $1/\beta>1/\gamma$. The
third and the fourth term in (\ref{semplific}) at large volume are
subdominant to the fifth and therefore they can be ignored:
\begin{equation}
V \sim \frac{K^{-1}_{11}}{\mathcal{V}^{2+2/\beta}}
+\frac{K^{-1}_{22}}{\mathcal{V}^{2+2/\gamma}}-\frac{\tau
_{2}}{\mathcal{V}^{2+1/\gamma}}+\frac{1}{\mathcal{V}^{3}},
\label{Semplific}
\end{equation}
If $1/\gamma>1$, then the $\mathcal{V}^{-3}$ term would be the
dominant one producing a run-away in the volume direction. Thus we
impose $1/\gamma\leq 1$. However we have already showed that the
situation with $1/\gamma< 1$ gives no LVS and so we deduce that we
need $1/\gamma=1$. Then we realise that the only possible
situation in which we can hope to fix $\tau_{2}$ large is when
either the first or the second term in (\ref{Semplific}) scales as
$\mathcal{V}^{-3}$. Now if the second term involving $K^{-1}_{22}$
were subleading with respect to the fourth term in
(\ref{Semplific}), then the first one should scale as
$\mathcal{V}^{-3}$. However at that point, knowing that
$K^{-1}_{11}$ will introduce a dependence on $\tau_{1}$, we would
not be able to stabilise $\tau_{1}$ large. Hence $K^{-1}_{22}$ has
to be in case (3): $K_{22}^{-1}=\mathcal{V}f_{1/2}(\tau
_{1},\tau_{2})$.

Now we have two different situations according to the fact that
$f_{1/2}$ indeed depends on both $\tau_{1}$ and $\tau_{2}$ or only
on $\tau_{2}$. The first possibility has already been studied with
the final conclusion that it produces no LVS. On the other hand,
when $K^{-1}_{22}$ depends only on $\tau_{2}$, i.e.
$K_{22}^{-1}\simeq\mathcal{V}\sqrt{\tau_{2}}$, we have that the
overall volume and $\tau_{2}$ are both stabilised by the interplay
of the second, the third and the fourth term in (\ref{Semplific}).
The first term is now subleading and can be used to fix $\tau_{1}$
if we write $K^{-1}_{11}\simeq
\mathcal{V}^{\alpha}\tau_{1}^{2-3\alpha/2}$ and then impose
$1/\beta=\alpha$ in order to make it scale as the fourth term in
(\ref{semplific}).

We point out that these results apply also to the direction (III)
where $1/\gamma>1/\beta$, if we swap $\gamma$ with $\beta$ and
$\tau_{1}$ with $\tau_{2}$. Let us finally summarise in the table
below what we have found for this case.

\textbf{Case (1): $K^{-1}_{12}\simeq\tau_{1}\tau_{2}$}

\begin{center}
\begin{tabular}{|c|c|c|c|c|}
\hline $K_{11}^{-1}$ & $K_{22}^{-1}$ & (I), (b) & (II), (b) &
(III), (b) \\ \hline 1 & 1 & NO & NO & NO \\ \hline 1 & 2 & NO &
NO & NO \\ \hline 1 & 3 & NO & NO & NO \\ \hline 1 & 4 & NO & NO &
NO \\ \hline 2 & 1 & NO & NO & NO \\ \hline 2 & 2 & NO & NO & NO
\\ \hline 2 & 3 & NO & NO & NO \\ \hline 2 & 4 & NO & NO & NO \\
\hline 3 & 1 & NO & NO & NO \\ \hline 3 & 2 & NO & NO & NO \\
\hline 3 & 3 & OK, $\gamma =1$ & NO & NO \\ \hline 3 & 4 & NO & NO
& OK, $\beta =1$, ($\ast \ast $) \\ \hline 4 & 1 & NO & NO & NO \\
\hline 4 & 2 & NO & NO & NO \\ \hline 4 & 3 & NO & OK, $\gamma
=1$, ($\ast $) & NO \\ \hline 4 & 4 & NO & NO & NO \\ \hline
\end{tabular}
\end{center}
($\ast $) $ \ K_{22}^{-1}\simeq \mathcal{V}\sqrt{\tau _{2}}$, $%
K_{11}^{-1}\simeq \mathcal{V}^{\alpha }\tau _{1}^{2-3\alpha /2}$ with $\frac{%
1}{\beta }=\alpha $
\newline
($\ast \ast $) $K_{11}^{-1}\simeq \mathcal{V}\sqrt{\tau _{1}}$, $%
K_{22}^{-1}\simeq \mathcal{V}^{\alpha }\tau _{2}^{2-3\alpha /2}$ with $\frac{%
1}{\gamma }=\alpha $
\bigskip

\textbf{Case (2):} $K^{-1}_{12}\simeq
\mathcal{V}^{\alpha}g_{2-3\alpha/2}(\tau_{1},\tau_{2})$,
$0<\alpha<1$

\bigskip
This case is more involved than the previous one since, depending
on the direction chosen for the large volume limit and the exact
value of the parameter $\alpha$, the axion minimum can not only be
in case (b), but also in (c) and (d). Let us start by considering
each case in detail:
\begin{enumerate}

\item[$\bullet$] axion minimum at (c) along direction (II) for
$\alpha=1/\gamma<2/3$ or $1/\gamma<\alpha<1$

We can easily conclude that no LVS is present given that, looking
at the general volume scaling of the scalar potential
(\ref{semPlifica}), we can notice that the fifth term would be
dominant with respect to the last one since we have always
$1/\gamma<1$. Thus the $\alpha'$ correction would be negligible at
large volume, so producing no LVS.

\item[$\bullet$] axion minimum at (d) along direction (III) for
$\alpha=1/\beta<2/3$ or $1/\beta<\alpha<1$

This situation looks like the previous one if we swap $\gamma$
with $\beta$ and $\tau_{1}$ with $\tau_{2}$, therefore we conclude
that no LARGE Volume minimum will be present.

\item[$\bullet$] axion minimum at (b) along direction (II) for
$\alpha<1/\gamma$ or $2/3<\alpha=1/\gamma<1$

First of all we realise that the situation with
$2/3<\alpha=1/\gamma<1$ does not give rise to any LVS because the
leading order $\alpha'$ correction would be negligible at large
volume. We shall therefore focus on the case $\alpha<1/\gamma$.
The scalar potential (\ref{finalPotential}) takes the form
\begin{eqnarray}
V &\sim &\frac{1}{\mathcal{V}^{2}}\left[
\sum\limits_{j=1}^{2}K^{-1}_{jj} a_{j}^{2} e^{-2a_{j}\tau_{j}}+2
a_{1}a_{2}g_{2-3\alpha/2}(\tau_{1},\tau_{2})\mathcal{V}^{\alpha}
e^{-(a_{1}\tau_{1}+a_{2}\tau_{2})}\right]  \notag \\
&&-4\frac{W_{0}}{\mathcal{V}^{2}}\left(a_{1}\tau
_{1}e^{-a_{1}\tau_{1}}+a_{2}\tau
_{2}e^{-a_{2}\tau_{2}}\right)+\frac{3}{4}\frac{\hat{\xi}
}{\mathcal{V}^{3}}W_{0}^{2}. \label{finalpotentialnn}
\end{eqnarray}
When we take the large volume limit (II) of (\ref{DIrections}),
(\ref{finalpotentialnn}) has the following volume scaling
\begin{equation}
V \sim \frac{K^{-1}_{11}}{\mathcal{V}^{2+2/\beta}}
+\frac{K^{-1}_{22}}{\mathcal{V}^{2+2/\gamma}}
+\frac{g_{2-3\alpha/2}(\tau_{1},\tau_{2})}
{\mathcal{V}^{2+1/\beta+1/\gamma-\alpha}} -\frac{\tau
_{1}}{\mathcal{V}^{2+1/\beta}}-\frac{\tau
_{2}}{\mathcal{V}^{2+1/\gamma}}+\frac{1}{\mathcal{V}^{3}}.
\label{semplificann}
\end{equation}
Setting $1/\gamma=1$ and recalling that in this direction
$1/\beta>1/\gamma$, the dominant terms in (\ref{semplificann})
become
\begin{equation}
V \sim \frac{K^{-1}_{11}}{\mathcal{V}^{2+2/\beta}}
+\frac{K^{-1}_{22}}{\mathcal{V}^{4}}-\frac{\tau
_{2}}{\mathcal{V}^{3}}+\frac{1}{\mathcal{V}^{3}}. \label{SEMPL}
\end{equation}
Now by noticing that equation (\ref{SEMPL}) has the same form of
(\ref{Semplific}) if we set $1/\gamma=1$, we can just repeat the
same consideration made before and obtain that $K^{-1}_{22}$ has
to be in case (3). Moreover if $K^{-1}_{22}$ depends on both
$\tau_{1}$ and $\tau_{2}$, then there is no LARGE Volume minimum,
but when $K^{-1}_{22}$ depends only on $\tau_{2}$, the first term
in (\ref{SEMPL}) is negligible at large volume and can be used to
fix $\tau_{1}$ if we make it compete with the fourth term in
(\ref{semplificann}) by writing $K^{-1}_{11}\simeq
\mathcal{V}^{\delta}\tau_{1}^{2-3\delta/2}$ and then imposing
$1/\beta=\delta$.

\item[$\bullet$] axion minimum at (b) along direction (III) for
$\alpha<1/\beta$ or $2/3<\alpha=1/\beta<1$

This situation looks like the previous one if we swap $\gamma$
with $\beta$ and $\tau_{1}$ with $\tau_{2}$, therefore we do not
need to discuss this case.

\item[$\bullet$] axion minimum at (b) along direction (I) for
$\alpha<1/\gamma$ or $2/3<\alpha=1/\gamma<1$

In this situation the full scalar potential still looks like
(\ref{finalpotentialnn}), but the volume scaling behaviour of its
terms now reads
\begin{equation}
V \sim \frac{K^{-1}_{11}}{\mathcal{V}^{2+2/\gamma}}
+\frac{K^{-1}_{22}}{\mathcal{V}^{2+2/\gamma}}
+\frac{g_{2-3\alpha/2}(\tau_{1},\tau_{2})}
{\mathcal{V}^{2+2/\gamma-\alpha}} -\frac{\tau
_{1}}{\mathcal{V}^{2+1/\gamma}}-\frac{\tau
_{2}}{\mathcal{V}^{2+1/\gamma}}+\frac{1}{\mathcal{V}^{3}}.
\label{semplificanno}
\end{equation}
For $2/3<\alpha=1/\gamma<1$ the last term in (\ref{semplificanno})
would be subdominant with respect to the fifth one, but we know
that its presence is crucial to find a minimum and so we can
conclude that this case admits no minimum. On the other hand for
$\alpha<1/\gamma$, setting $1/\gamma=1$, the dominant terms in
(\ref{semplificanno}) become
\begin{equation}
V \sim \frac{K^{-1}_{11}}{\mathcal{V}^{4}}
+\frac{K^{-1}_{22}}{\mathcal{V}^{4}}-\frac{\tau
_{1}}{\mathcal{V}^{3}}-\frac{\tau
_{2}}{\mathcal{V}^{3}}+\frac{1}{\mathcal{V}^{3}}. \label{SEMPLI}
\end{equation}
We immediately realise that (\ref{SEMPLI}) is absolutely similar
to (\ref{SEmplifica}). We can therefore repeat exactly the same
analysis and conclude that only if both $K^{-1}_{11}$ and
$K^{-1}_{22}$ is in case (3) we can have a LARGE volume minimum.
Let us finally summarise in the table below what we have found for
this case.
\end{enumerate}

\textbf{Case (2):
$K^{-1}_{12}=\mathcal{V}^{\alpha}g_{2-3\alpha/2}(\tau_{1},\tau_{2})$,
$0<\alpha<1$}

\begin{center}
\begin{tabular}{|c|c|c|c|c|c|c|}
\hline $K_{11}^{-1}$ & $K_{22}^{-1}$ & $\underset{\alpha \leq
1/\gamma }{(I),(b),}$ & $\underset{\alpha \leq 1/\gamma
}{(II),(b),}$  & $\underset{\alpha \geq 1/\gamma }{(II),(c),}$  &
$\underset{\alpha \leq 1/\beta }{(III),(b),}$ & $\underset{\alpha
\geq 1/\beta }{(III),(d),}$  \\ \hline 1 & 1 &
NO & NO & NO & NO & NO \\ \hline 1 & 2 & NO & NO & NO & NO & NO \\
\hline 1 & 3 & NO & NO & NO & NO & NO \\ \hline 1 & 4 & NO & NO &
NO & NO & NO \\ \hline 2 & 1 & NO & NO & NO & NO & NO \\ \hline 2
& 2 & NO & NO & NO & NO & NO \\ \hline 2 & 3 & NO & NO & NO & NO &
NO \\ \hline 2 & 4 & NO & NO & NO & NO & NO \\ \hline 3 & 1 & NO &
NO & NO & NO & NO \\ \hline 3 & 2 & NO & NO & NO & NO & NO \\
\hline 3 & 3 & OK, $\gamma =1$ & NO & NO & NO & NO \\ \hline 3 & 4
& NO & NO & NO & OK, $\beta=1$ ($\ast\ast$) & NO \\ \hline 4 & 1 &
NO & NO & NO & NO & NO
\\ \hline 4 & 2 & NO & NO & NO & NO & NO \\ \hline 4 & 3 & NO &
OK, $\gamma=1$ ($\ast$) & NO & NO & NO \\ \hline 4 & 4 & NO & NO &
NO & NO & NO
\\
\hline
\end{tabular}
\end{center}
($\ast $) $\ K_{22}^{-1}\simeq \mathcal{V}\sqrt{\tau _{2}}$, $%
K_{11}^{-1}\simeq \mathcal{V}^{\delta }\tau _{1}^{2-3\delta /2}$
with $\frac{1}{\beta }=\delta $
\newline
($\ast \ast $) $K_{11}^{-1}\simeq \mathcal{V}\sqrt{\tau _{1}}$, $%
K_{22}^{-1}\simeq \mathcal{V}^{\delta }\tau _{2}^{2-3\delta /2}$
with $\frac{1}{\gamma }=\delta $
\bigskip

\textbf{Case (3):} $K^{-1}_{12}=
\mathcal{V}f_{1/2}(\tau_{1},\tau_{2})$

\bigskip
We shall now consider each particular situation according to the
different possible locations of the axion minimum:
\begin{enumerate}

\item[$\bullet$] axion minimum at (c) along direction (II) for
$\gamma>1$

Thus at the minimum
\begin{equation}
\left\langle V_{AX}\right\rangle =2\left(\left\vert
X_{1}\right\vert-\left\vert Y_{12}\right\vert-\left\vert
X_{2}\right\vert\right). \label{axioneno}
\end{equation}
Therefore the full potential (\ref{finalPotential}) looks like
\begin{eqnarray}
V &\sim &\frac{1}{\mathcal{V}^{2}}\left[
\sum\limits_{j=1}^{2}K^{-1}_{jj} a_{j}^{2} e^{-2a_{j}\tau_{j}}-2
a_{1}a_{2}f_{1/2}(\tau_{1},\tau_{2})\mathcal{V}
e^{-(a_{1}\tau_{1}+a_{2}\tau_{2})}\right]  \notag \\
&&+4\frac{W_{0}}{\mathcal{V}^{2}}\left(a_{1}\tau
_{1}e^{-a_{1}\tau_{1}}-a_{2}\tau
_{2}e^{-a_{2}\tau_{2}}\right)+\frac{3}{4}\frac{\hat{\xi}
}{\mathcal{V}^{3}}W_{0}^{2}. \label{finalpotentialno}
\end{eqnarray}
When we take the large volume limit (II) of (\ref{DIrections}),
(\ref{finalpotentialno}) has the following volume scaling
\begin{equation}
V \sim \frac{K^{-1}_{11}}{\mathcal{V}^{2+2/\beta}}
+\frac{K^{-1}_{22}}{\mathcal{V}^{2+2/\gamma}}
-\frac{f_{1/2}(\tau_{1},\tau_{2})}
{\mathcal{V}^{1+1/\beta+1/\gamma}} +\frac{\tau
_{1}}{\mathcal{V}^{2+1/\beta}}-\frac{\tau
_{2}}{\mathcal{V}^{2+1/\gamma}}+\frac{1}{\mathcal{V}^{3}}.
\label{semplificano}
\end{equation}
Given that $1/\gamma<1$, the leading order $\alpha'$ correction
would be subleading in a Large Volume limit. However, we know that
its presence is crucial to find the minimum and so we conclude
that this case does not present any new Large Volume vacuum.

\item[$\bullet$] axion minimum at (d) along direction (III) for
$\beta>1$

This situation looks like the previous one if we swap $\gamma$
with $\beta$ and $\tau_{1}$ with $\tau_{2}$, therefore we conclude
that no LARGE Volume minimum is present.

\item[$\bullet$] axion minimum at (b) along direction (I) for
$0<\gamma\leq 1$

Thus at the minimum
\begin{equation}
\left\langle V_{AX}\right\rangle =2\left(\left\vert
Y_{12}\right\vert-\left\vert X_{1}\right\vert-\left\vert
X_{2}\right\vert\right),
\end{equation}
and so the full potential (\ref{finalPotential}) becomes (setting
$W_{0}=1$)
\begin{eqnarray}
V &\sim &\frac{1}{\mathcal{V}^{2}}\left[
\sum\limits_{j=1}^{2}K^{-1}_{jj} a_{j}^{2} e^{-2a_{j}\tau_{j}}+2
a_{1}a_{2}f_{1/2}(\tau_{1},\tau_{2})\mathcal{V}
e^{-(a_{1}\tau_{1}+a_{2}\tau_{2})}\right]  \notag \\
&&-\frac{4}{\mathcal{V}^{2}}\left(a_{1}\tau
_{1}e^{-a_{1}\tau_{1}}+a_{2}\tau
_{2}e^{-a_{2}\tau_{2}}\right)+\frac{3}{4}\frac{\hat{\xi}
}{\mathcal{V}^{3}}. \label{finalpOtentialnnl}
\end{eqnarray}
When we take the large volume limit (I) of (\ref{DIrections}),
(\ref{finalpOtentialnnl}) has the following volume scaling
\begin{equation}
V \sim \frac{K^{-1}_{11}}{\mathcal{V}^{2+2/\gamma}}
+\frac{K^{-1}_{22}}{\mathcal{V}^{2+2/\gamma}}
+\frac{f_{1/2}(\tau_{1},\tau_{2})} {\mathcal{V}^{1+2/\gamma}}
-\frac{\tau _{1}}{\mathcal{V}^{2+1/\gamma}}-\frac{\tau
_{2}}{\mathcal{V}^{2+1/\gamma}}+\frac{1}{\mathcal{V}^{3}}.
\label{semplificannolP}
\end{equation}
In the case where $1/\gamma>1$, the leading part of
(\ref{semplificannolP}) at Large Volume takes the form
\begin{equation}
V \sim \frac{K^{-1}_{11}}{\mathcal{V}^{2+2/\gamma}}
+\frac{K^{-1}_{22}}{\mathcal{V}^{2+2/\gamma}}+\frac{1}{\mathcal{V}^{3}}.
\label{SEMPLIlh}
\end{equation}
We have already checked explicitly that this situation does not
present any LARGE Volume minimum, therefore we shall focus only on
the case $1/\gamma=1$, for which the volume scaling
(\ref{semplificannolP}) becomes
\begin{equation}
V \sim \frac{K^{-1}_{11}}{\mathcal{V}^{4}}
+\frac{K^{-1}_{22}}{\mathcal{V}^{4}}
+\frac{f_{1/2}(\tau_{1},\tau_{2})} {\mathcal{V}^{3}} -\frac{\tau
_{1}}{\mathcal{V}^{3}}-\frac{\tau
_{2}}{\mathcal{V}^{3}}+\frac{1}{\mathcal{V}^{3}}. \label{simple}
\end{equation}
We immediately realise that as soon as either $K^{-1}_{11}$ or
$K^{-1}_{22}$ is in case (4) then the $\alpha'$ corrections would
be subleading in a large volume expansion. Thus we can reject this
possibility. Then we are left with three different situations.
Firstly when both $K^{-1}_{11}$ and $K^{-1}_{22}$ are subdominant
with respect to the last three terms in (\ref{simple}) (i.e. each
$K^{-1}_{jj}$ with $j=1,2$ is either in case (1) or (2)) the
scalar potential (\ref{finalpOtentialnnl}) takes the form
\begin{equation}
V=\frac{2}
{\mathcal{V}}a_{1}a_{2}f_{1/2}(\tau_{1},\tau_{2})e^{-(a_{1}\tau_{1}+a_{2}\tau_{2})}
-\frac{4}{\mathcal{V}^{2}}\left(a_{1}\tau_{1}
e^{-a_{1}\tau_{1}}+a_{2}\tau_{2}e^{-a_{2}\tau_{2}}\right)+\frac{3}{4}\frac{\hat{\xi}
}{\mathcal{V}^{3}}. \label{Sp}
\end{equation}
Since we have to find a minimum such that $a_{j}\tau_{j}\gg 1$ for
$j=1,2$, we can work at leading order in a
$\frac{1}{a_{1}\tau_{1}}$ and $\frac{1}{a_{2}\tau_{2}}$ expansion
and obtain that $\frac{\partial^{2}V}{\partial \tau_{1}^{2}}\simeq
0$ due to the presence of just one exponential in $\tau_{1}$ in
both the first and the second term in (\ref{Sp}). In fact, if we
are interested in the dependence of $V$ on just $\tau_{1}$,
(\ref{Sp}) can be rewritten as
\begin{equation}
V=c_{1}e^{-a_{1}\tau_{1}}\left(a_{2}f_{1/2}(\tau_{1},\tau_{2})e^{-a_{2}\tau_{2}}
-\frac{2\tau_{1}}{\mathcal{V}}\right) +c_{2}\equiv
c_{1}e^{-a_{1}\tau_{1}}g(\tau_{1})+c_{2}, \label{Spp}
\end{equation}
where $c_{1}$ and $c_{2}$ are constants and $g(\tau_{1})$ is the
sum of two homogeneous functions in $\tau_{1}$. Therefore at
leading order in a $\frac{1}{a_{1}\tau_{1}}$ expansion, we get:
\begin{equation}
\frac{\partial V}{\partial \tau_{1}}\simeq
-a_{1}c_{1}e^{-a_{1}\tau_{1}}g(\tau_{1})=0\Leftrightarrow
g(\tau_{1})=0,
\end{equation}
which implies
\begin{equation}
\left. \frac{\partial ^{2}V}{\partial \tau _{1}^{2}}\right\vert
_{\min }\simeq \left. a_{1}^{2}c_{1}e^{-a_{1}\tau _{1}}g(\tau
_{1})\right\vert_{\min }=0.
\end{equation}
Similarly we have $\frac{\partial^{2}V}{\partial
\tau_{2}^{2}}\simeq 0$, whereas
\begin{equation}
\left. \frac{\partial ^{2}V}{\partial \tau
_{1}\partial\tau_{2}}\right\vert_{\min }\simeq \left.
2a_{1}^{2}a_{2}^{2}f_{1/2}e^{-a_{1}\tau
_{1}}\frac{e^{-a_{2}\tau_{2}}}{\mathcal{V}}\right\vert_{\min
}\equiv c_{3}>0.
\end{equation}
Therefore considering $\mathcal{V}$ constant, the Hessian matrix
will look like
\begin{equation}
\mathcal{H}\simeq
\begin{pmatrix}
0 & c_{3} \\
c_{3} & 0
\end{pmatrix}
\Longrightarrow \det \mathcal{H}=-c_{3}^{2}<0,
\end{equation}
so implying that we can never have a minimum.

Secondly we have to contemplate the possibility that only one of
the first two terms in (\ref{simple}) is competing with the last
three ones while the other is negligible. However even this case
does not yield any new LVS due the asymmetry of the dependence of
the scalar potential in $\tau_{1}$ and $\tau_{2}$ that does not
allow us to stabilise the small K\"{a}hler moduli large enough. In
fact, let us assume for example that $K^{-1}_{11}$ is in case (1)
or (2) and hence is negligible, whereas
$K^{-1}_{22}\simeq\mathcal{V}\sqrt{\tau_{2}}$. Consequently we
obtain
\begin{eqnarray}
\frac{\partial V}{\partial \tau _{1}} &=&0\Longleftrightarrow
\mathcal{V}=
\frac{2\tau _{1}}{a_{2}f_{1/2}}e^{a_{2}\tau _{2}}, \label{qaz1} \\
\frac{\partial V}{\partial \tau _{2}} &=&0\Longleftrightarrow
\mathcal{V}= \frac{2\tau _{2}}{a_{1}f_{1/2}e^{-a_{1}\tau
_{1}}+a_{2}\sqrt{\tau _{2}} e^{-a_{2}\tau _{2}}}. \label{qaz2}
\end{eqnarray}
Now combining (\ref{qaz1}) with (\ref{qaz2}) we find:
\begin{equation}
a_{1}\tau _{1}f_{1/2}e^{a_{2}\tau_{2}-a_{1}\tau_{1}}+a_{2}\tau_{1}
\sqrt{\tau_{2}}=a_{2}\tau_{2}f_{1/2},
\end{equation}
which evaluated along the direction $a_{1}\tau_{1}\simeq
a_{2}\tau_{2}$ where the axion minimum is located, becomes
\begin{equation}
a_{1}\tau _{1}f_{1/2}+a_{2}\tau _{1}\sqrt{\tau _{2}}\simeq
a_{1}\tau _{1}f_{1/2}\Longleftrightarrow \sqrt{a_{1}a_{2}}\tau
_{1}^{3/2}\simeq 0,
\end{equation}
which is the negative result we mentioned above. Lastly when both
$K^{-1}_{11}$ and $K^{-1}_{22}$ is in case (3) all the terms in
(\ref{simple}) have the same volume scaling. It can be seen that,
regardless of the form of $f_{1/2}$, the LARGE Volume minimum is
always present.

\item[$\bullet$] axion minimum at (b) along direction (II) for
$0<\gamma\leq 1$

In this situation the full scalar potential still looks like
(\ref{finalpOtentialnnl}), but the volume scaling behaviour of its
terms now reads
\begin{equation}
V \sim \frac{K^{-1}_{11}}{\mathcal{V}^{2+2/\beta}}
+\frac{K^{-1}_{22}}{\mathcal{V}^{2+2/\gamma}}
+\frac{f_{1/2}(\tau_{1},\tau_{2})}
{\mathcal{V}^{1+1/\beta+1/\gamma}} -\frac{\tau
_{1}}{\mathcal{V}^{2+1/\beta}}-\frac{\tau
_{2}}{\mathcal{V}^{2+1/\gamma}}+\frac{1}{\mathcal{V}^{3}}.
\label{semplIficannl}
\end{equation}
Given that along the direction (II) $1/\beta>1/\gamma$, and the
axion minimum is present for $1/\gamma\geq 1$, the dominant terms
in (\ref{semplIficannl}) become
\begin{equation}
V \sim \frac{K^{-1}_{11}}{\mathcal{V}^{2+2/\beta}}
+\frac{K^{-1}_{22}}{\mathcal{V}^{4}}-\frac{\tau
_{2}}{\mathcal{V}^{3}}+\frac{1}{\mathcal{V}^{3}}, \label{qwa}
\end{equation}
where we have already set $\gamma=1$ because, as we argued before,
this is the only possible situation when we can hope to find a
LARGE Volume minimum. We notice now that whenever the first two
terms in (\ref{qwa}) are negligible at large volume, then we have
only one term in $V$ dependent on $\tau_{2}$ and so we can never
obtain a minimum at $a_{2}\tau_{2}\gg 1$. This happens if
$K^{-1}_{22}$ is either in case (1) or in case (2) and
$K^{-1}_{11}$ is in case (1), (2), (3) or even in case (4) if its
term in (\ref{qwa}) still goes like $\mathcal{V}^{\alpha}$,
$\alpha>3$. Moreover if $K^{-1}_{22}$ is in case (4) then it would
beat the last two terms in (\ref{qwa}) so giving no LVS.

On the other hand, when only the first term in (\ref{qwa}) is
negligible at large volume, we have the possibility to find a new
LVS if $K^{-1}_{22}$ is in case (3) and does not depend on
$\tau_{1}$, that is $K^{-1}_{22}\simeq
\mathcal{V}\sqrt{\tau_{2}}$. In fact, at leading order in a large
volume expansion the scalar potential looks like the one we
studied for the case with just one small modulus and we know that
the corresponding LARGE Volume minimum would be present. However
we have still to fix $\tau_{1}$. If $K^{-1}_{11}$ is in case (1),
(2) or (3) then the first term in (\ref{semplIficannl}) is always
subleading with respect to the third and the fourth one and
therefore it can be neglected. We can now focus on the third and
fourth term in (\ref{semplIficannl}) which have the same volume
scaling. Then combining the solution of $\frac{\partial
V}{\partial \tau_{1}}=0$ and $\frac{\partial V}{\partial
\tau_{2}}=0$ we end up with $\tau_{1}=\tau_{2}$ which is correct
if we choose $\beta a_{1}=a_{2}$, $\beta<1$. However we have still
to check the sign of $\frac{\partial^{2} V}{\partial
\tau_{1}^{2}}$ which turns out to be positive only if, writing
$f_{1/2}(\tau_{1},\tau_{2})\sim\tau_{1}^{\alpha}\tau_{2}^{1/2-\alpha}$
for arbitrary $\alpha$, we have $\alpha<1$.

The situation when $K^{-1}_{22}\simeq \mathcal{V}\sqrt{\tau_{2}}$
and $K^{-1}_{11}$ is in case (4) needs to be studied more
carefully. Writing
$K^{-1}_{11}\simeq\mathcal{V}^{\delta}\tau_{1}^{2-3\delta/2}$,
$\delta>1$, if $\delta>\frac{2}{\beta}+1$ then the first term in
(\ref{qwa}) beats all the other ones so giving no LVS. If
$\delta=\frac{2}{\beta}+1$, then the first term in (\ref{qwa})
scales as the other ones but we have already shown that this is
not an interesting situation. The only way to get a LVS is to
impose $\delta\leq \frac{1}{\beta}$ to make the first term in
(\ref{semplIficannl}) scale as the third and fourth term or to
render it subdominant with respect to them.

Finally if $K^{-1}$ is in case (4) and $K^{-1}_{22}$ is either in
case (1) or (2) then we don't find any new LVS since the second
term in (\ref{qwa}) would be subleading with respect to the other
ones so leaving just one term, the first one, which depends on
$\tau_{1}$.

\item[$\bullet$] axion minimum at (b) along direction (III) for
$0<\gamma<\beta\leq 1$

This situation looks like the previous one if we swap $\gamma$
with $\beta$ and $\tau_{1}$ with $\tau_{2}$, therefore we do not
need to discuss this case. Let us finally summarise in the table
below what we have found for this case.
\end{enumerate}
\textbf{Case (3):
$K^{-1}_{12}=\mathcal{V}f_{1/2}(\tau_{1},\tau_{2})$}
\begin{center}
\begin{tabular}{|c|c|c|c|c|c|c|}
\hline $K_{11}^{-1}$ & $K_{22}^{-1}$ & $\underset{0<\gamma \leq
1}{(I),(b),}$ & $ \underset{0<\gamma \leq 1}{(II),(b),}$  &
$\underset{\gamma
> 1}{(II),(c),}$
& $\underset{0<\gamma <\beta \leq 1}{(III),(b),}$  &
$\underset{\gamma \geq 1 }{(III),(d),}$  \\ \hline 1 & 1 & NO & NO
& NO & NO & NO \\ \hline 1 & 2 & NO & NO & NO & NO & NO
\\ \hline 1 & 3 & NO & OK, $\beta<\gamma=1$ ($\ast $)
& NO & NO & NO \\ \hline 1 & 4 & NO & NO & NO & NO & NO \\
\hline 2 & 1 & NO & NO & NO & NO & NO
\\ \hline 2 & 2 & NO & NO & NO & NO & NO \\ \hline
2 & 3 & NO & OK, $\beta <\gamma =1$ ($\ast $) & NO & NO & NO \\
\hline 2 & 4 & NO & NO & NO & NO & NO \\ \hline 3 & 1 & NO & NO &
NO & OK,
$\gamma <\beta =1$ ($\star $) & NO \\
\hline 3 & 2 & NO & NO & NO & OK, $\gamma<\beta =1$ ($\star$) & NO
\\ \hline 3 & 3 & OK, $\gamma =1$ & OK, $\beta<\gamma=1$ ($\ast $) & NO & OK,
$\gamma<\beta=1$ ($\star$) & NO
\\ \hline 3 & 4 & NO & NO & NO & OK, $\gamma <\beta =1$  ($\star\star$) & NO \\
\hline 4 & 1 & NO & NO & NO & NO & NO
\\ \hline 4 & 2 & NO & NO & NO & NO & NO \\ \hline 4 & 3 & NO &
OK, $\beta <\gamma=1$ ($\ast\ast$) & NO & NO & NO \\ \hline 4 & 4
& NO & NO &
NO & NO & NO \\
\hline
\end{tabular}
\end{center}

($\ast $) $K_{22}^{-1}\simeq \mathcal{V}\sqrt{\tau _{2}}$,
$f_{1/2}\sim\tau_{1}^{\alpha}\tau_{2}^{1/2-\alpha}$ with
$\alpha<1$

\bigskip
($\ast\ast$) $K_{22}^{-1}\simeq \mathcal{V}\sqrt{\tau_{2}}$,
$K^{-1}_{11}\simeq\mathcal{V}^{\delta}\tau_{1}^{2-3\delta/2}$ and
$f_{1/2}\sim\tau_{1}^{\alpha}\tau_{2}^{1/2-\alpha}$ with
$\alpha<1$ for $1<\delta<\frac{1}{\beta}$ and $\forall\alpha$ for
$\delta=\frac{1}{\beta}$

\bigskip
($\star $) $K_{11}^{-1}\simeq\mathcal{V} \sqrt{\tau_{1}}$,
$f_{1/2}\sim\tau_{2}^{\alpha}\tau_{1}^{1/2-\alpha}$ with
$\alpha<1$

\bigskip
($\star\star$) $K_{11}^{-1}\simeq \mathcal{V}\sqrt{\tau_{1}}$,
$K^{-1}_{22}\simeq\mathcal{V}^{\delta}\tau_{2}^{2-3\delta/2}$ and
$f_{1/2}\sim\tau_{2}^{\alpha}\tau_{1}^{1/2-\alpha}$ with
$\alpha<1$ for $1<\delta<\frac{1}{\gamma}$ and $\forall\alpha$ for
$\delta=\frac{1}{\gamma}$

\bigskip

\textbf{Case (4):} $K^{-1}_{12}\simeq
\mathcal{V}^{\alpha}h_{2-3\alpha/2}(\tau_{1},\tau_{2})$,
$\alpha>1$

\bigskip
Let us focus on each particular situation according to the
different possible positions of the axion minimum:
\begin{enumerate}

\item[$\bullet$] axion minimum at (c) along direction (II) for
$\alpha>1/\gamma$

Thus at the minimum
\begin{equation}
\left\langle V_{AX}\right\rangle =2\left(\left\vert
X_{1}\right\vert-\left\vert Y_{12}\right\vert-\left\vert
X_{2}\right\vert\right). \label{axionenl}
\end{equation}
Therefore the full scalar potential (\ref{finalPotential}) reads
\begin{eqnarray}
V &\sim &\frac{1}{\mathcal{V}^{2}}\left[
\sum\limits_{j=1}^{2}K^{-1}_{jj} a_{j}^{2} e^{-2a_{j}\tau_{j}}-2
a_{1}a_{2}h_{2-3\alpha/2}(\tau_{1},\tau_{2})\mathcal{V}^{\alpha}
e^{-(a_{1}\tau_{1}+a_{2}\tau_{2})}\right]  \notag \\
&&+4\frac{W_{0}}{\mathcal{V}^{2}}\left(a_{1}\tau
_{1}e^{-a_{1}\tau_{1}}-a_{2}\tau
_{2}e^{-a_{2}\tau_{2}}\right)+\frac{3}{4}\frac{\hat{\xi}
}{\mathcal{V}^{3}}W_{0}^{2}. \label{finalpotentialnl}
\end{eqnarray}
When we take the large volume limit (II) of (\ref{DIrections}),
(\ref{finalpotentialnl}) has the following volume scaling
\begin{equation}
V \sim \frac{K^{-1}_{11}}{\mathcal{V}^{2+2/\beta}}
+\frac{K^{-1}_{22}}{\mathcal{V}^{2+2/\gamma}}
-\frac{h_{2-3\alpha/2}(\tau_{1},\tau_{2})}
{\mathcal{V}^{2+1/\beta+1/\gamma-\alpha}} +\frac{\tau
_{1}}{\mathcal{V}^{2+1/\beta}}-\frac{\tau
_{2}}{\mathcal{V}^{2+1/\gamma}}+\frac{1}{\mathcal{V}^{3}}.
\label{semplificanl}
\end{equation}
Recalling that in this direction $1/\beta>1/\gamma$, the dominant
terms in (\ref{semplificanl}) are
\begin{equation}
V \sim \frac{K^{-1}_{11}}{\mathcal{V}^{2+2/\beta}}
+\frac{K^{-1}_{22}}{\mathcal{V}^{2+2/\gamma}}
-\frac{h_{2-3\alpha/2}(\tau_{1},\tau_{2})}
{\mathcal{V}^{2+1/\beta+1/\gamma-\alpha}}-\frac{\tau
_{2}}{\mathcal{V}^{2+1/\gamma}}+\frac{1}{\mathcal{V}^{3}}.
\label{eqproof}
\end{equation}
We know that the presence of the last term in (\ref{eqproof}) is
crucial to find the LARGE Volume minimum, so in order not to make
it subleading with respect to the fourth one, we need to have
$1/\gamma\geq 1$. If $1/\gamma=1$, the new volume scaling looks
like
\begin{equation}
V \sim \frac{K^{-1}_{11}}{\mathcal{V}^{2+2/\beta}}
+\frac{K^{-1}_{22}}{\mathcal{V}^{4}}
-\frac{h_{2-3\alpha/2}(\tau_{1},\tau_{2})}
{\mathcal{V}^{3+1/\beta-\alpha}}-\frac{\tau
_{2}}{\mathcal{V}^{3}}+\frac{1}{\mathcal{V}^{3}}, \label{eqprooff}
\end{equation}
with $1/\beta\geq \alpha$ to keep the last term in
(\ref{eqprooff}). Now setting $1/\beta=\alpha>1$, (\ref{eqprooff})
reduces to
\begin{equation}
V \sim \frac{K^{-1}_{11}}{\mathcal{V}^{2+2\alpha}}
+\frac{K^{-1}_{22}}{\mathcal{V}^{4}}
-\frac{h_{2-3\alpha/2}(\tau_{1},\tau_{2})}
{\mathcal{V}^{3}}-\frac{\tau
_{2}}{\mathcal{V}^{3}}+\frac{1}{\mathcal{V}^{3}}.
\label{eqproofff}
\end{equation}
By studying the expression (\ref{eqproofff}), we realise that
there are only three possible situations in which the necessary
but not sufficient conditions to stabilise $a_{j}\tau_{j}\gg 1$,
$j=1,2$, and not to neglect the leading order $\alpha'$
corrections, can be satisfied. The first one is when $K^{-1}_{22}$
is in case (3) and depends also on $\tau_{1}$:
$K^{-1}_{22}=\mathcal{V}f_{1/2}(\tau_{1},\tau_{2})$. In addition,
the first term in (\ref{eqproofff}) is subleading with respect to
the other ones given that $K^{-1}_{11}$ is in case (1) or (2) or
(3) or even in case (4) but still being subleading. However we can
again show that this case does not lead to any new LVS by noticing
that $\partial V/\partial \tau_{1}=0$ combined with $\partial
V/\partial \tau_{2}=0$ gives rise to a differential equation for
$f_{1/2}$ whose solution is not homogeneous.

The second situations takes place when the second term in
(\ref{eqproofff}) is subleading with respect to the others. This
occurs when $K^{-1}_{22}$ is either in case (1) or (2) and
$K^{-1}_{11}$ is in case (4). Moreover if
$K^{-1}_{11}\simeq\mathcal{V}^{\delta}\tau_{1}^{2-3\delta/2}$ we
have to impose $\delta=2\alpha-1$. However this case would not
work because the minimisation equation $\frac{\partial V}{\partial
\tau_{2}}=0$ would produce a negative volume:
\begin{equation}
\mathcal{V}^{\alpha}=-\frac{2\tau_{2}}{a_{1}h_{2-3\alpha/2}}e^{a_{1}\tau_{1}}.
\end{equation}
Finally we have to contemplate the possibility that all the terms
in (\ref{eqproofff}) have the same volume scaling. This can happen
only if $K^{-1}_{22}\simeq\mathcal{V}\sqrt{\tau_{2}}$ and
$K^{-1}_{22}\simeq\mathcal{V}^{\delta}\tau_{1}^{2-3\delta/2}$ with
$\delta=(2\alpha-1)>1$. Even this case can be seen to produce no
LVS. In fact, it is possible to integrate out the overall volume
from one of the usual minimisation equations, so being left with
two equations in $\tau_{1}$ and $\tau_{2}$. Then given that we
know that we are looking for a minimum located at $\beta
a_{1}\tau_{1}\simeq a_{2}\tau_{2}$, making this substitution, we
end up with two equations in just $\tau_{1}$ which can be seen to
disagree.

On the other hand, for $1/\beta>\alpha$, we would be left with
\begin{equation}
V \sim \frac{K^{-1}_{11}}{\mathcal{V}^{2+2/\beta}}
+\frac{K^{-1}_{22}}{\mathcal{V}^{4}}-\frac{\tau
_{2}}{\mathcal{V}^{3}}+\frac{1}{\mathcal{V}^{3}}. \label{Eeq}
\end{equation}
Now by noticing that equation (\ref{Eeq}) has the same form of
(\ref{Semplific}) if we set $1/\gamma=1$, we can just repeat the
same consideration made before and obtain that $K^{-1}_{22}$ has
to be in case (3). Moreover if $K^{-1}_{22}$ depends on both
$\tau_{1}$ and $\tau_{2}$, then we have no LARGE Volume minimum.
On the other hand, when $K^{-1}_{22}$ depends only on $\tau_{2}$,
the first term in (\ref{Eeq}) is now negligible at large volume
and can be used to fix $\tau_{1}$ if we make it compete with the
third term in (\ref{eqprooff}) by writing $K^{-1}_{11}\simeq
\mathcal{V}^{\delta}\tau_{1}^{2-3\delta/2}$ and then imposing
$1/\beta+\alpha=1+\delta$.

On the contrary, if $1/\gamma>1$, (\ref{eqproof}) takes the form
\begin{equation}
V \sim \frac{K^{-1}_{11}}{\mathcal{V}^{2+2/\beta}}
+\frac{K^{-1}_{22}}{\mathcal{V}^{2+2/\gamma}}
-\frac{h_{2-3\alpha/2}(\tau_{1},\tau_{2})}
{\mathcal{V}^{2+1/\beta+1/\gamma-\alpha}}+\frac{1}{\mathcal{V}^{3}}.
\label{EQproof}
\end{equation}
Now for $(1/\beta+1/\gamma-\alpha)>1$, (\ref{EQproof}) at leading
order in a large volume expansion, reduces to
\begin{equation}
V \sim \frac{K^{-1}_{11}}{\mathcal{V}^{2+2/\beta}}
+\frac{K^{-1}_{22}}{\mathcal{V}^{2+2/\gamma}}
+\frac{1}{\mathcal{V}^{3}},
\end{equation}
which has already been proved to produce no LVS. On the other
hand, for $(1/\beta+1/\gamma-\alpha)<1$, the $\alpha'$ correction
would be negligible at large volume, so forcing us to impose
$(1/\beta+1/\gamma-\alpha)=1$ and obtain:
\begin{equation}
V \sim \frac{K^{-1}_{11}}{\mathcal{V}^{2+2/\beta}}
+\frac{K^{-1}_{22}}{\mathcal{V}^{2+2/\gamma}}
-\frac{h_{2-3\alpha/2}(\tau_{1},\tau_{2})}
{\mathcal{V}^{3}}+\frac{1}{\mathcal{V}^{3}}. \label{EQpf}
\end{equation}
However we can show explicitly that there is no LARGE Volume
minimum. In fact, setting $K^{-1}_{11}\simeq
\mathcal{V}^{\eta}\tau_{1}^{2-3\eta/2}$ with
$\eta=\frac{2}{\beta}-1$, and $K^{-1}_{22}\simeq
\mathcal{V}^{\delta}\tau_{2}^{2-3\delta/2}$ with
$\delta=\frac{2}{\gamma}-1$, and then substituting the solutions
of $\partial V/\partial \tau_{1}=0$ and $\partial V/\partial
\tau_{2}=0$ in $\partial V/\partial \mathcal{V}=0$, one finds that
it is never possible to fix the small K\"{a}hler moduli large
enough to be able to neglect the higher order instanton
contributions to $W$. If $h_{2-3\alpha/2}$ did not depend on both
$\tau_{1}$ and $\tau_{2}$ but just on one of them, this negative
result would not be altered as the term involving
$h_{2-3\alpha/2}$ depends always on both the two small moduli via
the two exponentials $e^{a_{1}\tau_{1}}e^{a_{2}\tau_{2}}$(see
(\ref{finalpotentialnl})).

\item[$\bullet$] axion minimum at (d) along direction (III) for
$\alpha>1/\beta$

This situation looks like the previous one if we swap $\gamma$
with $\beta$ and $\tau_{1}$ with $\tau_{2}$, therefore we do not
need to discuss this case.

\item[$\bullet$] axion minimum at (b) along direction (I) for
$1<\alpha\leq 1/\gamma$

Thus at the minimum
\begin{equation}
\left\langle V_{AX}\right\rangle =2\left(\left\vert
Y_{12}\right\vert-\left\vert X_{1}\right\vert-\left\vert
X_{2}\right\vert\right), \label{axionennl}
\end{equation}
and so the full potential (\ref{finalPotential}) becomes
\begin{eqnarray}
V &\sim &\frac{1}{\mathcal{V}^{2}}\left[
\sum\limits_{j=1}^{2}K^{-1}_{jj} a_{j}^{2} e^{-2a_{j}\tau_{j}}+2
a_{1}a_{2}h_{2-3\alpha/2}(\tau_{1},\tau_{2})\mathcal{V}^{\alpha}
e^{-(a_{1}\tau_{1}+a_{2}\tau_{2})}\right]  \notag \\
&&-4\frac{W_{0}}{\mathcal{V}^{2}}\left(a_{1}\tau
_{1}e^{-a_{1}\tau_{1}}+a_{2}\tau
_{2}e^{-a_{2}\tau_{2}}\right)+\frac{3}{4}\frac{\hat{\xi}
}{\mathcal{V}^{3}}W_{0}^{2}. \label{finalpotentialnnl}
\end{eqnarray}
When we take the large volume limit (I) of (\ref{DIrections}),
(\ref{finalpotentialnnl}) has the following volume scaling
\begin{equation}
V \sim \frac{K^{-1}_{11}}{\mathcal{V}^{2+2/\gamma}}
+\frac{K^{-1}_{22}}{\mathcal{V}^{2+2/\gamma}}
+\frac{h_{2-3\alpha/2}(\tau_{1},\tau_{2})}
{\mathcal{V}^{2+2/\gamma-\alpha}} -\frac{\tau
_{1}}{\mathcal{V}^{2+1/\gamma}}-\frac{\tau
_{2}}{\mathcal{V}^{2+1/\gamma}}+\frac{1}{\mathcal{V}^{3}}.
\label{semplificannol}
\end{equation}
Due to the fact that $1<\alpha\leq 1/\gamma$, the third, the
fourth and the fifth term in (\ref{semplificannol}) are suppressed
with respect to the remaining ones by an appropriate power of the
volume and so we can neglect them. Thus the leading part of
(\ref{semplificannol}) takes the form
\begin{equation}
V \sim \frac{K^{-1}_{11}}{\mathcal{V}^{2+2/\gamma}}
+\frac{K^{-1}_{22}}{\mathcal{V}^{2+2/\gamma}}+\frac{1}{\mathcal{V}^{3}}.
\label{SEMPLIl}
\end{equation}
We have already checked explicitly that this situation does not
present any LARGE Volume minimum.

\item[$\bullet$] axion minimum at (b) along direction (II) for
$1<\alpha\leq 1/\gamma$

In this situation the full scalar potential still looks like
(\ref{finalpotentialnnl}), but the volume scaling behaviour of its
terms now reads
\begin{equation}
V \sim \frac{K^{-1}_{11}}{\mathcal{V}^{2+2/\beta}}
+\frac{K^{-1}_{22}}{\mathcal{V}^{2+2/\gamma}}
+\frac{h_{2-3\alpha/2}(\tau_{1},\tau_{2})}
{\mathcal{V}^{2+1/\beta+1/\gamma-\alpha}} -\frac{\tau
_{1}}{\mathcal{V}^{2+1/\beta}}-\frac{\tau
_{2}}{\mathcal{V}^{2+1/\gamma}}+\frac{1}{\mathcal{V}^{3}}.
\label{semplificannl}
\end{equation}
Given that along the direction (II) $1/\beta>1/\gamma$, and the
axion minimum is present for $1<\alpha\leq 1/\gamma$, the dominant
terms in (\ref{semplificannl}) become
\begin{equation}
V \sim \frac{K^{-1}_{11}}{\mathcal{V}^{2+2/\beta}}
+\frac{K^{-1}_{22}}{\mathcal{V}^{2+2/\gamma}}+\frac{1}{\mathcal{V}^{3}}.
\label{SEMPLl}
\end{equation}
Thus we conclude that this case does not show any LVS, as we have
already showed.

\item[$\bullet$] axion minimum at (b) along direction (III) for
$1<\alpha\leq 1/\beta<1/\gamma$

This situation looks like the previous one if we swap $\gamma$
with $\beta$ and $\tau_{1}$ with $\tau_{2}$, therefore we do not
need to discuss this case. Let us finally summarise in the table
below what we have found for this case.
\end{enumerate}

\textbf{Case (4):
$K^{-1}_{12}=\mathcal{V}^{\alpha}h_{2-3\alpha/2}(\tau_{1},\tau_{2})$,
$\alpha>1$}
\begin{center}
\begin{tabular}{|c|c|c|c|c|c|c|}
\hline $K_{11}^{-1}$ & $K_{22}^{-1}$ & $\underset{1<\alpha \leq
1/\gamma }{(I),(b),} $ & $\underset{1<\alpha \leq 1/\gamma
}{(II),(b),}$ & $\underset{\alpha
> 1/\gamma }{(II),(c),}$ & $\underset{1<\alpha <1/\beta <1/\gamma }{(III),(b),%
}$ & $\underset{\alpha >1/\beta }{(III),(d),}$ \\ \hline 1 & 1 &
NO & NO & NO & NO & NO \\ \hline 1 & 2 & NO & NO & NO & NO & NO \\
\hline 1 & 3 & NO & NO & NO & NO & NO \\ \hline 1 & 4 & NO & NO &
NO & NO & NO \\ \hline 2 & 1 & NO & NO & NO & NO & NO \\ \hline 2
& 2 & NO & NO & NO & NO & NO \\ \hline 2 & 3 & NO & NO & NO & NO &
NO \\ \hline 2 & 4 & NO & NO & NO & NO & NO \\ \hline 3 & 1 & NO &
NO & NO & NO & NO \\ \hline 3 & 2 & NO & NO & NO & NO & NO \\
\hline 3 & 3 & NO & NO & NO & NO & NO \\ \hline 3 & 4 & NO & NO &
NO & NO & OK, $\beta =1$ ($\ast \ast $) \\ \hline 4 & 1 & NO & NO
& NO & NO & NO \\ \hline 4 & 2 & NO & NO & NO & NO & NO \\ \hline
4 & 3 & NO & NO & OK, $\gamma =1$ ($\ast $) & NO & NO \\ \hline 4
& 4 & NO & NO & NO & NO & NO \\ \hline
\end{tabular}
\end{center}
($\ast $) $\ K_{22}^{-1}\simeq \mathcal{V}\sqrt{\tau _{2}}$, $%
K_{11}^{-1}\simeq \mathcal{V}^{\delta }\tau _{1}^{2-3\delta /2}$
with $\frac{1}{\beta}+\alpha=1+\delta$, $\frac{1}{\beta}>\alpha$
\newline
($\ast \ast $) $K_{11}^{-1}\simeq \mathcal{V}\sqrt{\tau _{1}}$, $%
K_{22}^{-1}\simeq \mathcal{V}^{\delta }\tau _{2}^{2-3\delta /2}$
with $\frac{1}{\gamma}+\alpha=1+\delta$, $\frac{1}{\gamma}>\alpha$
\bigskip

Therefore we realise that the positive results represent cases
where all the $N_{small}$ small K\"{a}hler moduli plus a
particular combination, which is the overall volume, are
stabilised. It is then straightforward to see that at this stage
there will be $(h_{1,1}(X)-N_{small}-1)$ flat directions. This
terminates our proof of the LARGE Volume Claim.
\end{proof}

\subsection{General Picture}
\label{genAnal}

We shall try now to draw some conclusions from the previous LARGE
Volume Claim. This can be done by noticing that it is possible to
understand the topological meaning of two of the four cases for
the form of the elements of the inverse K\"{a}hler matrix.

Let us focus on the K\"{a}hler modulus $\tau_{1}$. From the
general expression of the inverse K\"{a}hler matrix for an
arbitrary Calabi-Yau (\ref{inversaAlfa}), we deduce that in this
case, dropping all the coefficients
\begin{equation}
K_{11}^{-1}\simeq-\mathcal{V}k_{11i}t^{i}+\tau _{1}^{2}.
\label{inversa11}
\end{equation}
Case (1) states that $K^{-1}_{11}\simeq \tau_{1}^2$, therefore the
quantity $k_{11i}t^{i}$ has to vanish. This is definitely true if
$k_{11i}=0, \forall i=1,...,h_{1,1}(X)$, that is if the volume is
linear in $t_{1}$, the 2-cycle volume corresponding to $\tau_{1}$.
This is the definition of a three-fold with a K3 fibration
structure over the base $t_{1}$ \cite{oguiso}. Thus we realise
that Calabi-Yau K3 fibrations correspond to case (1). More
precisely if the three-fold is a single fibration only
$K^{-1}_{11}$ will be in case (1) but not $K^{-1}_{22}$. On the
contrary, double K3 fibrations will have both $K^{-1}_{11}$ and
$K^{-1}_{22}$ in case (1). Thus we have proved that

\begin{claim}
$K^{-1}_{11}\sim\tau_{1}^{2}\Leftrightarrow\tau_{1}$ is a K3 fiber
over the base $t_{1}$.
\end{claim}

One could wonder whether this reasoning is correct being worried
about possible field redefinitions since we showed that they can
change the intersection numbers. However this argument is indeed
correct because, as we have explained above, when one restricts
himself to changes of basis which do not alter the form of the
superpotential (\ref{form of W}), the form of the elements of the
inverse K\"{a}hler matrix do not change as the physics depends
only on them and we know that it should not be modified by changes
of basis. Therefore it suffices to calculate $K^{-1}_{11}$ in one
frame where the geometrical interpretation is clear.

The same procedure can be followed to prove that
$K^{-1}_{11}\sim\mathcal{V}\sqrt{\tau_{1}}$ if and only if
$\tau_{1}$ is a blow-up mode resolving a point-like singularity.
The blow-up of a singularity at a point $P$ is obtained by
removing the point $P$ and replacing it with a projective space
like $\mathbb{C}P^{1}$. This procedure introduces an extra
divisor, called \textit{exceptional}, with the corresponding extra
K\"{a}hler modulus that is what we call a blow-up mode. An
exceptional divisor $D_1$ is such that it has only its triple
self-intersection number non-vanishing \cite{fulton}
\begin{equation}
D_1\cdot D_i\cdot D_j\neq 0\textit{ \ only \ if \ }i=j=1.
\end{equation}
\newline
\newline
\begin{figure}[ht]
\begin{center}
\epsfig{file=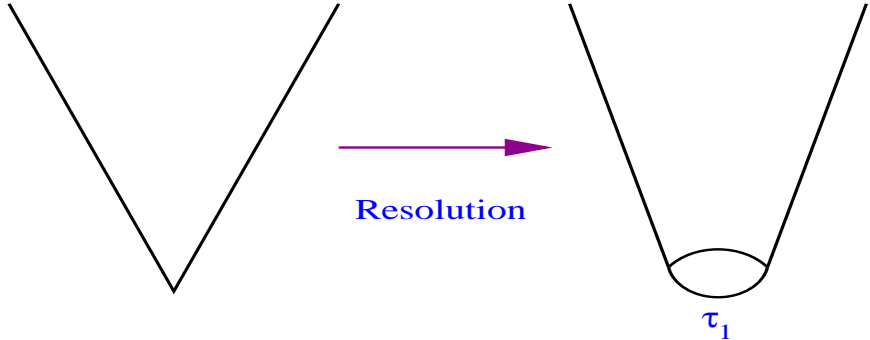, height=45mm,width=115mm}
\caption{Blow-up cycle $\tau_{1}$ resolving a point-like
singularity.}
\end{center}
\end{figure}

Therefore if $\tau_{1}$ is a blow-up then, in a suitable basis, we
can always write the volume as
\begin{equation}
\mathcal{V}=f_{3/2}(\tau_{j})-\tau_{1}^{3/2},\text{ \ }j\neq 1
\label{blow-up}
\end{equation}
where $f_{3/2}(\tau_{j})$ is an homogeneous function of degree
3/2. It is then clear that these blow-up modes are purely
\textit{local} effects, since the change in the volume of the
Calabi-Yau as the blow-up cycle is collapsed, only goes as the
volume of the cycle, with no dependence on the overall volume. In
fact, a change of $\tau_{1}$, $\delta\tau_{1}$, would generate a
change of the total volume of the form
\begin{equation}
\delta \mathcal{V}=\frac{\partial \mathcal{V}}{\partial \tau
_{1}}\delta \tau _{1}=-\frac{3}{2}\sqrt{\tau _{1}}\delta \tau
_{1}. \label{point}
\end{equation}
Let us approximate the volume (\ref{blow-up}) as
$\mathcal{V}\simeq f_{3/2}(\tau_{j})$ and calculate the K\"{a}hler
matrix
\bigskip
\begin{equation}
\frac{\partial K}{\partial \tau _{1}}\simeq \frac{\sqrt{\tau
_{1}}}{\mathcal{ V}}\text{ \ }\Longrightarrow \text{ \ }\left\{
\begin{array}{c}
K_{11}=\frac{\partial ^{2}K}{\partial \tau _{1}^{2}}\simeq
\frac{1}{\sqrt{\tau _{1}}
\mathcal{V}}, \\
K_{1j}=\frac{\partial ^{2}K}{\partial \tau _{j}\partial \tau
_{1}}\simeq -\frac{ \sqrt{\tau
_{1}}}{\mathcal{V}^{2}}\frac{\partial \mathcal{V}}{\partial \tau
_{j}}\simeq o\left( \frac{1}{\mathcal{V}^{5/3}}\right) ,\text{
}j\neq 1.
\end{array}
\right. \label{argument}
\end{equation}
It turns out that $\frac{K_{1j}}{K_{11}}\simeq
o\left(\frac{1}{\mathcal{V}^{2/3}}\right)\ll 1$, $j\neq 1$ and so
we can immediately deduce the leading order term in the "11"
element of the inverse K\"{a}hler matrix by simply taking the
inverse of $K_{11}$, which gives case (3) or more explicitly
$K_{11}^{-1}\simeq\mathcal{V}\sqrt{\tau_{1}}$. But does
$K^{-1}_{11}\sim\mathcal{V}\sqrt{\tau_{1}}$ imply a form of the
volume as $\mathcal{V}=f_{3/2}(\tau_{j})-\tau_{1}^{3/2}$ with
$j\neq 1$? We shall prove now that this is indeed the case. Let us
focus on $N_{small}=1$ without loss of generality. Then
\begin{equation}
K_{ij}\equiv\frac{\partial^{2} K}{\partial
\tau_{i}\partial\tau_{j}}=\frac{2}{\mathcal{V}}\left(\frac{1}{\mathcal{V}}\frac{\partial
\mathcal{V}}{\partial \tau_{i}}\frac{\partial
\mathcal{V}}{\partial \tau_{j}}
-\frac{\partial^{2}\mathcal{V}}{\partial\tau_{i}\partial
\tau_{j}}\right),\text{ \ \ for \ }i,j=1,2. \label{IFFF}
\end{equation}
We can then invert the K\"{a}hler matrix and find (by denoting
$\frac{\partial\mathcal{V}}{\partial\tau_{j}}\equiv\mathcal{V}_{j}$
and similarly for the second derivative)
\begin{equation}
K^{-1}_{11}=\frac{\mathcal{V}}{2}\left[\frac{\frac{1}{\mathcal{V}}\mathcal{V}_{2}^{2}-\mathcal{V}_{22}}{\left(\frac{1}{
\mathcal{V}}\mathcal{V}_{1}^{2}-\mathcal{V}_{11}\right)\left(\frac{1}{
\mathcal{V}}\mathcal{V}_{2}^{2}-\mathcal{V}_{22}\right)-\left(\frac{1}{
\mathcal{V}}\mathcal{V}_{1}\mathcal{V}_{2}-\mathcal{V}_{12}\right)^{2}}\right].
\end{equation}
Now if we impose that at leading order
$K^{-1}_{11}=c_{1}\mathcal{V}\sqrt{\tau_{1}}$ with
$c_{1}\in\mathbb{R}\smallsetminus\{0\}$, we get that at leading
order
\begin{equation}
\left[\frac{\frac{1}{\mathcal{V}}\mathcal{V}_{2}^{2}-\mathcal{V}_{22}}{\left(\frac{1}{
\mathcal{V}}\mathcal{V}_{1}^{2}-\mathcal{V}_{11}\right)\left(\frac{1}{
\mathcal{V}}\mathcal{V}_{2}^{2}-\mathcal{V}_{22}\right)-\left(\frac{1}{
\mathcal{V}}\mathcal{V}_{1}\mathcal{V}_{2}-\mathcal{V}_{12}\right)^{2}}\right]=2c_{1}\sqrt{\tau_{1}}.
\label{IFF}
\end{equation}
Now using the homogeneity property of the volume in terms of the
4-cycle moduli,
$\tau_{1}\mathcal{V}_{1}+\tau_{2}\mathcal{V}_{2}=\frac{3}{2}\mathcal{V}$,
we derive
\begin{equation}
\mathcal{V}_{2}=\frac{3\mathcal{V}}{2\tau
_{2}}-\frac{\mathcal{V}_{1}\tau _{1}}{ \tau _{2}},\text{ \ \ \ \
}\mathcal{V}_{12}=\frac{\mathcal{V}_{1}}{2\tau _{2}
}-\frac{\mathcal{V}_{11}\tau _{1}}{\tau _{2}},\text{ \ \ \ \
}\mathcal{V} _{22}=\frac{3\mathcal{V}-4\mathcal{V}_{1}\tau
_{1}+4\mathcal{V}_{11}\tau _{1}^{2}}{4\tau _{2}^{2}}. \label{IFF1}
\end{equation}
Now plugging (\ref{IFF1}) back in (\ref{IFF}), we find that at
leading order
\begin{equation}
3\mathcal{V}\left(1+2c_{1}\sqrt{\tau_{1}}\mathcal{V}_{11}\right)
-2\mathcal{V}_{1}\left(2\tau_{1}+c_{1}\sqrt{\tau_{1}}\mathcal{V}_{1}\right)=0.
\label{IFF2}
\end{equation}
We can now write the general form of $\mathcal{V}_{1}$ as
$\mathcal{V}_{1}=c_{2}\mathcal{V}^{\alpha}\tau_{1}^{\frac{(1-3\alpha)}{2}}$
with $c_{2}\in\mathbb{R}\smallsetminus\{0\}$ and $\alpha\leq 1$.
It follows then that
\begin{equation}
\mathcal{V}_{11}
=c_{2}\frac{\partial}{\partial\tau_{1}}\left(\mathcal{V}^{\alpha}
\tau_{1}^{\frac{(1-3\alpha)}{2}}\right)=\alpha
c_{2}^{2}\mathcal{V}^{2\alpha-1}\tau_{1}^{1-3\alpha}+c_{2}\frac{(1-3\alpha)}
{2}\mathcal{V}^{\alpha}\tau_{1}^{-\frac{(3\alpha+1)}{2}},
\label{IFF7}
\end{equation}
which for $\alpha<1$, $\alpha\neq 1/3$, at leading order reduces
to
\begin{equation}
\mathcal{V}_{11}=c_{2}\frac{(1-3\alpha)}
{2}\mathcal{V}^{\alpha}\tau_{1}^{-\frac{(3\alpha+1)}{2}},
\label{IFF8}
\end{equation}
while for $\alpha=1/3$ reads
\begin{equation}
\mathcal{V}_{11}=\frac{c_{2}^{2}}{3}\mathcal{V}^{-1/3},
\label{IFF80}
\end{equation}
 and for $\alpha=1$ becomes
\begin{equation}
\mathcal{V}_{11}=c_{2}\frac{\mathcal{V}}{\tau_{1}^{2}}(c_{2}-1).
\label{IFF9}
\end{equation}
Now if $1/2<\alpha\leq 1$, then (\ref{IFF2}) at leading order
looks like
\begin{equation}
3\mathcal{V}_{11}
=c_{2}^{2}\mathcal{V}^{2\alpha-1}\tau_{1}^{(1-3\alpha)}.
\label{IFF4}
\end{equation}
For $1/2<\alpha< 1$, using (\ref{IFF8}), (\ref{IFF4}) gives
\begin{equation}
3c_{2}\frac{(1-3\alpha)}
{2}\mathcal{V}^{\alpha}\tau_{1}^{-\frac{(3\alpha+1)}{2}}
=c_{2}^{2}\mathcal{V}^{2\alpha-1}\tau_{1}^{(1-3\alpha)},
\label{IFF10}
\end{equation}
which at leading order reduces to
\begin{equation}
3c_{2}\frac{(1-3\alpha)}
{2}\mathcal{V}^{\alpha}\tau_{1}^{-\frac{(3\alpha+1)}{2}}=0,
\label{IFF11}
\end{equation}
with the solution $\alpha=1/3$ that is in contradiction with the
fact that we are considering $\alpha>1/2$. For $\alpha=1$, using
(\ref{IFF9}), (\ref{IFF4}) becomes
\begin{equation}
3(c_{2}-1) =c_{2}, \label{IFF12}
\end{equation}
but the solution $c_{2}=3/2\Rightarrow
\tau_{1}\mathcal{V}_{1}=\frac{3}{2}\mathcal{V}$, using the
homogeneity property of the volume,
$\tau_{1}\mathcal{V}_{1}+\tau_{2}\mathcal{V}_{2}=\frac{3}{2}\mathcal{V}$,
would imply $\mathcal{V}_{2}=0$ and so we have to reject it. On
the contrary if $\alpha= 1/2$, then (\ref{IFF2}) at leading order
takes the form
\begin{equation}
6c_{1}\sqrt{\tau_{1}}\mathcal{V}_{11} =(2c_{1}c_{2}^{2}-3),
\label{IFF5}
\end{equation}
and by means of (\ref{IFF8}), this expression at leading order
becomes
\begin{equation}
c_{1}c_{2}\mathcal{V}^{1/2}\tau_{1}^{-\frac{3}{4}} =0,
\label{IFF15}
\end{equation}
that clearly admits no possible solution. Finally if $\alpha<1/2$,
then (\ref{IFF2}) at leading order becomes
\begin{equation}
3\mathcal{V}\left(1+2c_{1}\sqrt{\tau_{1}}\mathcal{V}_{11}\right)=0.
\label{IFF16}
\end{equation}
Due to (\ref{IFF8}), (\ref{IFF16}) for $\alpha\neq 1/3$ reads
\begin{equation}
3\mathcal{V}\left(1+c_{1}c_{2}(1-3\alpha)
\mathcal{V}^{\alpha}\tau_{1}^{-\frac{3\alpha}{2}}\right)=0,
\label{IFF13}
\end{equation}
whereas using (\ref{IFF80}), (\ref{IFF16}) for $\alpha=1/3$ takes
the form
\begin{equation}
0=3\mathcal{V}\left(1+2c_{1}\sqrt{\tau_{1}}\frac{c_{2}^{2}}{3}\mathcal{V}^{-1/3}\right)\simeq
3\mathcal{V}, \label{IFF160}
\end{equation}
which is impossible to solve. Now focusing on (\ref{IFF13}), if
$\alpha<0$ we do not find any solution, whereas if $0<\alpha<1/2$,
(\ref{IFF13}) reduces to
\begin{equation}
3c_{1}c_{2}(1-3\alpha)
\mathcal{V}^{\alpha+1}\tau_{1}^{-\frac{3\alpha}{2}}=0,
\label{IFF130}
\end{equation}
which is solved by $\alpha=1/3$ that is in disagreement with the
fact that we are considering $\alpha\neq 1/3$. Lastly for
$\alpha=0$, (\ref{IFF13}) looks like
\begin{equation}
3\mathcal{V}\left(1+c_{1}c_{2}\right)=0,
\end{equation}
which admits a solution of the form $c_{2}=-\frac{1}{c_{1}}$.
Therefore we have $\mathcal{V}_{1}=-\frac{\sqrt{\tau_{1}}}{c_{1}}$
and $\mathcal{V}_{11}=-\frac{1}{2c_{1}\sqrt{\tau_{1}}}$ that imply
an overall volume of the form
$\mathcal{V}=\lambda_{2}\tau_{2}^{3/2}-\lambda_{1}\tau_{1}^{3/2}$.
It is easy now to generalise this result for $N_{small}>1$ given
that we have shown that

\begin{claim}
$K_{11}^{-1}\sim \mathcal{V}\sqrt{\tau _{1}}\Leftrightarrow
\mathcal{V}=f_{3/2}(\tau _{j})-\tau _{1}^{3/2}$ with $j\neq 1$
$\Leftrightarrow $ $\tau _{1}$ is the only blow-up mode resolving
a point-like singularity.
\end{claim}

We point out also that the stressing that the blow-up has to
resolve a point-like singularity is exactly related to the fact
that it has to be a purely local effect. In fact, the resolution
of a hyperplane or line-like singularity would \textit{not} be a
local effect, even though it would still enable us to take a
sensible large volume limit by sending $\tau_{b}$ large and
keeping $\tau_{1}$ small. In this case, it is plausible to expect
an expression for the overall volume of the form
\begin{equation}
\mathcal{V}=\tau_{b}^{3/2}-\tau_{b}\tau_{1}^{1/2}-\tau_{1}^{3/2}.
\label{linea}
\end{equation}
If we approximate the volume as $\mathcal{V}\simeq
\tau_{b}^{3/2}$, we can see that the change of $\mathcal{V}$ with
the increase of the cycle size $\tau_{1}$ does not depend on
powers of $\tau_{1}$ alone as in (\ref{point}) but it looks like
\begin{equation}
\delta \mathcal{V}=\frac{\partial \mathcal{V}}{\partial \tau
_{1}}\delta \tau _{1}\simeq-\frac{1}{2}\frac{\tau_{b}}{\sqrt{\tau
_{1}}}\delta \tau _{1}\simeq
-\frac{1}{2}\frac{\mathcal{V}^{2/3}}{\sqrt{\tau _{1}}}\delta \tau
_{1}.
\end{equation}
Moreover the case (\ref{linea}) gives rise to an inverse
K\"{a}hler metric of the form $K_{11}^{-1}\simeq
\mathcal{V}^{1/3}\tau_{1}^{3/2}$ which does not satisfy the
condition of the LARGE Volume Claim exactly because $\tau_{1}$ is
not resolving a point-like singularity.

Let us show now that if we have $N_{small}=2$ with one small
modulus $\tau_{2}$ which is a local blow-up mode then the cross
term $K^{-1}_{12}$ has to be in case (1):
$K^{-1}_{12}\sim\tau_{1}\tau_{2}$. Without loss of generality we
can consider just one large modulus $\tau_{3}$ and so the volume
will look like
$\mathcal{V}=f_{3/2}(\tau_{3},\tau_{1})-\tau_{2}^{3/2}$. The
computation of the K\"{a}hler metric gives an expression like
(\ref{IFFF}) but now for $i,j=1,3$ with in addition:
\begin{equation}
K_{22}=\frac{3}{2\mathcal{V}\sqrt{\tau_{2}}},\text{ \ \ \
}K_{2j}=-\frac{3\sqrt{\tau_{2}}}{\mathcal{V}^{2}}\mathcal{V}_{j},\text{
\ with }j=1,3,
\end{equation}
and the "12" element of the inverse K\"{a}hler metric in full
generality reads
\begin{equation}
K^{-1}_{12}=\frac{\tau_{2}\mathcal{V}^{2}\left(\mathcal{V}_{1}\mathcal{V}_{13}
-\mathcal{V}_{1}\mathcal{V}_{33}\right)}
{\left(3\tau_{2}^{3/2}-\mathcal{V}\right)\left(\mathcal{V}_{11}\mathcal{V}_{3}^{2}
+\mathcal{V}_{33}\mathcal{V}_{1}^{2}
-2\mathcal{V}_{1}\mathcal{V}_{3}\mathcal{V}_{13}\right)-\mathcal{V}^{2}\left(\mathcal{V}_{13}^{2}
-\mathcal{V}_{11}\mathcal{V}_{33}\right)}. \label{K-112}
\end{equation}
Using again the homogeneity property of the volume, we can find
the following relations
\begin{gather*}
\mathcal{V}_{3}=\frac{1}{\tau _{3}}\left[ \frac{3}{2}\left(
\mathcal{V}+\tau _{2}^{3/2}\right) -\tau
_{1}\mathcal{V}_{1}\right] ,\text{ \ \ \ \ \ \ \ \ \ \
}\mathcal{V}_{13}=\frac{1}{\tau _{3}}\left[
\frac{\mathcal{V}_{1}}{2}
-\tau _{1}\mathcal{V}_{11}\right] , \\
\mathcal{V}_{33}=\frac{1}{4\tau _{3}^{2}}\left[ 3\left(
\mathcal{V}+\tau _{2}^{3/2}\right) +4\tau _{1}\left( \tau
_{1}\mathcal{V}_{11}-\mathcal{V} _{1}\right) \right] ,
\end{gather*}
which substituted back in (\ref{K-112}) give the final result
\begin{equation}
K^{-1}_{12}=\frac{2\mathcal{V}^{2}\tau_{1}\tau_{2}}{2\mathcal{V}^{2}+6\mathcal{V}\tau_{2}^{3/2}-9\tau_{2}^{3}}\simeq
\tau_{1}\tau_{2}.
\end{equation}
Similarly one can check the correctness of Claim 3 by finding at
leading order $K^{-1}_{22}\sim\mathcal{V}\sqrt{\tau_{2}}$. Let us
summarise this result in the following

\begin{claim}
If $N_{small}=2$ and $K_{22}^{-1}\sim
\mathcal{V}\sqrt{\tau_{2}}\Rightarrow K_{12}^{-1}\sim
\tau_{1}\tau_{2}$.
\end{claim}

We immediately realise that this Claim rules out the possible
LARGE Volume minima along the directions (II) and (III) for the
case (2), (3) and (4). However following arguments similar to the
ones presented to prove Claim 3, one can show that if
$K_{22}^{-1}\sim \mathcal{V}\sqrt{\tau_{2}}$ and so
$K_{12}^{-1}\sim \tau_{1}\tau_{2}$, $K_{11}^{-1}$ can never be in
case (4). Hence also the new would-be LVS along the directions
(II) and (III) for the case (1) have to be rejected because
mathematically inconsistent. Claim 4 also implies that the LVS
along the direction (I) for case (2) and (3) is viable only if
$K^{-1}_{jj}\sim\mathcal{V}h^{(j)}_{1/2}(\tau_{1},\tau_{2})$ with
$\frac{\partial^{2}h^{(j)}}{\partial\tau_{1}\partial\tau_{2}}\neq
0$ $\forall j=1,2$. In fact if
$\frac{\partial^{2}h^{(j)}}{\partial\tau_{1}\partial\tau_{2}}$
were vanishing, then Claim 4 would imply $K^{-1}_{12}$ in case (1)
and not (2) or (3).

In reality we understand these two cases better by realising that
we can go further in our connection of the topological features of
the Calabi-Yau with the elements of $K^{-1}$. In fact, one could
wonder what happens when a singularity is not resolved by just one
blow-up cycle but by several independent local blow-ups. A
concrete example where this happens, is the resolution of the
singularity at the origin of the quotient $\mathbb{C}^2/G$, where
$G$ is a finite subgroup of $SU(2)$ acting linearly on
$\mathbb{C}^2$. This resolution replaces the singularity by
several $\mathbb{C}P^1$'s which correspond to new K\"{a}hler
moduli whose number is determined by the group $G$. For example,
if $G=\mathbb{Z}_n$, one gets $n-1$ such $\mathbb{C}P^1$'s which
play the r\^{o}le of the simple roots of the Lie algebra $A_{n-1}
= su(n)$. After resolving the singularity of $\mathbb{C}^2/G$, one
obtains an example of an ALE space \cite{vafa}.
\newline
\begin{figure}[ht]
\begin{center}
\epsfig{file=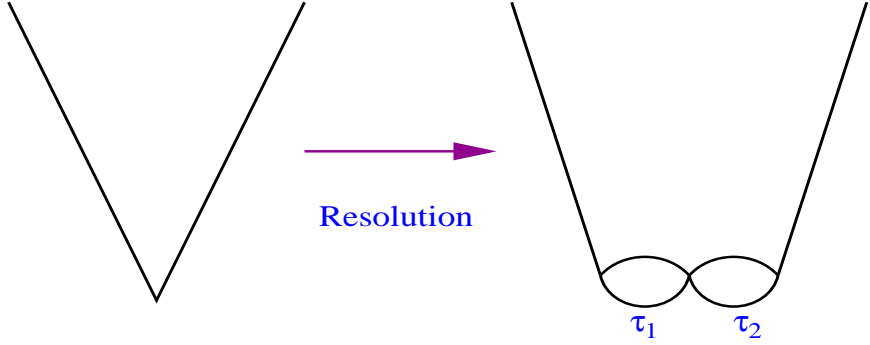, height=45mm,width=115mm}
\caption{Resolution by two independent blow-ups $\tau_{1}$ and
$\tau_{2}$.}
\end{center}
\end{figure}

Focusing on the case $N_{small}=2$, in a suitable basis, the
overall volume will take the general form
\begin{equation}
\mathcal{V}=\tau_{b}^{3/2}-g_{3/2}(\tau_{1},\tau_{2}),
\label{blow-ups}
\end{equation}
with $g_{3/2}(\tau_{1},\tau_{2})\neq
\tau_{1}^{3/2}+\tau_{2}^{3/2}$ since in that special case
$\tau_{1}$ and $\tau_{2}$ would be blow-up cycles resolving two
different point-like singularities. If the total volume is given
by (\ref{blow-ups}) then the scaling with the volume of the
elements of the K\"{a}hler metric is (denoting $\partial
g_{3/2}/\partial \tau_{j}\equiv f_{j}$ and similarly for the
second derivative):
\begin{eqnarray*}
K_{bb} &\sim &\frac{1}{\mathcal{V}\sqrt{\tau _{b}}}+\frac{\tau
_{b}}{ \mathcal{V}^{2}}\sim \frac{1}{\mathcal{V}^{4/3}},\text{ \ \
\ \ }K_{12}\sim
\frac{g_{1}g_{2}}{\mathcal{V}^{2}}+\frac{g_{12}}{\mathcal{V}}\sim
\frac{1}{
\mathcal{V}},\text{\ \ } \\
K_{jb} &\sim &\frac{\sqrt{\tau _{b}}}{\mathcal{V}^{2}}g_{j}\sim
\frac{1}{ \mathcal{V}^{5/3}},\text{ \ \ \ \ \ \ }K_{jj}\sim
\frac{g_{j}^{2}}{\mathcal{V}^{2}}+\frac{g_{jj}}{\mathcal{V}}\sim
\frac{1}{\mathcal{V}},\text{ \ \ }j=1,2
\end{eqnarray*}
therefore producing
\begin{equation}
K_{ij}\sim \left(
\begin{tabular}{ccc}
\cline{1-2} \multicolumn{1}{|c}{$\mathcal{V}^{-1}$} &
$\mathcal{V}^{-1}$ &
\multicolumn{1}{|c}{$\mathcal{V}^{-5/3}$} \\
\multicolumn{1}{|c}{$\mathcal{V}^{-1}$} & $\mathcal{V}^{-1}$ &
\multicolumn{1}{|c}{$\mathcal{V}^{-5/3}$} \\ \cline{1-2}
$\mathcal{V}^{-5/3}$ & $\mathcal{V}^{-5/3}$ & $\mathcal{V}^{-4/3}$
\end{tabular}
\right) \label{DirK}
\end{equation}
where we have highlighted with a box the submatrix with the
leading powers of the volume. We have just to invert this
submatrix to get $K^{-1}_{jj}$ for $j=1,2$ and $K^{-1}_{12}$ which
turn out to be given by
\begin{eqnarray}
K_{11}^{-1} &\sim &\mathcal{V}\left( \frac{g_{22}}{2\Delta
}\right) = \mathcal{V}h_{1/2}^{(1)}(\tau _{1},\tau _{2}),\text{ \
\ \ }K_{22}^{-1}\sim \mathcal{V}\left( \frac{g_{11}}{2\Delta
}\right) =\mathcal{V}
h_{1/2}^{(2)}(\tau _{1},\tau _{2}),  \notag \\
K_{12}^{-1} &\sim &\mathcal{V}\left( \frac{g_{12}}{2\Delta
}\right) = \mathcal{V}f_{1/2}(\tau _{1},\tau _{2}),\text{ \ \ \ \
where }\Delta \equiv g_{11}g_{22}-g_{12}^{2}.  \label{espa}
\end{eqnarray}
Following arguments similar to the ones used to prove Claim 3 we
can also show that starting from
$K^{-1}_{11}\sim\mathcal{V}h^{(1)}_{1/2}(\tau_{1},\tau_{2})$ with
$h^{(1)}$ really dependent on both the small moduli, the form of
the volume has to be (\ref{blow-ups}). A good intuition for this
result is that by setting $\tau_{1}=\tau_{2}$, this is the only
way to recover Claim 3. Therefore we have shown that

\begin{claim}
$K_{11}^{-1}\sim \mathcal{V}h^{(1)}_{1/2}(\tau_{1},\tau_{2})$ and
$K_{22}^{-1}\sim \mathcal{V}h^{(2)}_{1/2}(\tau_{1},\tau_{2})
\Leftrightarrow
\mathcal{V}=f_{3/2}(\tau_{j})-g_{3/2}(\tau_{1},\tau_{2})$ with
$j\neq 1,2$ $\Leftrightarrow $ $\tau_{1}$ and $\tau_{2}$ are two
independent blow-up modes resolving the same point-like
singularity,
\end{claim}

along with

\begin{claim}
If $K_{11}^{-1}\sim \mathcal{V}h^{(1)}_{1/2}(\tau_{1},\tau_{2})$
and $K_{22}^{-1}\sim
\mathcal{V}h^{(2)}_{1/2}(\tau_{1},\tau_{2})\Rightarrow
K_{12}^{-1}\sim \mathcal{V}f_{1/2}(\tau_{1},\tau_{2})$.
\end{claim}

In generality we can conclude that whenever $K^{-1}_{jj}\simeq
\mathcal{V} h^{(j)}_{1/2}(\tau_{1},\tau_{2},...,\tau_{N_{small}})$
then $\tau_{j}$ is a blow-up resolving a point-like singularity.
Moreover, if $h^{(j)}_{1/2}$ depends only on $\tau_{j}$, then
$\tau_{j}$ will be the only blow-up cycle resolving the
singularity and $K^{-1}_{ij}\sim\tau_{i}\tau_{j}$ $\forall i\neq
j=1,...,N_{small}$; on the contrary, if $h^{(j)}_{1/2}$ depends on
several 4-cycle moduli, say $\tau_{j}$ and $\tau_{k}$ for $j\neq
k$, the singularity is resolved by all those independent 4-cycles
with $K^{-1}_{jk}\sim\mathcal{V}f_{1/2}(\tau_{j},\tau_{k})$ and
$K^{-1}_{il}\sim\tau_{i}\tau_{j}$ $\forall i\neq
l=1,...,N_{small}$ for $l=j,k$.

These considerations imply that the would-be LVS along direction
(I) for case (2) is mathematically inconsistent. Thus for
$N_{small}=2$ we are left with just two cases that give rise to a
LARGE Volume minimum located at $\mathcal{V}\sim
e^{a_{1}\tau_{1}}\sim e^{a_{2}\tau_{2}}$:

\begin{enumerate}
\item $K^{-1}_{12}\sim \tau_{1}\tau_{2}$,
$K^{-1}_{jj}\sim\mathcal{V}\sqrt{\tau_{j}}$ $\forall j=1,2$ where
$\tau_{1}$ and $\tau_{2}$ are local blow-up modes resolving two
different point-like singularities;

\item $K^{-1}_{12}\sim \mathcal{V}f_{1/2}(\tau_{1},\tau_{2})$,
$K^{-1}_{jj}\sim\mathcal{V}h^{(j)}_{1/2}(\tau_{1},\tau_{2})$
$\forall j=1,2$ where $\tau_{1}$ and $\tau_{2}$ are two
independent blow-up modes resolving the same point-like
singularity.
\end{enumerate}

The only difference between these two cases is that the first one
works always whereas the second one gives a LVS only if, writing
the volume as
$\mathcal{V}=f_{3/2}(\tau_{j})-g_{3/2}(\tau_{1},\tau_{2})$ with
$j\neq 1,2$, the homogeneous function $g_{3/2}$ is symmetric in
$\tau_{1}$ and $\tau_{2}$. This can be seen easily by comparing
the solution of the two minimisation equations $\frac{\partial
V}{\partial \tau_{1}}=0$ and $\frac{\partial V}{\partial
\tau_{2}}=0$, then substituting the solution we are looking for,
that is $a_{1}\tau_{1}\simeq a_{2}\tau_{2}$, recalling
(\ref{espa}) and lastly finding that we do not get a contradiction
only if $g_{3/2}$ is symmetric.

\end{document}